\documentclass[onecolumn, numberedappendix]{aastex631}
\usepackage{graphicx}%{color}
\usepackage{color}

\newcommand{\av}{$A_V$}

\newcommand{\eg}{{\it e.g.}}
\newcommand{\etal}{et~al.}

\newcommand{\ks}{$K_{\rm s}$}
\newcommand{\vmkz}{$(V-K_{\rm s})_0$}
\newcommand{\vmk}{$(V-K_{\rm s})$}

\newcommand{\mum}{$\mu$m}

\begin{document}

\title{Rotation of Low-Mass Stars in Upper Centaurus Lupus and Lower Centaurus Crux with {\it TESS}}

\reportnum{Version from \today}

\author[0000-0001-6381-515X]{L.~M.~Rebull}
\affiliation{Infrared Science Archive (IRSA), IPAC, 1200 E.\
California Blvd., California Institute of Technology, Pasadena, CA
91125, USA; rebull@ipac.caltech.edu}
\author[0000-0003-3595-7382]{J.~R.~Stauffer}
\affiliation{Spitzer Science Center (SSC), IPAC, 1200 E.\ California
Blvd., California Institute of Technology, Pasadena, CA 91125, USA}
\author{L.~A.~Hillenbrand}
\affiliation{Astronomy Department, California Institute of
Technology, Pasadena, CA 91125, USA}
\author[0000-0002-3656-6706]{A.~M.~Cody}
\affiliation{SETI Institute, 189 Bernardo Avenue, Suite 200, Mountain View, CA 94043, USA}
\author[0000-0002-0493-1342]{Ethan Kruse}
\affiliation{NASA Goddard Space Flight Center, Greenbelt, MD 20771, USA}
\author[0000-0003-0501-2636]{Brian P. Powell}
\affiliation{NASA Goddard Space Flight Center, Greenbelt, MD 20771, USA}

\begin{abstract}
We present stellar rotation rates derived from Transiting
Exoplanet Survey Satellite ({\it TESS}) light curves for stars in
Upper Centaurus-Lupus (UCL; $\sim$136 pc, $\sim$16 Myr) and Lower
Centaurus-Crux (LCC; $\sim$115 pc, $\sim$17 Myr). We find 
spot-modulated periods ($P$) for $\sim$90\% of members. The range
of light curve and periodogram shapes echoes that found for other
clusters with {\it K2}, but fewer multi-period stars may be an
indication of different noise characteristics of {\it TESS}, or a
result of the source selection methods here.  The distribution of $P$
as a function of color as a proxy for mass fits nicely in between that
for both older and younger clusters observed by {\it K2}, with fast
rotators found among both the highest and lowest masses probed here,
and a well-organized distribution of M star rotation rates. About 13\%
of the stars have an infrared (IR) excess, suggesting a circumstellar
disk; this is well-matched to expectations, given the age of the
stars. There is an obvious pile-up of disked M stars at $P\sim$2
days, and the pile-up may move to shorter $P$ as the mass decreases.
There is also a strong concentration of disk-free M stars at $P\sim$2 days,
hinting that perhaps these stars have recently freed themselves from
their disks. Exploring the rotation rates of stars in UCL/LCC has the
potential to help us understand the beginning of the end of the
influence of disks on rotation, and the timescale on which the star
responds to unlocking.
\end{abstract}

\section{Introduction}
\label{sec:intro} %\textcolor{red}{sec:intro}

The Scorpius-Centaurus OB Association is the closest  association to
the Sun including massive star formation, and as such,  is both
important for our studies of young stars, and difficult to study in
its  entirety because it  subtends large angles on the sky, more than
4700 square degrees. In the context of Hipparcos results, de Zeeuw
\etal\ (1999)  summarized the literature starting with Kapteyn (1914).
Blaauw (1964) broke Sco-Cen into three sub-groups, namely Upper
Scorpius or Upper Sco (USco), Upper Centaurus-Lupus (UCL), and Lower
Centaurus-Crux (LCC). de Zeeuw \etal\ (1999) defined boundaries in
Galactic coordinates (see Figure~\ref{fig:where}) for USco
($l\sim343-360\arcdeg$, $b\sim10-30\arcdeg$), LCC
($l\sim285-312\arcdeg$, $b\sim-10-21\arcdeg$), and UCL
($l\sim312-350\arcdeg$, $b\sim0-25\arcdeg$, with a `bite' taken out of
it to allow for USco\footnote{There aren't points belonging to UCL
`hidden' under the USco box in Figure~\ref{fig:where}; the vertices of
the UCL box are (all in $l,b$, in degrees): 350, 0; 312, 0; 312, 25; 343, 25; 343,
10; 350, 10; and 350, 0.}).  It has been difficult to identify members
of these groups not only because they cover  a large swath of sky, but
because this region is in the Galactic plane and close to the Galactic
center, so there is a high surface density of non-member (NM) stars.

In recent years, all-sky surveys as well as instruments that 
can quickly cover large areas of sky have enabled more complete
catalogs of members. Gaia (Gaia Collaboration 2016, 2018), in
particular, has greatly expanded our knowledge of these clusters. 
Many groups have recently used Gaia DR2 (Gaia Collaboration 2018) to
select USco/UCL/LCC  members (e.g., Zari \etal\ 2018; Goldman \etal\
2018; Damiani \etal\ 2019; Kounkel \& Covey 2019; Kerr \etal\ 2021;
and references therein). There are known age and distance
differences among the three subgroups -- USco is at $\sim$143 pc and
$\sim$11 Myr,  UCL is at $\sim$136 pc and $\sim$16 Myr, and  LCC is at
$\sim$115 pc and $\sim$17 Myr (Wright \& Mamajek 2018 for distances
and Pecaut \etal\ 2012 for ages, though the ages are still
controversial and USco may be as young as 3 Myr -- see Rebull \etal\
2018 and references therein). We have taken USco here to be $\sim$8 Myr
based on David \etal\ (2019). There is likely additional substructure
within these three broad groupings (see, \eg, Damiani \etal\ 2019 or
Kerr \etal\ 2021 and references therein), although in the context of
this paper, we have limited the groups under consideration to UCL and
LCC (and USco from our earlier work).

We have recently been working to understand how the rotational
evolution of low-mass stars changes as a function of both age and
mass. {\it K2} (Howell \etal\ 2014) revolutionized our understanding of
this.  We have published thousands of {\it K2} rotation rates from the
Pleiades ($\sim$125 Myr; Rebull \etal\ 2016a,b, Stauffer \etal\ 2016b;
papers I, II, and III, respectively), Praesepe ($\sim$790 Myr; Rebull
\etal\ 2017; paper IV),  USco/$\rho$ Oph ($\sim$8 Myr and $\sim$1 Myr,
respectively; Rebull \etal\ 2018; paper V), and Taurus/Taurus
Foreground ($\sim$3-5 Myr and  $\sim$30 Myr, respectively; Rebull
\etal\ 2020,2021; Paper VI).  Others have also published legions of 
stellar rotation rates using {\it Kepler} and {\it K2} (e.g., 
Rampalli \etal\ 2021, Popinchalk \etal\ 2021, Curtis \etal\ 2020, 
and references therein).

Disk lifetimes are mass (and wavelength)  dependent but are
likely between $\sim$2 and $\sim$20 Myr (see,
e.g., Ribas \etal\ 2014, 2015), with longer-lived disks around
lower-massed stars. USco has a disk fraction of $\sim$10-25\% (see,
e.g., Luhman \& Mamajek 2012) whereas UCL/LCC has a disk fraction of a
few percent (see, e.g.,  Goldman \etal\ 2018). Exploring the rotation
rates of stars in UCL/LCC has the potential to help us understand the
beginning of the end  of the influence of disks on rotation, and  the
timescale on which the star responds to unlocking. Since there are
several recent papers identifying many thousand members of UCL/LCC,
and since NASA's Transiting Exoplanet Survey Satellite  ({\it TESS}; Ricker
\etal\ 2015) is surveying the whole sky for variability,  now is an
opportune time to explore the rotation rates of  low-mass stars in
UCL/LCC to see how they fit in with the rotation rates already known.

We have deliberately performed our analyses of all of our space-based 
light curves (LCs) and supporting data for stars in all of these
clusters, now including UCL/LCC, in as homogeneous a fashion 
as possible in order to best compare the rotation data across 
these clusters.

In preparation for discussing the distribution of rotation rates in
UCL/LCC, we start by assembling the sample of member stars and
defining the member subsamples we use throughout the rest of the paper
in Sec.~\ref{sec:sampledefinition}. The initial sample is constructed
from several papers presenting members, which we then winnow into
four sets of targets: gold (the vast majority of the member sample
in the end, $\sim$80\%), silver, and bronze members, and rejected 
(discarded) targets.  These sets are highlighted in tables and figures
throughout the paper; the gold sample is the best possible sample, and
all the periods ($P$) measured for these stars are high-confidence.
Sec.~\ref{sec:supportingdata} describes in more detail all the
supporting data amassed to enable this analysis, and 
Sec.~\ref{sec:tess} describes the {\it TESS} data, how we  deal with
confusion and contamination, and how we find and interpret periods in
the LCs.  Section~\ref{sec:ucllccwhole} considers the UCL/LCC sample
as a whole and the member subsamples within it.
Sec.~\ref{sec:rotationdistribclusters} presents the distribution of
periods with color as a proxy for mass for UCL/LCC in context with 
other clusters, including specifically considering the stars with
disks. Finally, we summarize in Sec.~\ref{sec:concl}. There
are several Appendices on rejected stars, literature periods, unusual
LCs, and timescales.

\section{Sample Definition}
\label{sec:sampledefinition}%\textcolor{red}{sec:sampledefinition}

We start with a set of literature members (Sec.~\ref{sec:initiallist})
and in the end (Sec.~\ref{sec:finalsample}) have four sets of targets
that are assembled based on membership confidence and reliability of
the linkage between the LC and the star in question: gold (the vast
majority of the member sample, $\sim$80\%), silver, bronze, and 
rejected (discarded) targets. Gold members have  \ks\ photometry (and
either $V$ measured explicitly or an inferred \vmk),  distances
(Bailer-Jones \etal\ 2018) in the right regime ($<$300 pc), and have
no indications of source confusion in their {\it TESS} LC (see
discussion below in Sec.~\ref{sec:confusioncontamination}).  (It also
turns out that the periodic gold members all have highly reliable
periods, though that was not imposed upon the sample.)  The silver
members have all relevant photometry, and only one reason to be
concerned about them,
which could be a distance that may be too far (see Sec.~\ref{sec:distances})
or a suggestion that the
LC might be contaminated (see Sec.~\ref{sec:confusioncontamination}. 
The bronze members have all relevant
photometry and two reasons to be concerned.  Targets that have three or
more reasons to be concerned, or are obviously subject to source confusion,
are rejected.

Because we will be referring to these subsamples throughout the 
paper, we also take the opportunity here to start to put the sample in
context (Sec.~\ref{sec:incontext}) and present all the statistics in
one place -- see  Table~\ref{tab:summarystats} for numbers of stars
and sample fractions.

\subsection{Initial List of Candidate Members} 
\label{sec:initiallist} %\textcolor{red}{sec:initiallist}

As we have for our {\it K2} papers (Papers I-VI), we start with an
expansive, encompassing list ($\sim$5300 stars), assembled from
the literature as follows.

Because UCL and LCC are relatively nearby, many groups have used Gaia
DR2 (Gaia Collaboration 2018) to select members  (e.g., Zari \etal\
2018; Goldman \etal\ 2018; Damiani \etal\ 2019; Kounkel \& Covey
2019).  Pecaut \& Mamajek (2016) add higher-mass probable UCL/LCC
members.   For our analysis,  we chose to merge the member lists from
Zari \etal\ (2018), Damiani \etal\ (2019), and Pecaut \& Mamajek
(2016) to generate our set of UCL and LCC candidates.  Note that 
these papers are based largely on Gaia DR2, not EDR3 
(Gaia Collaboration 2021). These lists
contain many duplicates with each other,  but also many stars that
appear in only one list (see Table~\ref{tab:summarystats}).  For
Pecaut \& Mamajek (2016),  we took all stars from their tables 7 or 9
identified as being part of UCL or LCC; note that the the 
classical boundaries for UCL+LCC from de Zeeuw \etal\ (1999) were
imposed by Pecaut \& Mamajek.  For Zari \etal\ (2018), we 
started with their all-sky pre-main sequence sample and 
took only those within
the de Zeeuw \etal\ classical UCL+LCC boundaries.
Damiani \etal\ (2019) breaks the population in this region 
among several subgroups,
including clusters that are not UCL/LCC.
We took those tagged by them as  ``LCC'' or ``D2b'' (part of LCC) and within the
classical LCC boundaries as members of LCC; for UCL, we took those
tagged ``UCL'' or ``D1'' (part of UCL) and within the classical UCL boundaries as
members of UCL.  While Damiani \etal\ explored a region beyond
the classical boundaries, their figure 8 indicates that within 
these boundaries, we are capturing the overwhelming majority 
of members.
The collection of all of those literature members
results in $\sim$5300 stars thought to be members of UCL/LCC. We
matched this list of stars to numbers from the {\it TESS} Input Catalog
(TIC; Stassun \etal\ 2018, 2019). Not all of the stars in this 
initial member catalog have TIC numbers, and not all of the stars 
with TIC numbers have {\it TESS} LCs (see Sec.~\ref{sec:tess}). In order
for the star to be included in this analysis, however, at minimum, 
there must be a TIC number, and a LC.

\begin{figure}[htb!]
\epsscale{1}
\plotone{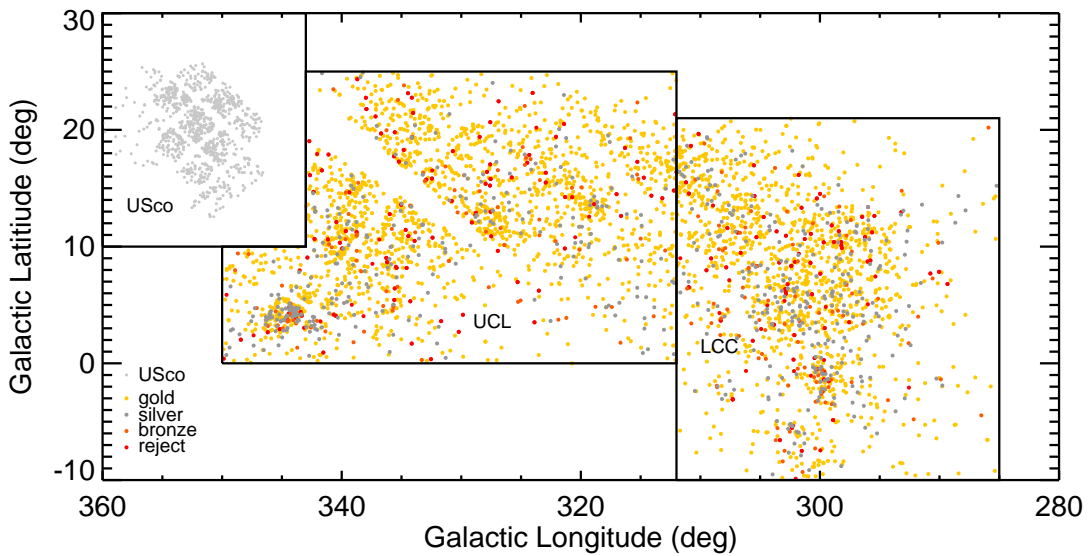}
\caption{Location of targets in Galactic coordinates. The black lines
denote the classical boundaries between USco, UCL, and LCC according
to de Zeeuw \etal\ (1999); see the text. Note that there aren't points
belonging to UCL `hidden' under the USco box. The gold, silver, and
bronze members  (see Sec.~\ref{sec:membership}) are indicated, along
with the rejected (discarded) sources (red). The gaps in the spatial 
distribution are an artifact of {\it TESS}' observation strategy.
Targets from USco with rotation periods (Paper V) are shown in light
grey in the upper left; gaps between the {\it K2} chips are readily
apparent.  Note that these regions are in the Galactic plane and
near the Galactic center, and as a result, given the large pixel size
in {\it TESS},  source confusion is a concern. Note also that the UCL/LCC
member sample is overwhelmingly gold ($\sim$80\%). }
\label{fig:where}
\end{figure}

\subsection{The Sample In Context}
\label{sec:incontext} %\textcolor{red}{sec:incontext}

As part of putting this large sample into context, we have included in
Table~\ref{tab:summarystats} the numbers and sample fractions
originating in the three literature membership studies. Note that
the sample fractions given there refer to the final sample of 4101
sources presented in Sec.~\ref{sec:finalsample}, as opposed to the
$\sim$5300 stars described thus far. Figure~\ref{fig:where} shows the
distribution of our USco stars with rotation periods from {\it K2}
observations (Paper V) in context with the targets from the present
paper in UCL/LCC. The classical boundaries between USco, UCL, and
LCC (de Zeeuw \etal\ 1999) are shown. Note that these regions are in
the Galactic plane and near the Galactic center, and as a result,
source confusion is a concern, especially given the size of {\it TESS}'
pixels (see Sec.~\ref{sec:tess}).  Additionally, IC 2602 is
in the lower right of the box defining LCC, the box defining UCL
includes the Lupus clouds, and the USco box includes $\rho$ Oph in the
lower center (see, e.g., Figure 8 in  Damiani \etal\ 2019); a few
objects from Lupus are incorporated in our member list.  Finally, we
note that a few of the targets in our list appear in the literature as
possible members of TWA, $\eta$ Cha, $\epsilon$ Cha, or USco. 
There are so many objects that are legitimate members of UCL/LCC 
that the very few objects that may not be members of UCL/LCC are
unlikely to make a significant difference in our analysis. 

\begin{figure}[htb!]
\epsscale{1}
\plotone{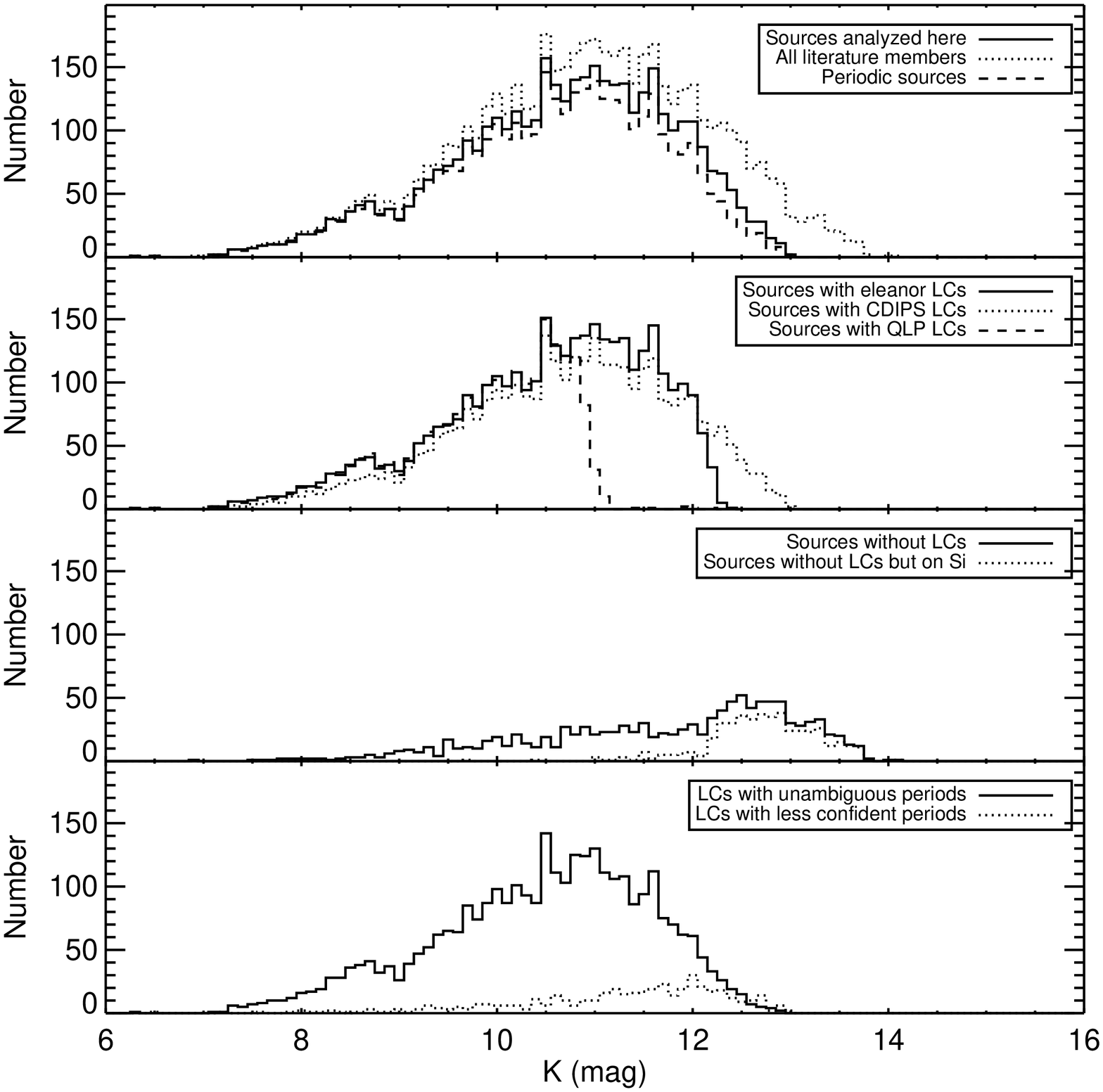}
\caption{Histograms of brightness (in \ks) of targets in  various
subsets of data.  (Note that y-axis range is the same in all panels.)
Top panel: sources analyzed here (solid line) in comparison to all the
literature sources identified as likely members of UCL/LCC (dotted
line) and the sources identified here as periodic (dashed line). 
Sources that are missing LCs are biased towards fainter sources.
Second panel: sources with \texttt{eleanor} LCs (solid line), CDIPS
LCs (dotted line), and QLP LCs (dashed line). QLP LCs are biased
towards  brighter sources, as expected; CDIPS LCs can reach fainter
magnitudes than \texttt{eleanor} LCs.  Third panel: sources without
LCs (solid line) and sources without LCs that could, theoretically,
exist as the target was on silicon (``on Si'')  during {\it TESS}'
first year of operations. Targets that are missing LCs are strongly
biased towards fainter stars. Fourth panel: sources with LCs that have
unambiguous periods (solid line) and LCs where the periods are less
confident (dotted line). Targets with less confident periods are on
average fainter. Based on these histograms, we conclude most
importantly that the selection effect imposed by our requirement that
there be a LC does not unduly bias our results; we are missing some of
the fainter targets, which is unsurprising.}
\label{fig:wherebrightness}
\end{figure}

Figure~\ref{fig:wherebrightness} includes a histogram of \ks\
(see Sec.~\ref{sec:litphotom}) for this entire set of $\sim$5300 
stars, for comparison to other subsets presented below (see, e.g., 
Sec.~\ref{sec:tess}). The 
literature distribution peaks at \ks$\sim$11 or 12, which is {\em very} 
roughly M4-M5. The faintest literature sources are \ks$\sim$13.

\subsection{Final Culled Sample}
\label{sec:membership} %\textcolor{red}{sec:membership}
\label{sec:finalsample} %\textcolor{red}{sec:finalsample}

The details of winnowing the sample are included later in the paper
(Sections~\ref{sec:supportingdata} and \ref{sec:tess}),
but in summary, the gold sample is the best possible member sample;
there is nothing that we can see that stands in the way of
working with these stars in this analysis. To the best of our ability
to determine, the LC goes with the star we think it does, the star
seems to be a legitimate member of UCL or LCC, and the star
has all relevant supporting data. Periodic gold stars all
have periods in which we have high confidence, although this was 
not imposed upon the gold sample.  Of the members, the
vast majority ($\sim$80\%) are gold members. The silver membership
sample has targets that can be worked with in this analysis, but there
is one reason to be concerned that maybe the star is not the best
possible sample; perhaps there is some question as to whether the star
is a  member based on distance, or whether the LC is uncontaminated
such that it really corresponds to the star we think it does (see
Sec.~\ref{sec:confusioncontamination}).  The bronze sample has targets
with two reasons to be concerned. Rejected targets encompass those
that have three or more reasons to be concerned, or incontrovertible
evidence that at least at this time, we can't tie a given LC to that
star, or we are missing $V$ or \ks\ photometry (or a LC) such that it
isn't possible to include that star in subsequent analysis.

In the remainder of this paper, when we use the term ``entire
sample,'' we mean all of the targets with LCs and \vmk.  When we use
the term ``all members,'' we mean the gold+silver+bronze samples
together.

\floattable
\begin{deluxetable}{rccccccc}
\tabletypesize{\scriptsize}
\tablecaption{Summary of statistics on UCL/LCC sample\tablenotemark{a}\label{tab:summarystats}}
\tablewidth{0pt}
\tablehead{\colhead{property} & \colhead{Initial} & \colhead{all {\it TESS}} & \colhead{Gold}
&\colhead{Silver}& \colhead{Bronze}& \colhead{ all}& 
\colhead{ NM or
rejected} \\ [-0.4cm]
\colhead{} & \colhead{literature} & \colhead{sources} & 
\colhead{members} & \colhead{members}  & 
\colhead{members}& \colhead{members}& \colhead{sources} }
\startdata
count&5264&        4101 &         2978 &          542 &          194 &         3714 &          387\\
listed in Zari+ &3170 (0.60)&        2715 (0.66) &        2007 (0.67) &         335 (0.62) &         124 (0.64) &        2466 (0.66) &         249 (0.64)  \\
listed in Damiani+&3887 (0.74)&        2895 (0.71) &        2025 (0.68) &         428 (0.79) &         152 (0.78) &        2605 (0.70) &         290 (0.75)  \\
listed in Pecaut\&Mamajek16&391 (0.10)&         351 (0.09) &         289 (0.10) &          16 (0.03) &           7 (0.04) &         312 (0.08) &          39 (0.10)  \\
\hline
has eleanor LC&&        3627 (0.88) &        2646 (0.89) &         437 (0.81) &         180 (0.93) &        3263 (0.88) &         364 (0.94)  \\
has CDIPS LC&&        3384 (0.83) &        2477 (0.83) &         437 (0.81) &         160 (0.82) &        3074 (0.83) &         310 (0.80)  \\
has QLP LC&&        2173 (0.53) &        1683 (0.57) &         201 (0.37) &          69 (0.36) &        1953 (0.53) &         220 (0.57)  \\
used eleanor LC as best&&        2533 (0.62) &        1789 (0.60) &         329 (0.61) &         135 (0.70) &        2253 (0.61) &         280 (0.72)  \\
used CDIPS LC as best&&        1287 (0.31) &         976 (0.33) &         180 (0.33) &          45 (0.23) &        1201 (0.32) &          86 (0.22)  \\
used QLP LC as best&&         281 (0.07) &         213 (0.07) &          33 (0.06) &          14 (0.07) &         260 (0.07) &          21 (0.05)  \\
\hline
$V$ and \ks\ measured&&        2934 (0.72) &        2249 (0.76) &         334 (0.62) &         118 (0.61) &        2701 (0.73) &         233 (0.60)  \\
\vmk\ via GaiaDR1 $G-K_s$&&         241 (0.06) &         185 (0.06) &          29 (0.05) &          14 (0.07) &         228 (0.06) &          13 (0.03) \\
\vmk\ via GaiaDR2 $G-K_s$&&         430 (0.10) &         242 (0.08) &         105 (0.19) &          29 (0.15) &         376 (0.10) &          54 (0.14)  \\
SED-interpolated $V$&&         468 (0.11) &         302 (0.10) &          74 (0.14) &          33 (0.17) &         409 (0.11) &          59 (0.15)  \\
\av\ from $JHK_s$&&        1657 (0.40) &        1176 (0.39) &         255 (0.47) &          78 (0.40) &        1509 (0.41) &         148 (0.38)  \\
\av\ from SpTy&&          97 (0.02) &          76 (0.03) &           3 (0.01) &           1 (0.01) &          80 (0.02) &          17 (0.04)  \\
median \av\ taken&&        2318 (0.57) &        1726 (0.58) &         284 (0.52) &         115 (0.59) &        2125 (0.57) &         193 (0.50)  \\
\hline
has detected IRAC-1 &&         299 (0.07) &         190 (0.06) &          62 (0.11) &          18 (0.09) &         270 (0.07) &          29 (0.07)  \\
has detected WISE-1 &&        3904 (0.95) &        2940 (0.99) &         500 (0.92) &         177 (0.91) &        3617 (0.97) &         287 (0.74)  \\
has detected WISE-3 &&        3742 (0.91) &        2846 (0.96) &         458 (0.85) &         161 (0.83) &        3465 (0.93) &         277 (0.72)  \\
has detected WISE-4 &&        1146 (0.28) &         873 (0.29) &         109 (0.20) &          44 (0.23) &        1026 (0.28) &         120 (0.31)  \\
has clear IR excess&&         354 (0.09) &         242 (0.08) &          61 (0.11) &          20 (0.10) &         323 (0.09) &          31 (0.08)  \\
has possible IR excess&&         189 (0.05) &         135 (0.05) &          26 (0.05) &          12 (0.06) &         173 (0.05) &          16 (0.04)  \\
has any IR excess&&         543 (0.13) &         377 (0.13) &          87 (0.16) &          32 (0.16) &         496 (0.13) &          47 (0.12) \\
\hline
obvious source confusion&&         367 (0.09) &           0 (0.00) &           0 (0.00) &           0 (0.00) &           0 (0.00) &         367 (0.95)  \\
distance$>$300 pc&&         128 (0.03) &           0 (0.00) &          85 (0.16) &          25 (0.13) &         110 (0.03) &          18 (0.05)  \\
TIC contamination $>$1.6&&         434 (0.11) &           0 (0.00) &         160 (0.30) &         178 (0.92) &         338 (0.09) &          96 (0.25)  \\
mean flux suggests contam&&         589 (0.14) &           0 (0.00) &         290 (0.54) &         185 (0.95) &         475 (0.13) &         114 (0.29)  \\
\hline
periodic&&        3640 (0.89) &        2693 (0.90) &         440 (0.81) &         145 (0.75) &        3278 (0.88) &         362 (0.94)  \\
single $P$&&         555 (0.14) &        2326 (0.78) &         364 (0.67) &         126 (0.65) &        2816 (0.76) &         269 (0.70)  \\
multiple $P$&&         555 (0.14) &         367 (0.12) &          76 (0.14) &          19 (0.10) &         462 (0.12) &          93 (0.24)  \\
periodic+clear IRx&&         299 (0.07) &         204 (0.07) &          51 (0.09) &          15 (0.08) &         270 (0.07) &          29 (0.07)  \\
multiperiodic+clear IRx&&          42 (0.01) &          28 (0.01) &           7 (0.01) &           1 (0.01) &          36 (0.01) &           6 (0.02)  \\
burster&&           2 (0.00) &           2 (0.00) &           0 (0.00) &           0 (0.00) &           2 (0.00) &           0 (0.00)  \\
dipper&&          56 (0.01) &          39 (0.01) &           7 (0.01) &           0 (0.00) &          46 (0.01) &          10 (0.03)  \\
dipper+clear IRx&&          49 (0.01) &          39 (0.01) &           5 (0.01) &           0 (0.00) &          44 (0.01) &           5 (0.01)  \\
dipper+clear IRx+periodic&&          48 (0.01) &          38 (0.01) &           5 (0.01) &           0 (0.00) &          43 (0.01) &           5 (0.01) \\
\hline
double-dip&&         400 (0.10) &         318 (0.11) &          24 (0.04) &           9 (0.05) &         351 (0.09) &          49 (0.13) \\
moving double-dip&&          28 (0.01) &          20 (0.01) &           0 (0.00) &           2 (0.01) &          22 (0.01) &           6 (0.02)  \\
shape changer&&         205 (0.05) &         132 (0.04) &          29 (0.05) &           7 (0.04) &         168 (0.05) &          37 (0.10)  \\
scallop/clouds?\tablenotemark{b}&&          99 (0.02) &          75 (0.03) &           8 (0.01) &           2 (0.01) &          85 (0.02) &          14 (0.04)  \\
beater&&          95 (0.02) &          60 (0.02) &           4 (0.01) &           1 (0.01) &          65 (0.02) &          30 (0.08)  \\
complex peak&&          15 (0.00) &          10 (0.00) &           3 (0.01) &           1 (0.01) &          14 (0.00) &           1 (0.00)  \\
resolved, close peaks&&         264 (0.06) &         170 (0.06) &          35 (0.06) &          11 (0.06) &         216 (0.06) &          48 (0.12)\\
resolved, distant peaks&&         364 (0.09) &         231 (0.08) &          54 (0.10) &          14 (0.07) &         299 (0.08) &          65 (0.17) \\
pulsator&&          12 (0.00) &           6 (0.00) &           5 (0.01) &           1 (0.01) &          12 (0.00) &           0 (0.00)  \\
\enddata
\tablenotetext{a}{Numbers in table are raw number of stars meeting
the stated criterion/criteria, followed by the sample fraction within 
the column in parentheses. For example, 68\% of the gold member sample 
appears as members in Zari \etal\ (2018); 69\% of the entire member
sample appears in Damiani \etal\ (2019). }
\tablenotetext{b}{This category includes the scallop shell, persistent flux dip,
and transient flux dip categories; see Papers I-VI and Stauffer \etal\ (2021).}
\end{deluxetable}

%\clearpage

\section{Supporting Data}
\label{sec:supportingdata} %\textcolor{red}{sec:supportingdata}

In this section, we amass supporting data from the literature,
including photometry (Sec.~\ref{sec:litphotom}) and spectral types
(Sec.~\ref{sec:spty}). As for our earlier papers, we wish to use
\vmkz\ as a proxy for mass, and we continue to do that here, largely
using the same approach.  The main reason we want to use \vmkz\ is to
enable comparisons with our other clusters; even though Gaia data are
available for essentially all of the entire UCL/LCC sample, Gaia data are not
available for all of the stars in the other clusters in Papers I-VI.
We describe how we deredden our photometry (where relevant) in
Sec.~\ref{sec:dereddening}.  We identify IR excesses which we
interpret as circumstellar disks (Sec.~\ref{sec:disks}). Finally, we
describe the distances we used here (Sec.~\ref{sec:distances}).

\subsection{Literature and Derived Photometry}
\label{sec:litphotom} %\textcolor{red}{sec:litphotom}

Based on coordinates from Gaia DR2, for each of our target stars,  we
obtained corresponding near and mid-IR photometry from 2MASS
(Skrutskie \etal\ 2003, 2006), DENIS (Epchtein  \etal\ 1999, 
DENIS team 1999), WISE/AllWISE
(Wright \etal\ 2010ab),  CatWISE (Eisenhardt \etal\ 2020, 
CatWISE team 2020), unWISE
(Meisner \etal\ 2019), Spitzer (Werner \etal\ 2004)
SEIP\footnote{http://irsa.ipac.caltech.edu/data/SPITZER/Enhanced/SEIP/overview.html} 
(Capak 2013), 
and AKARI (Murakami \etal\ 2007, AKARI team 2010ab).  We obtained optical broadband
photometry from Gaia DR1 (Gaia Collaboration 2016ab) and DR2 (Gaia
Collaboration 2018ab), Pan-STARRS DR1 (Chambers \etal\ 2016), APASS
(Henden \etal\ 2016), NOMAD (Zacharias \etal\ 2005), the Southern
Proper Motion Program (SPM; Girard \etal\ 2011), and the GSC-II
(Lasker \etal\ 2008). Pecaut \& Mamajek (2016) also provide optical
magnitudes.  We used a typical source matching radius of one
arcsecond.

In part as a check on source merging across catalogs,  we used all of
the photometry to generate a spectral energy distribution (SED) for
each source.  If the photometric data from one catalog were obviously 
inconsistent with the rest of the SED, then we removed the data points 
from that catalog for that source on the assumption that the positional match
failed. We also used the IRSA Finder Chart tool (as well as IRSA Viewer) 
to investigate such source mismatches. 

As part of this process, we investigated the WISE and Spitzer (if
relevant) images of the target. The WISE objects that were identified
as detections but which did not appear obvious in the image and whose
detections were not consistent with the rest of the SED were changed
to be upper limits. If Spitzer data existed and the source was visible, but 
photometry from SEIP was not available, we performed standard aperture
photometry on the SEIP mosaics. About 10\% of the entire
list of sources were missing photometry 
and were patched in this fashion. 

We need \vmk\ for our target stars to use as a proxy for mass when 
comparing to other clusters from our Papers I-VI. Essentially all of the stars in
our entire sample have measured \ks\ (see Fig.~\ref{fig:wherebrightness}), so
we need to find or calculate $V$. A substantial fraction of the
targets have measured $V$ magnitudes in the literature. In cases where
$V$ magnitudes were obtainable from  more than one place (APASS,
NOMAD, GSC-II, and/or Pecaut \& Mamajek 2016), we calculated an
average $V$ (see Table~\ref{tab:summarystats}).  Other methods we have
used in our other papers  (\vmk\ from Gaia DR1 or $V$ from
interpolation of the  assembled SED) were necessary in some targets. A
few targets missing data from Gaia DR1 but having DR2 necessitated
that we derive a new relationship between \vmk\ and Gaia
DR2\footnote{Where $x=G-$\ks,  $V-$\ks=  $-1.383 +  3.554x +
(-1.442)x^2 + 0.347x^3 +(-0.026)x^4$.  This works well enough for
$G-$\ks$<$5 and $V-$\ks$<$7.}. Nearly all of the targets thus have a
measured or inferred \vmk; see Table~\ref{tab:summarystats}.  
However, specifically because we have had to infer \vmk\ values
in many cases, it likely contributes scatter to the distribution.

Stars missing \ks\ or \vmk\ cannot be analyzed in the same way as the
rest of the entire sample, so those stars are effectively rejected.
Most of those missing \ks\ are blends with nearby stars,
so the {\it TESS} LC is certain to be subject to source confusion anyway
(see Sec.~\ref{sec:tess}).

Table~\ref{tab:bigperiods} includes, for members, the relevant 
supporting photometric data, including the observed or interpolated
\vmk, and the IR excess assessments (from Section~\ref{sec:disks}), 
plus the periods we derive (from Section~\ref{sec:tess}).  A similar
table with all the stars we had to reject appears in
Appendix~\ref{app:nm}.

\subsection{Spectral Types}
\label{sec:spty} %\textcolor{red}{sec:spty}

We obtained spectral types from 
Comer\'on \etal\ (2009),
Skiff (2014), 
Jang-Condell \etal\ (2015), 
Galli \etal\ (2015),
Mellon \etal\ (2017),
Faherty \etal\ (2018), 
Goldman \etal\ (2018)
Nicholson \etal\ (2018),
Bowler \etal\ (2019), 
Cruzal\`ebes \etal\ (2019),
Moolekamp \etal\ (2019; largely photometric types), 
and Luhman (2022a);
finally, if no other types were available, we also consulted 
SIMBAD\footnote{http://simbad.u-strasbg.fr/simbad/} 
for any additional spectral types. 
In the end, we have spectral types for less than 20\% of the member sample.
We have few stars with types earlier than G0, and few later than M6.

\subsection{Dereddening}
\label{sec:dereddening} %\textcolor{red}{sec:dereddening}

Neither UCL nor LCC have substantial or patchy reddening (see, e.g.,
Pecaut \etal\ 2012, Mellon \etal\ 2017).  As in our earlier papers, we
placed the stars on a $J-H$ vs.\ $H-K_s$ diagram, then
shifted them back along the reddening law derived by Indebetouw \etal\
(2008) to the expected $JHK_s$ colors for young stars from Pecaut \&
Mamajek (2013) or the T~Tauri locus from Meyer \etal\ (1997), and
converted to $E(V-K_s)$ via $A_K = 0.114 A_V$ (Cardelli \etal\
1989).  Using this approach, nearly 80\% of the member sample has
essentially no reddening. 

\begin{figure}[htb!]
\epsscale{1.}
\plotone{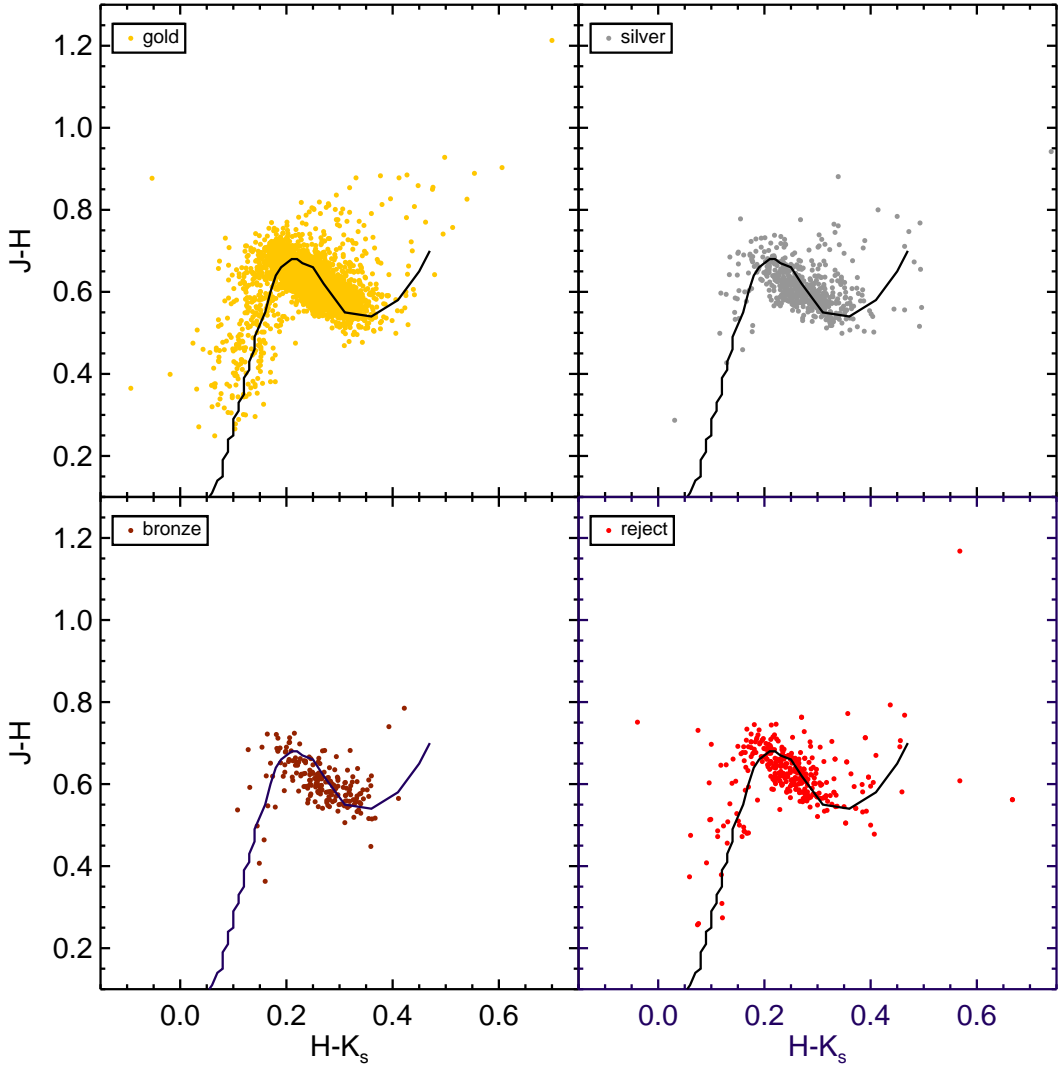}
\caption{$J-H$ vs.~$H-K_s$ (observed)  for: upper left: gold members
(see Sec.~\ref{sec:membership} for membership);   upper right: silver
members; lower left: bronze members; lower right: the rejected
targets.  These plots show both that there is not  much reddening on
average towards these targets (Sec.~\ref{sec:dereddening}),  and
moreover that the stars that are  being dropped are not particularly
or obviously different than the ones that are retained. }
\label{fig:jhk}
\end{figure}

Figure~\ref{fig:jhk} shows a $JHK_s$  color-color diagram for the
sample, demonstrating that there is not much reddening for most
targets, and for that matter, few optically  thick disks in the
near-IR (also see Sec.~\ref{sec:disks}). (It also shows that all the
gold/silver/bronze subsamples are largely similar;  we have not
preferentially discarded red, or reddened, stars, except for those too
faint to have {\it TESS} LCs.)

We can also start with the spectral type (Sec.~\ref{sec:spty}) and
compare the observed colors to those expected for that type from
Pecaut \& Mamajek (2013), from which the reddening can be derived.
This approach is less effective here in UCL/LCC because there are so
few spectral types known; even among those,
there is not much reddening, in general, towards these stars.

We determined or assigned a reddening of 0 to more than half the
stars; we took the reddening derived from the $JHK_s$ colors in about
40\% of the entire sample (see Table~\ref{tab:summarystats}).  Values of
\vmk\ are included in Table~\ref{tab:bigperiods} for the members and
in Appendix~\ref{app:nm} for the rejected sources. However, as in our
earlier papers, to emphasize the net uncertainty, the ``vmk0'' column
in Table~\ref{tab:bigperiods} has been rounded to the nearest 0.1 mag.
The values used in plots here can be recovered by using the $E(V-K_s)$
(``ev-k'') and  $(V-K_s)_{\rm observed}$ (``vk-used'') columns. 
Table~\ref{tab:bigperiods} (and its analogous Table~\ref{tab:bignm}
for NM) include a 2-digit code indicating the origin of the \vmk\
value and the method by which the \vmk\ was dereddened to \vmkz\ (see
Table~\ref{tab:bigperiods} or \ref{tab:bignm} for specific
definitions). 

%\clearpage

\subsection{IR excesses}
\label{sec:disks} %\textcolor{red}{sec:disks}

In the process of SED inspection above (Section~\ref{sec:litphotom}),
we determined which stars were likely to have unambiguous IR excesses
(and therefore likely circumstellar disks in the case of young stars),
and noted the wavelength at which the IR excess begins. We noted those
with obvious IR excesses (``high confidence'') separately from those
where an IR excess might be present (``low confidence'' or
``possible''; Table~\ref{tab:bigperiods}). A high  confidence IR
excess might be one where the IR excess is large and detected at more
than one wavelength, and/or by more than one instrument; a
low-confidence IR excess might be one where the excess is in  only one
band and where $\chi$=(IR color$_{\rm observed}$ $-$ IR  color$_{\rm
expected}$) / (error in IR color) is between 3 and 5. Our source
inspection process (Sec.~\ref{sec:litphotom}) 
ensures that all W3 and W4 detections are secure, so an
excess that appears solely at W3 might indeed be real.  Just a few
percent of the targets have high-confidence IR excesses, and fewer
have low-confidence IR excesses (see Table~\ref{tab:summarystats}). 
This is consistent with prior
determinations of disk fractions in this region (see, e.g., Goldman
\etal\ 2018). Checking against the literature (e.g., Luhman 2022b,
Cotten \& Song 2016, Mittal \etal\ 2015, Chen \etal\ 2014, Carpenter
\etal\ 2009) suggests that our selection mechanism has found the
obvious disks, but has indeed missed debris disks with very small IR
excesses, which is expected given our approach. To securely and
systematically identify  subtle excesses would require spectral types
and modeling beyond  the scope of the present paper.

\begin{figure}[htb!]
\epsscale{1.}
\plotone{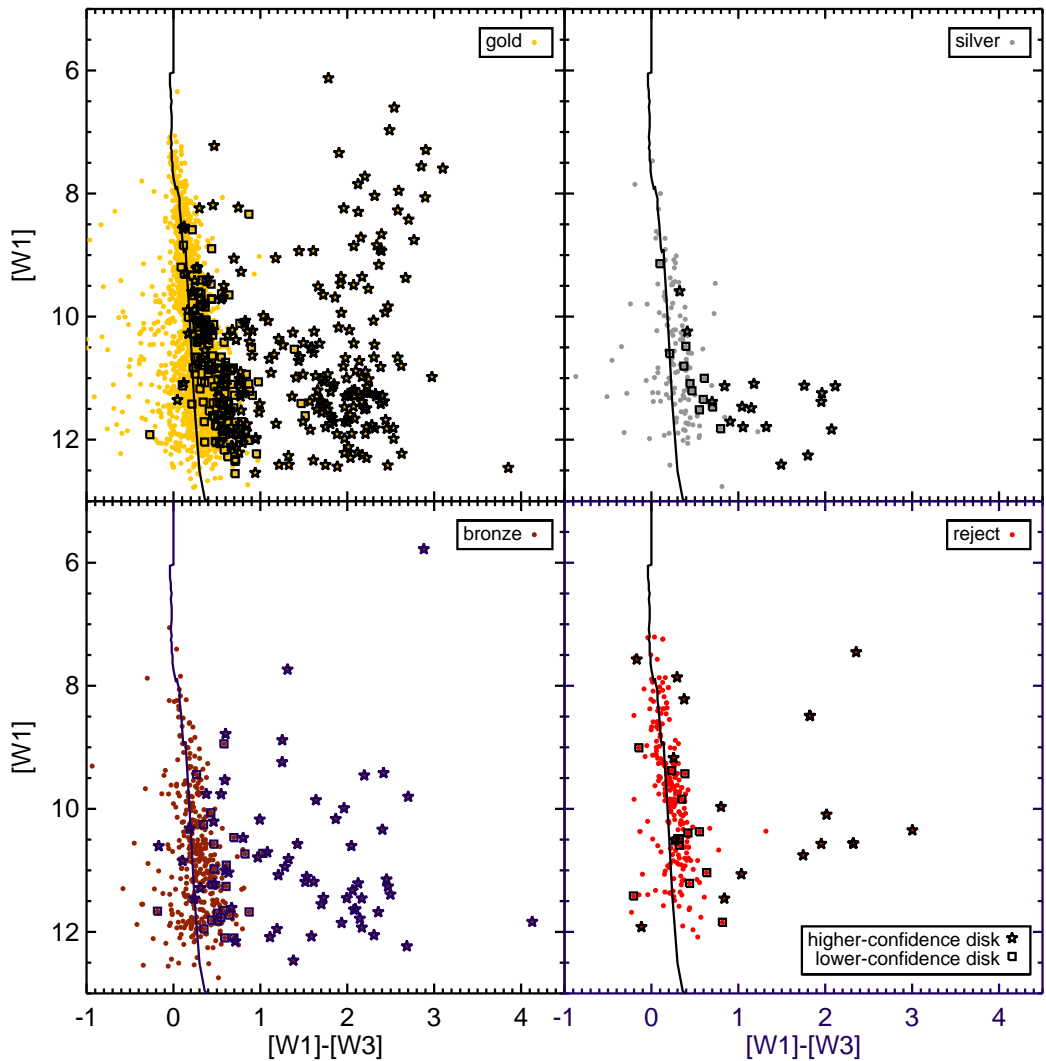}
\caption{[W1] vs.~[W1]$-$[W3] for the gold, silver, bronze, and
rejected targets (see Sec.~\ref{sec:membership}).  High-confidence
disks have an additional star and lower-confidence disks have an
additional square. The black line is the expected disk-free colors
from Pecaut \& Mamajek (2013).  The points that are not identified as
disks but still appear in this plot to have significant [W1]$-$[W3]
are stars for which the only indication  of an excess is a marginal
[W3] detection with a larger-than-typical error bar -- that is, they
do not have a significant excess. Conversely, stars with [W1]$-$[W3]
near 0 but are selected as disks are those that have an IR excess at
wavelengths $>$12 \mum. Most of the stars with disks are in the gold
sample. Disked stars are not preferentially being dropped from the
member sample.}
\label{fig:disks2}
\end{figure}

Spitzer data do not cover the whole region, but WISE data do.  WISE
detections, even of photospheres,  at 12 \mum\ are common at this
distance, but 22 \mum\ limits are also common. Detections from other
IR surveys at $\geq$5 \mum\ are relatively rare. 
Figure~\ref{fig:disks2} shows the [W1] vs.\ [W1]$-$[W3] for the stars
discussed here. Most of the excesses obvious by eye from this plot 
are selected as high- or low-confidence excesses. The points that are
not  identified as disks but still appear in this plot to have
significant [W1]$-$[W3]  are stars for which the only indication of an
excess is a marginal  [W3] detection with a larger-than-typical error
bar -- that is,  they do not have a significant excess.

Fig.~\ref{fig:disks2} and Table~\ref{tab:summarystats} also
show that the gold sample has most of the stars with disks
(of the stars with unambiguous disks, 68\% are in the 
gold sample); stars with
disks are apparently less likely to be at the wrong distance (which
makes sense, since  young stars in this part of the sky with the right
photospheric brightness are more likely than field stars to have IR
excesses) or have source confusion issues. We have not preferentially
discarded disks; if anything, we seem to have have preferentially
retained them.

However, our approach for identifying disks is unambiguously biased
towards the sources with large excesses. Many of the sources carefully
studied with Spitzer (e.g., Chen \etal\ 2014) were known well before
Gaia to be UCL/LCC members (or simply young), and therefore have much
more supporting data in the literature than more recently identified
members. The larger net that we have cast has selected stars that 
for the most part do not have spectral types (Sec.~\ref{sec:spty}), and,
particularly without that constraint, doing careful assessments of
small IR excesses is beyond the scope of the present paper.

\subsection{Distances}
\label{sec:distances} %\textcolor{red}{sec:distances}

The distances we used here are those based on Gaia DR2 provided by
Bailer-Jones \etal\ (2018). Since the membership lists we used as
input from Zari \etal\ (2018) and Damiani \etal\ (2019) both
extensively used data from Gaia DR2, we used 
distances derived from DR2\footnote{Comparing to Bailer-Jones \etal\ 
(2021) which uses EDR3, fewer than 40 stars have 
distances significantly enough different
than those from Bailer-Jones \etal\ (2018) that they would be treated
differently here.}. We expected the members selected in this
fashion to all have distances appropriate for UCL/LCC, $\sim$100-200
pc, but for  nearly two hundred stars, the distances retrieved from
Bailer-Jones \etal\ are $>$300 pc. These apparently distant sources 
are not biased towards, say, those members added from Pecaut \&
Mamajek (2016), where that analysis was completed pre-Gaia. 
The distances appear in Table~\ref{tab:bigperiods} 
for all members and in Appendix~\ref{app:nm} for the 
discarded (rejected) sample. 

\begin{figure}[htb!]
\epsscale{0.75}
\plotone{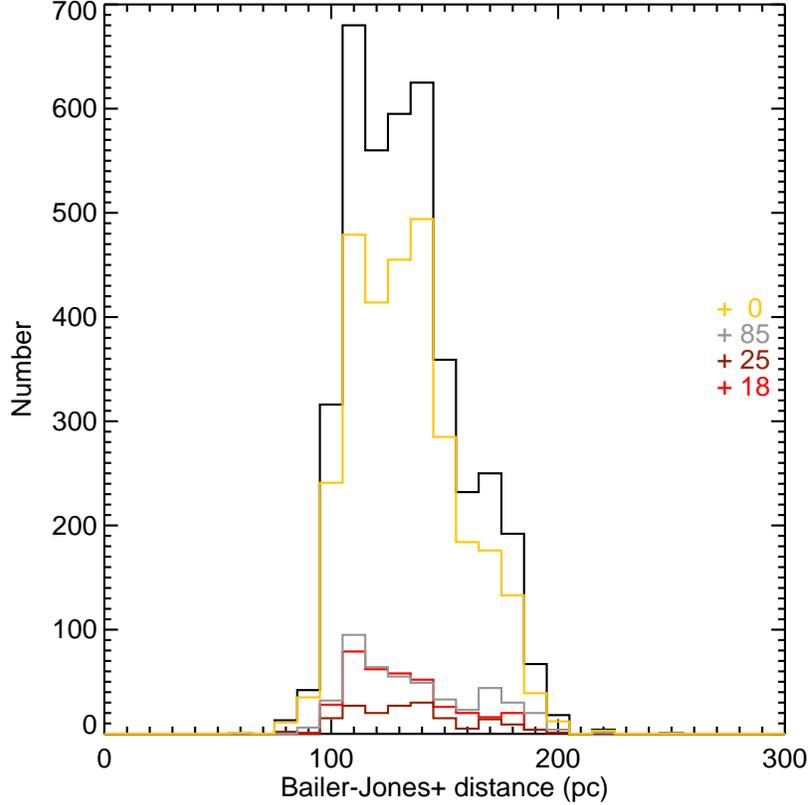}
\caption{Histogram of distances from Bailer-Jones \etal\ (2018)  from
the entire sample (black); the discarded targets (red);  bronze,
silver, and gold members (see Sec.~\ref{sec:membership}).  There are
additional targets with a Bailer-Jones+ distance $>$300, and those are
indicated by the correspondingly colored numbers on the far right
hand side. Most of the sample (and all of the gold sample) straddles
the range of expected distances for UCL/LCC, $\sim$100-200 pc.}
\label{fig:distances}
\end{figure}

Figure~\ref{fig:distances} has histograms of the Bailer-Jones \etal\
(2018) distances for the entire sample as well as the gold, silver,
and bronze samples, with an indication of how many stars in those
three samples have distances $>$300 pc. All of the gold sample is
$<$300 pc by definition. A substantial fraction (see
Table~\ref{tab:summarystats}) of the silver sample includes stars that
would be in the gold sample, except for their Bailer-Jones \etal\
(2018) distances. We did not summarily discard these stars because
these stars with $>$300 pc distances were also identified in the
literature from Gaia proper motions as belonging to UCL/LCC (see
Sec.~\ref{sec:initiallist}).  All of these Gaia analyses cannot be
simultaneously correct. By way of a specific example,  a star known to
Simbad as Sz 127 (=TIC 255255634) was selected by Zari \etal\ (2018)
as a member of UCL/LCC. It appears in several papers (e.g., Galli
\etal\ 2013) as  a member of Lupus, so historically, it has been
regarded as quite close  to us (less than a kpc). It has an IR excess,
which supports its youth. Simple inversion of the Gaia DR2 parallax
yields $\sim$160 pc;  Bailer-Jones \etal\ (2018) lists $\sim$1640 pc.
Because the distances can be wrong, therefore, if any given star has a
Bailer-Jones \etal\ (2018) distance that places it too far away
($>$300 pc), then that is just one mark against it, as opposed to a
reason to reject it entirely. 

Our entire sample is biased towards things with Gaia DR2 data, but (in
contrast with the other clusters in Papers I-VI), UCL/LCC is
sufficiently close and the stars sufficiently not subject to
reddening that the bias towards Gaia data is likely to still mean that
the sample is representative of the true distribution, at least for
sufficiently bright members (earlier than mid-M).  Additionally, by
discarding (or demoting) those stars that are apparently too far
away, simply because they don't represent a large fraction of the
sample, we are unlikely to introduce significant bias even if those
apparently distant sources really are 100-200 pc away. 
Kolmogorov-Smirnov (KS) and Anderson-Darling (AD) tests suggest that
all the sub-samples in Fig.~\ref{fig:distances}, truncated to $<$300
pc, are similar to each other; gold and silver are the most different,
with silver having fractionally more of the slightly closer sources.

%\clearpage

%\floattable
\begin{deluxetable}{ccp{13cm}}
\tabletypesize{\scriptsize}
%%\rotate
\tablecaption{Contents of Table: Periods and Supporting Data for
UCL/LCC Members with Viable Light Curves\label{tab:bigperiods}}
\tablewidth{0pt}
\tablehead{\colhead{Number} & \colhead{Column} & \colhead{Contents}}
\startdata
1 & TIC & Number in the {\it TESS} Input Catalog (TIC)\\
2 & coord & Coordinate-based (right ascension and declination, J2000) name for target \\
3 & othername & Alternate name for target \\
4 & gaiaid & Gaia DR2 ID \\
5 & distance & Distance from Bailer-Jones \etal\ (2018) in parsecs\\
6 & member & membership sample (gold, silver, or bronze) \\
7 & Kmag & \ks\ magnitude (in Vega mags), if observed\\
8 & vmk-used & \vmk\ used, in Vega mags (observed or inferred; see text)\\
9 & evmk & $E(V-K_s)$ adopted for this star (in mags; see \S~\ref{sec:dereddening}) \\
10 & Kmag0 & dereddened $K_{s,0}$ magnitude (in Vega mags), as inferred (see \S\ref{sec:dereddening})\\
11 & vmk0 & $(V-K_s)_0$, dereddened $V-K_s$ (in Vega mags), as inferred (see \S~\ref{sec:dereddening}; rounded to nearest 0.1 to emphasize the relatively low accuracy)\\
12 & color\_uncertcode & two digit uncertainty code denoting origin of \vmk\ and \vmkz\
(see \S\ref{sec:litphotom} and \ref{sec:dereddening}):
First digit (origin of \vmk): 
1=$V$ measured directly from the literature (including SIMBAD) and $K_s$ from 2MASS; 
2=$V$ from the literature (see \S\ref{sec:litphotom}) and $K_s$ from 2MASS;
3=\vmk\ inferred from Gaia DR1 $G$ and $K_s$ from 2MASS (see \S\ref{sec:litphotom});
4=\vmk\ inferred from Pan-STARRS1 $g$ and $K_s$ from 2MASS (see \S\ref{sec:litphotom});
6=$V$ inferred from well-populated optical SED and $K_s$ from 2MASS (see \S\ref{sec:litphotom});
7=\vmk\ inferred from Gaia DR2 $G$ and $K_s$ from 2MASS (see \S\ref{sec:litphotom});
-9= no measure of \vmk.
Second digit (origin of $E(V-K_s)$ leading to \vmkz): 
1=dereddening from $JHK_s$ diagram (see \S\ref{sec:dereddening});
2=dereddening back to \vmkz\ expected for spectral type;
3=dereddening from SED fits; 
4=used median $E(V-K_s)$=0 (see \S\ref{sec:dereddening});
-9= no measure of  $E(V-K_s)$ \\
13 & P1 & Primary period, in days (taken to be rotation period in cases where there is $>$ 1 period)\\
14 & P2 & Secondary period, in days\\
15 & P3 & Tertiary period, in days\\
16 & P4 & Quaternary period, in days\\
17 & p\_uncertcode & uncertainty code for period -- is there any reason to worry about 
this period? Values are `n' (no worry, full confidence; by far the most common
value), `(n)' (no period), 
and then `n?' and `y?' are progressively less confident periods.\\
18 & IRexcess & Whether an IR excess is present or not (see \S\ref{sec:disks})\\
19 & IRexcessStart & Minimum wavelength at which the IR excess is detected or 
the limit of our knowledge of where there is no excess (see \S\ref{sec:disks}) \\
20 & SEDslope & best-fit slope to all detections between 2 and 25 microns \\
21 & SEDclass & SED class (I, flat, II, or III) based on the SED slope between 2 and 25 microns \\
22 & dipper & LC matches dipper characteristics (see \S\ref{sec:interpofperiods})\\
23 & burster & LC matches burster characteristics (see \S\ref{sec:interpofperiods})\\
24 & single/multi-P &  single or multi-period star \\
25 & dd &  LC and power spectrum matches double-dip characteristics (see \S\ref{sec:interpofperiods})\\
26 & ddmoving & LC and power spectrum matches moving double-dip characteristics (see \S\ref{sec:interpofperiods})\\
27 & shapechanger & LC matches shape changer characteristics (see \S\ref{sec:interpofperiods})\\
28 & beater &  LC has beating visible (see \S\ref{sec:interpofperiods})\\
29 & complexpeak & power spectrum has a complex, structured peak and/or has a wide peak (see \S\ref{sec:interpofperiods})\\
30 & resolvedclose & power spectrum has resolved close peaks (see \S\ref{sec:interpofperiods})\\
31 & resolveddist & power spectrum has resolved distant peaks (see \S\ref{sec:interpofperiods})\\
32 & pulsator & power spectrum and LC match pulsator characteristics (see \S\ref{sec:interpofperiods})\\
33 & scallop &  LC matches scallop or flux dip characteristics (see \S\ref{sec:interpofperiods} and App.~\ref{app:weirdos}) \\
34 & EB &  LC has characteristics of eclipsing binary (see \S\ref{sec:interpofperiods} and App.~\ref{app:weirdos}) \\
\enddata
\end{deluxetable}

%\clearpage

\section{{\it TESS} data} 
\label{sec:tess} %\textcolor{red}{sec:tess}

In this section, we obtain {\it TESS} LC versions from three different
pipelines  (Sec.~\ref{sec:initialTESSlcs}). Source confusion and
contamination of  LCs is a major concern with {\it TESS} LCs, and we discuss
this in some detail in Sec.~\ref{sec:confusioncontamination}, with
some of the mitigation strategies for this developed in the context of
looking for periods.  The search for periods is presented in
Sec.~\ref{sec:lookingforperiods}.  In Sec.~\ref{sec:comparelit}, we
show that we are obtaining periods at least as reliably as others in
the literature, especially for $P<20$ days. We discuss 
the interpretation of these periodic signals 
(Sec.~\ref{sec:interpofperiods}), and we finish with a summary
of the limitations on the range of periods to which we are sensitive
(Sec.~\ref{sec:rangeofperiods}).

\subsection{{\it TESS} LC Versions} 
\label{sec:initialTESSlcs} %\textcolor{red}{sec:initialTESSlcs}

A primary driver for inclusion of targets in this analysis is that
there be a corresponding {\it TESS} LC from the first year of the
mission.  {\it TESS} covers most of the sky, but not all of it.  We
used {\it TESS}-Point (Burke \etal\ 2020) to assess which of our
initial set of stars could possibly have an observed LC  in the first
year of operations. 

We created \texttt{eleanor} (Feinstein \etal\ 2019) 30-minute cadence LCs 
for most of the candidate members of UCL/LCC from TESS LCs from sectors
9-12. 
This approach provides
three versions: PCA, principal  component analysis; COR, corrected;
and RAW.  Some LCs were not successfully extracted, and thus some
targets do not have \texttt{eleanor} versions. Additional 30-minute cadence LCs
from both the Cluster Difference Imaging Photometric Survey (CDIPS;
Bouma \etal\ 2019;  
\dataset[https://doi.org/10.17909/t9-ayd0-k727]{https://doi.org/10.17909/t9-ayd0-k727}) 
and the MIT Quick-Look Pipeline 
(QLP; Huang \etal\ 2020ab; 
\dataset[https://doi.org/10.17909/t9-r086-e880]{https://doi.org/10.17909/t9-r086-e880})
were available via MAST, the Mikulski 
Archive for Space
Telescopes.  Table~\ref{tab:summarystats} includes the numbers and
sample fractions of those targets having LCs from each of the three
data reductions.

In Figure~\ref{fig:where}, the obvious stripes of missing sources are
primarily an artifact of {\it TESS}'s observation strategy.  There is
also an effective cutoff at both the bright and faint ends,  where
{\it TESS} saturates or there is an insufficient signal-to-noise
ratio. The stars that are missing do not have LCs either because they
were not observed in the first year of the mission (``not on
silicon''), or largely because they are too faint.  The fact that we
are missing some fainter stars is unsurprising. 
Figure~\ref{fig:wherebrightness} includes the \ks\ brightnesses for
the targets analysed here, with separate histograms for sources with
\texttt{eleanor}, CDIPS, and QLP LCs. 

There is no reason to think that the unobserved wedges on the sky
would introduce an astrophysical bias. {\it TESS} has unavoidable
bright and faint cutoffs. In practice, the completeness for UCL/LCC
falls dramatically for spectral types earlier than late F (no spots to
cause periodic modulation) and later than roughly M5 (too faint to
have a LC).  However, there may be a subtle bias resulting from the
fact that  we do not have all the {\it TESS} LC versions for every
source.  Specifically, the \texttt{eleanor} LCs have many more
outliers that could mask periods in the LCs, and for many sources are
our only choice.  In an attempt to counteract the specific issues
introduced by photometric outliers, we have spent considerable time
on each source to make sure that the best possible LC we have used
does not have obvious outliers, and we subsequently identify periods
based on cleaned LCs. We are also missing LCs for a few faint sources,
even when the target is observable (see
Fig.~\ref{fig:wherebrightness}), which  biases our sample of periods
against spectral types later than $\sim$M4-M5.

\subsection{Confusion and Contamination}
\label{sec:confusioncontamination} %\textcolor{red}{sec:confusioncontamination}

Because the {\it TESS} pixels are $20\arcsec\times20\arcsec$, source
contamination is a concern, especially given that UCL/LCC are in the
Galactic plane and located towards the Galactic center.  In Papers
I-VI, based on {\it K2} data, assessment of cluster membership was the
dominant issue in assembling our final best set of targets 
and LCs to be analyzed. In the
case of UCL/LCC, membership concerns play a more minor role, while
source contamination becomes a much more significant concern 
when assembling the final best sample for analysis.

Because each LC was inspected by hand in order to look for periods
(Sec.~\ref{sec:lookingforperiods}), in RA order, there were LCs that
we noticed immediately were identical to another nearby (in
projection) source's LC in the set of candidate members. In the past,
with the {\it K2} LCs (Papers I-VI), we have been reasonably
successful at teasing apart, through careful data reduction,  the
periods corresponding to each of any potentially confused targets,
though we were unable to do so in a few cases.  However, because the
{\it TESS} pixels are so large, we have had to  abandon in most cases
here, the hope of attaching individual periods to confused targets
with much confidence. Thus, any stars where the LCs were identical to
another nearby UCL/LCC star were discarded, without further analysis
or discussion, except for the rare occurrence where the stars were of
much different brightnesses. In those cases, the period was attached
to the brighter star, and the fainter star was discarded.

The total numbers of stars so affected are given in 
Table~\ref{tab:summarystats}, and they make up the majority of the
discarded (rejected) sources.

\begin{figure}[htb!]
\epsscale{0.75}
\plotone{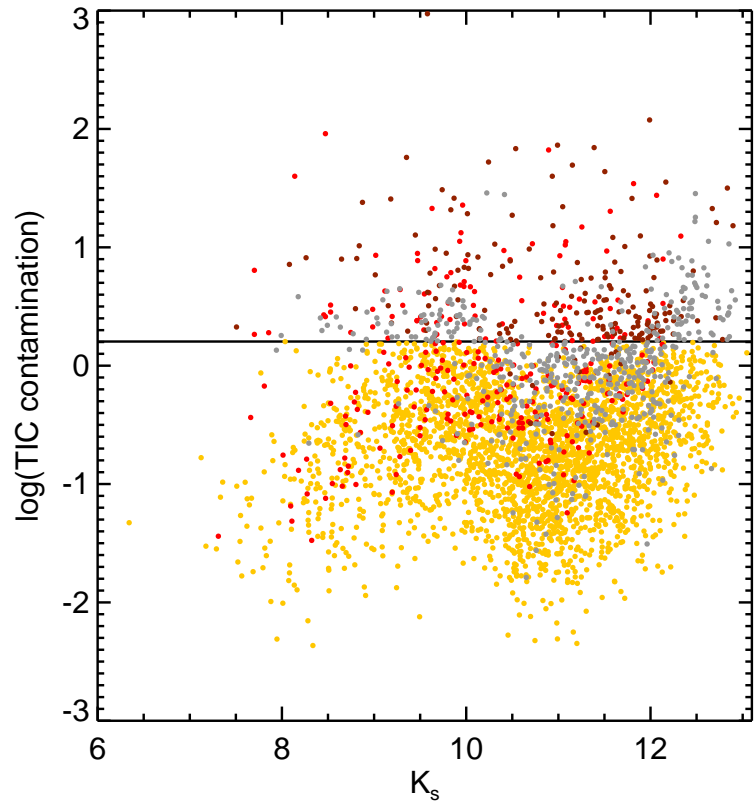}
\caption{The log of the contamination ratio from the TIC (Stassun
\etal\ 2018, 2019) plotted against \ks. The discarded targets are
red;  bronze, silver, and gold members are colored correspondingly. 
The horizontal line is at  a contamination ratio of 1.6, a number we
arrived at by inspection (of this and other similar plots) as a
reasonable cut-off between the LC is ``likely  contaminated'' and
``likely not or at least less likely  to be contaminated.'' Gold
members (by definition) have a contamination  ratio $<$1.6.}
\label{fig:ticcontam}
\end{figure}

Stassun \etal\ (2018, 2019) assembled, as part of the TIC, a metric
that attempts to quantify the degree to which the {\it TESS} LC is likely
to be contaminated by nearby stars and reported it as a contamination
ratio.  This was assembled pre-launch, and makes a number of
assumptions, so we did not rely solely on this value, but it was
included in our assessment of whether to believe the {\it TESS} LC.
Figure~\ref{fig:ticcontam}  shows the distribution of this
contamination ratio for our targets  as a function of brightness.
Based on this and other similar plots, we decided that a TIC
contamination ratio of 1.6 was an appropriate cutoff; targets with a
contamination ratio $>$1.6 have a mark against them.  As for our treatment of
distances above, the contamination flag could be unreliable, and so 
we downgrade (add a mark against) but do not a 
priori discard sources having a 
high TIC contamination ratio.

\begin{figure}[htb!]
\epsscale{1}
\plottwo{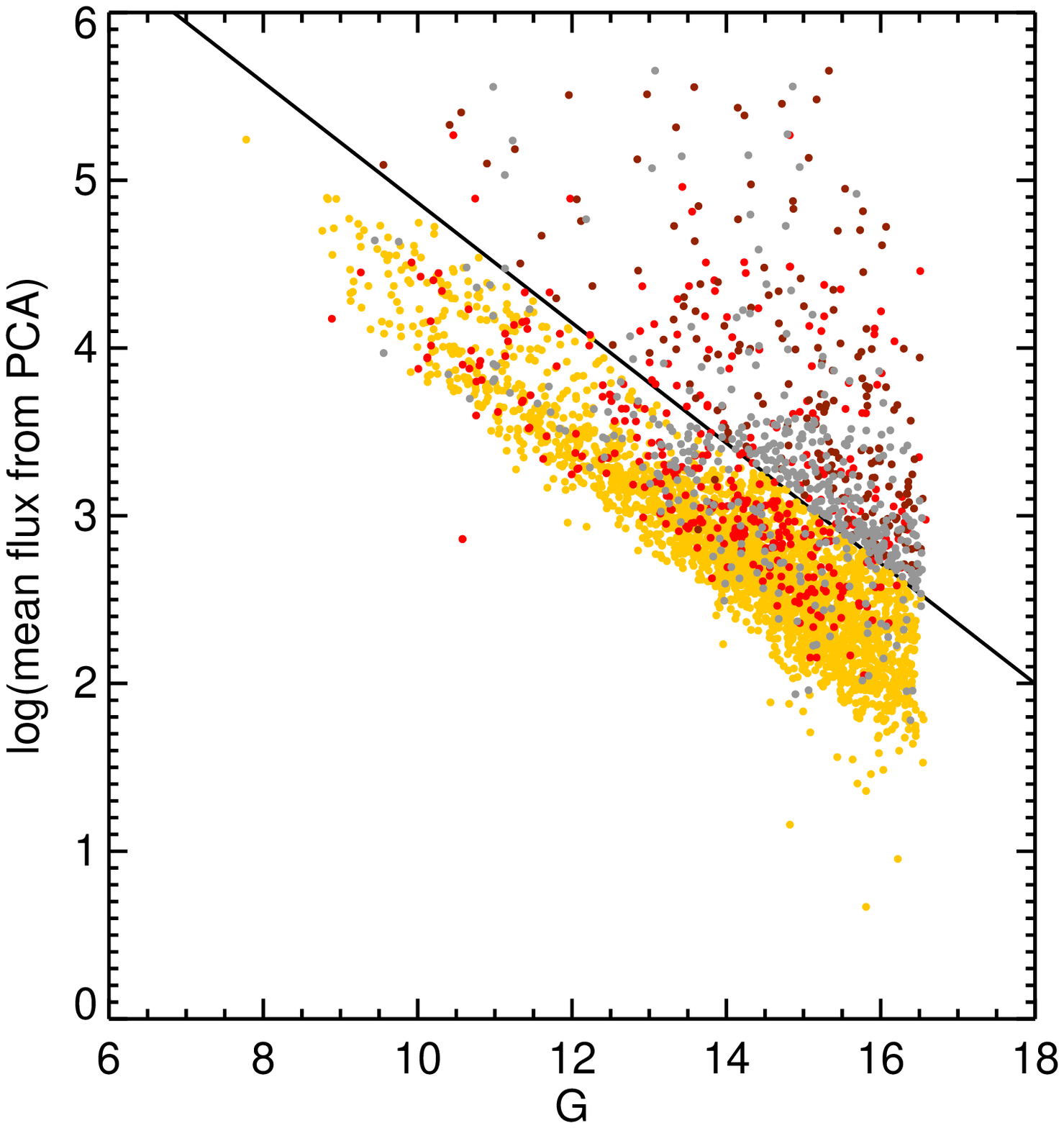}{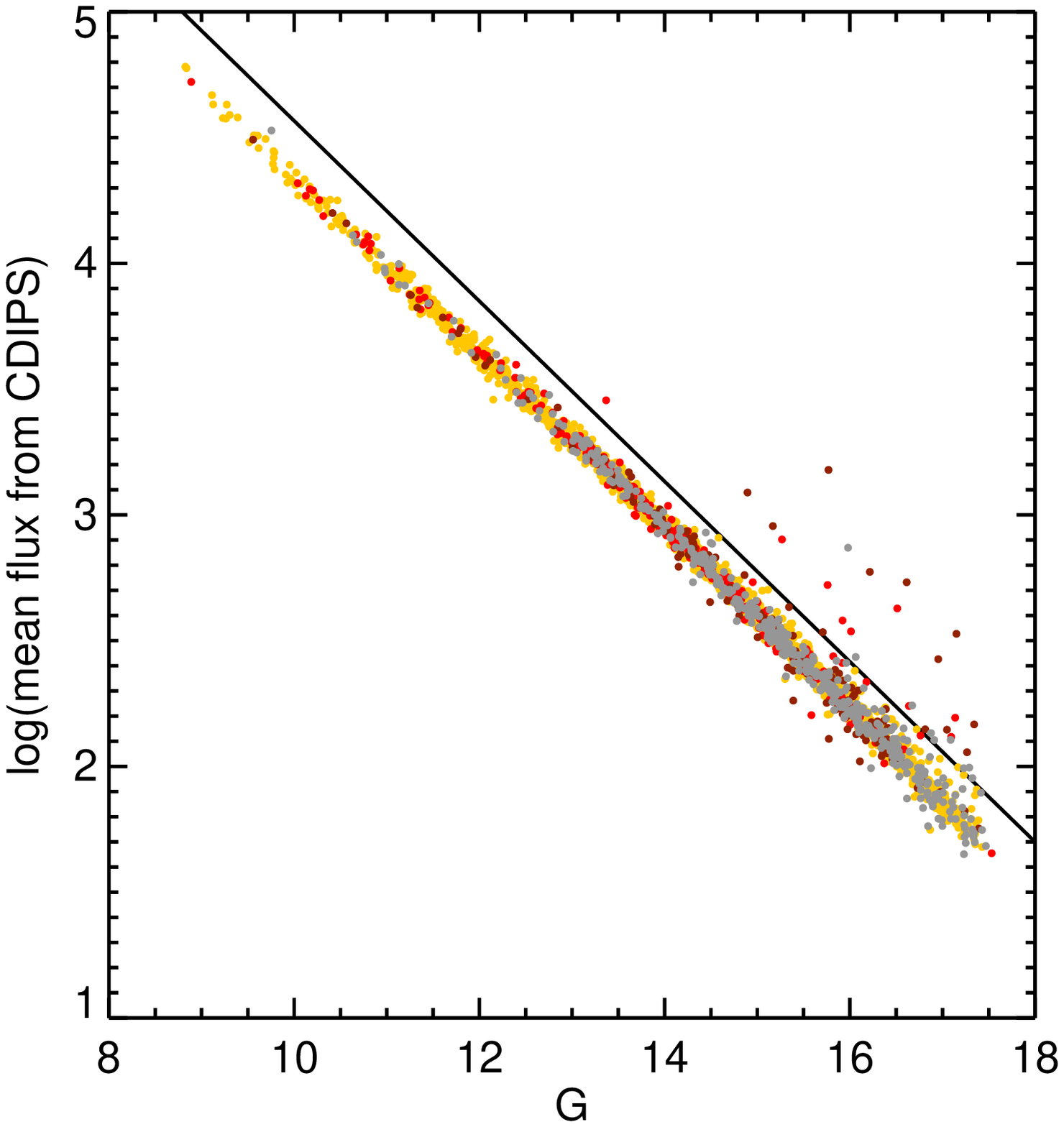}
\caption{The log of the mean flux from the PCA \texttt{eleanor} LCs
plotted against Gaia $G$ (left) and the log of the mean flux from the
CDIPS LCs plotted against Gaia $G$ (right). The discarded targets are
red;  bronze, silver, and gold members are colored correspondingly.
The black line on the left is given by log(mean PCA flux) =
$-$0358$\times G$ + 8.45 on the left.  The black line divides the
stars whose $G$ is much different than the mean PCA LC flux from those
whose $G$ seems appropriate given their mean PCA flux; stars above
this line are likely to be contaminated by neighbors.  Gold members
(by definition) have a mean PCA flux consistent  with their $G$. Not
every one of our targets has an \texttt{eleanor} LC, so the right plot
repeats the analysis using CDIPS; the black line is the same on the
right, just shifted down by 0.3 dex.  The scatter is worse for the
\texttt{eleanor} LCs (left plot) than for the CDIPS LCs (right plot)
because there are  far more outliers in the \texttt{eleanor} LCs.}
\label{fig:gcontam}
\end{figure}

Stauffer \etal\ (2021), in looking for UCL/LCC members that were
uncontaminated scallop shell stars, used a quick assessment of 
contamination. They compared the mean PCA flux in the LC to the Gaia
DR2  $G$ mag. Figure~\ref{fig:gcontam} does this for the ensemble in
order to identify those LCs whose mean flux seems inconsistent with
the target's measured $G$ mag, suggesting that the LC is contaminated
by flux from nearby stars. The black line in the left panel is given
by  log(mean PCA flux) = $-$0358$\times G$ + 8.45.  Targets above the
black line in  Fig.~\ref{fig:gcontam} are ones  that are tagged as
having LCs that are likely contaminated, and those targets accumulate
another mark against them. This calculation can also be subject to
error, so this criterion is also not enough,  on its own, to discard
the source. However, a lot of sources that have a high TIC
contamination ratio also have an  anomalously high PCA flux mean. Most
of the sources in the  bronze member bin have both a high TIC
contamination ratio and a high {\it TESS} flux compared to the $G$
magnitude (see Table~\ref{tab:summarystats}). 

Because not all of our targets have \texttt{eleanor} LCs, we do not
have a mean  PCA flux for all targets. Figure~\ref{fig:gcontam} also
includes a similar analysis for the CDIPS LCs. This distribution is
far better behaved than it is for the PCA LCs; the black line in this
plot has the same slope as that in the PCA plot. 
There are several reasons why the scatter is worse in the PCA
version of this plot than in the CDIPS version. Because the LCs that emerge from
the CDIPS analysis have far fewer outliers than the \texttt{eleanor}
LCs, the mean CDIPS flux is a more accurate representation of the
flux from the star. The CDIPS analysis was only performed on a subset
of stars that are (a) cluster members, (b) show signs of youth, and
(c) have $G<$16 (Bouma et al. 2019). The \texttt{eleanor} LCs were
calculated from lists of RA/Dec and likely therefore were extracted 
even for stars that could be too faint/contaminated to even expect a
reasonable LC. Stars that are above the black line in the right panel
of Figure~\ref{fig:gcontam} are tagged as possibly  contaminated, even
if they were below the line in the left panel.  We note for
completeness that the QLP LCs arrive normalized, so the approach
adopted for the other LC sets will not work.  Therefore, stars that
only have a QLP LC cannot be identified as possibly contaminated using
this approach.

In summary, the issues arising from {\it TESS}' large pixels are
significant. We lose several sources because we cannot identify with
certainty which star is the origin of the variation seen in the
extracted LC, and/or because effectively the same LC is returned for
more than one target in our list. However, there is no reason to think
that a significant astrophysical bias would be introduced based on
projected distance to a neighbor star bright enough to affect the LC.
Later, more sophisticated LC extraction approaches might recover the
LCs for some targets. We
demoted stars to silver (one mark against), bronze
(two marks against), or rejected (three or more marks against) based on how 
likely we thought it was that the star was too far away (Sec.~\ref{sec:distances})
or the LC was contaminated.
Despite our best efforts to omit stars where
source confusion is significant, it's likely that a few sources
subject to source confusion still remain in our data, even the gold
subsample. We anticipate, however, that they are a small minority of
the sources used here, particularly in the gold subsample.

\subsection{Finding Periods in {\it TESS} Data}
\label{sec:lookingforperiods} %\textcolor{red}{sec:lookingforperiods}

As in our earlier papers, we selected the  `best available' light
curve version, this time from the products provided by \texttt{eleanor}, CDIPS,
and QLP, when available.  To identify significant periods, 
we used the Lomb-Scargle (Scargle 1982)
approach as implemented by the NASA Exoplanet Archive Periodogram
Service\footnote{https://exoplanetarchive.ipac.caltech.edu/cgi-bin/Pgram/nph-pgram}
(Akeson \etal\ 2013). We also used the Infrared Science Archive (IRSA)
Time Series
Tool\footnote{http://irsa.ipac.caltech.edu/irsaviewer/timeseries},
which employs the same underlying code as the Exoplanet Archive
service, but allows for interactive period selection. The period range
searched was 0.05 days (1.2 hours) through 20 days. Stars with only
one {\it TESS} sector have about a 30 day campaign 
($\sim$24.1-26.9 days for these sectors), suggesting that the
maximum plausible period should be about 15 days. About 12\% of  the
targets have more than one {\it TESS} sector, so $P>$15 days could be
retrievable in those cases. In practice, no periods $>$ 17.1 days were
retained as plausible or reasonable.

Each LC was separately inspected by hand. Although an automatic 
5$\sigma$ clipping was imposed, many LCs had significant numbers of 
extreme outliers. More aggressive automatic clipping both failed to 
remove all the outliers as well as inappropriately rejected points in
LCs that did not have so many outliers. Such problematic outliers 
were manually removed and the Lomb-Scargle analysis was redone as necessary. 
The period analysis was also performed with a maximum $P$
of 2 days in an attempt to limit the influence of additional
occasional $\sim$6-7 day timescales in the data that we assume are
either instrumental or introduced in data reduction. 
Stars that appeared to have a period near that regime were subject to
particular scrutiny to ensure that they were plausible periods.

We retained up to four viable periods for each target. 
For the {\it K2} clusters analyzed in Papers I-VI, about 20\% of
the stars are multi-periodic, but only $\sim$12\% of the members here
are multi-periodic. There is no astrophysical reason to assume that
the stars in UCL/LCC would be intrinsically less likely to be
multi-periodic than the stars in Papers I-VI (which include both older
and younger stars). We thus assume that the {\it TESS} light
curves are more difficult to analyze using our approach,
perhaps due to increased noise and/or shorter individual campaigns. 
However, this could be a selection effect.
In the {\it K2} analyses, M stars with multiple periods often turned out to
be binaries. The Gaia analyses feeding into our membership
lists may have, through the selections made in those papers, 
biased the samples against binaries (and higher order multiples), so it may
be the case that our initial member list is fundamentally
less likely to include binaries. Therefore, if most multiple periods are 
due to stellar multiplicity, our source list is therefore less likely to exhibit
multiple periods.

The periods we derive appear in Table~\ref{tab:bigperiods} 
for all members and in Appendix~\ref{app:nm} for the discarded 
(rejected) targets. Note that we have included periods for all
the periodic LCs, but for the discarded targets, the periods 
may very well not correspond to the stars as listed.

For about 10\% of the entire sample, there were periods that could be
ambiguous; for example, there seemed to be a strong peak in the power
spectrum, but the phased light curve was not as convincing as the
remaining $\sim$90\% of the sample. These are indicated in the 
corresponding data tables (Tables~\ref{tab:bigperiods} and
\ref{tab:bignm}). They are all fainter targets; see the last panel in
Fig.~\ref{fig:wherebrightness} for a histogram comparing the
brightness of the high-confidence periods to that of the
lower-confidence periods. None of these less-confident periods are
found in the gold sample -- we note that no star was removed from the 
gold sample on the basis of having a questionable period.

\subsection{Comparison to Literature Periods}
\label{sec:comparelit} %\textcolor{red}{sec:comparelit}

\begin{figure}[htb!]
\epsscale{.88}
\plotone{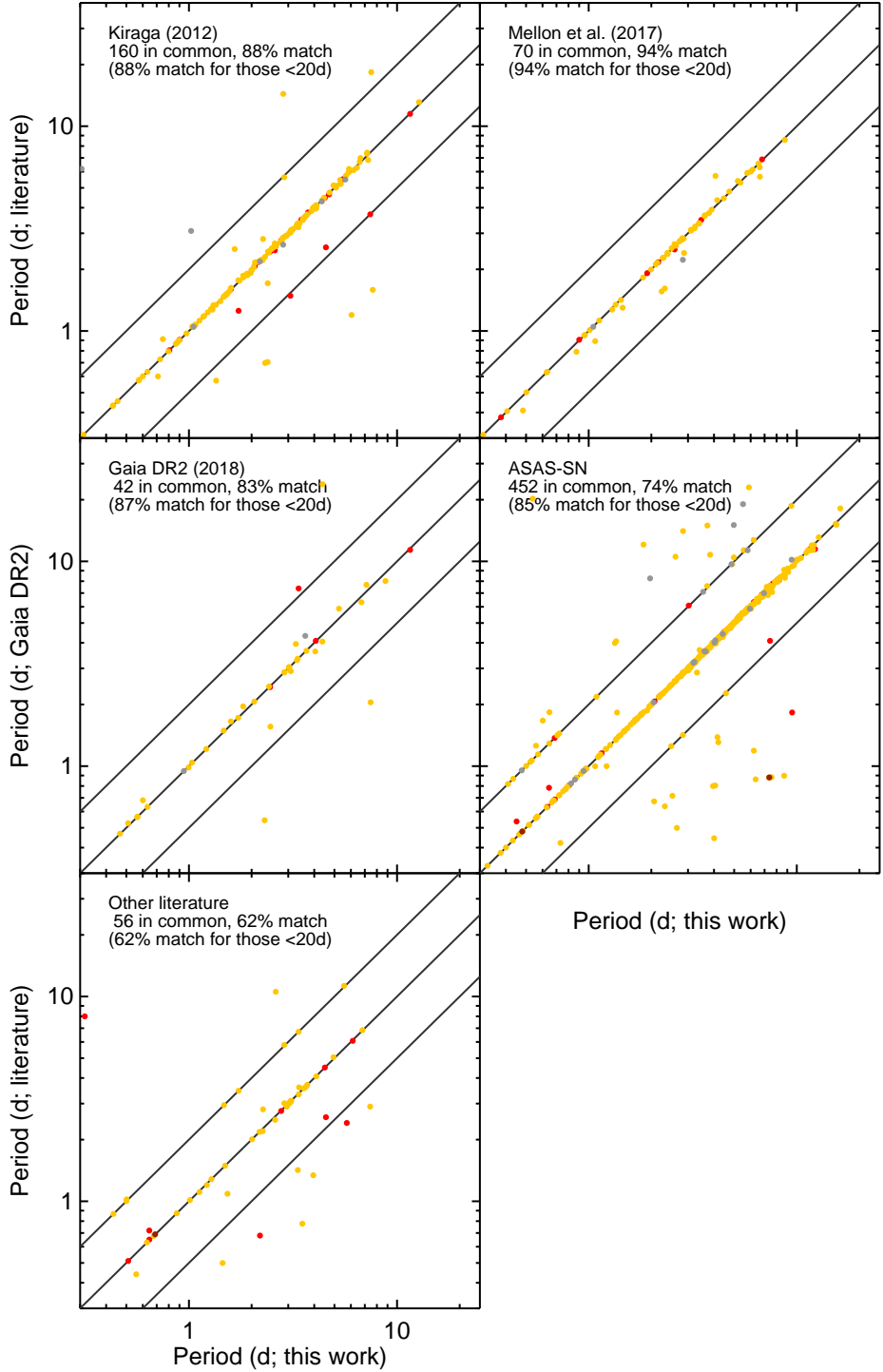}
\caption{Comparison of periods derived here with those found in the literature. 
Upper left: Kiraga (2012) periods; upper right: Mellon \etal\ (2017) periods; 
center left: Gaia DR2 periods; 
center right: ASAS-SN periods; lower left: all other literature 
(see text and Table~\ref{tab:comparelit}). Periods $>$20 days not shown. 
In all panels, gold, silver, and bronze members are indicated, along with
discarded sources (red); note that some of the discarded sources are rejected
because the period in the LC cannot be securely tied to an individual source. 
Most periods agree well; percentages quoted in figure indicate 
those periods matching within 20\%, overall and just for those 
periods reported in the literature to be $<$20 days. Note that most points
in these plots fall on the 1:1 line.}
\label{fig:comparelit}
\end{figure}

As a check on our ability to match stars with the correct LCs 
as well as retrieve accurate rotation periods, we
compare our periods with those in the literature. Four studies monitoring 
targets in the optical have significant overlap with our targets, and
several individual targets appear in a scattering of other papers. 
Figure~\ref{fig:comparelit} presents a comparison of the measured
rotation periods, where all stars in common are included, regardless
of whether the source is ultimately dropped in our analysis.  
(Appendix~\ref{app:lit} includes a list of 
the measured periods for individual comparison, plus notes.) 

Kiraga (2012) reported periods from the All Sky Automatic Survey
(ASAS). Because the positions as originally reported are not very
precise,  we took the counterparts as identified in Kiraga (2012), 
rather than matching anew by position. We have 162 stars in common.
Mellon \etal\ (2017) has 70 stars in common with us; there are a few
(see Appendix~\ref{app:lit}) for which they discarded the $P$
they obtained, but we independently recovered it, suggesting that the
period is real.  Gaia DR2 (Gaia Collaboration \etal\ 2018) included
rotation periods, and  we have 43 stars in common.  The ASAS-SN
catalog of variable stars (Jayasinghe \etal\ 2018) reports periods for
456 stars in common with us.  Other rotation period data in the 
literature can also provide a comparison (Zu\~niga-Fern\'andez \etal\
2021,  Ripepi \etal\ 2019, Nicholson \etal\ 2018, Drake \etal\ 2017,
Samus' \etal\ 2017,  Distefano \etal\ 2016, Siwak \etal\ 2016,
Desidera \etal\ 2015,  K\'osp\'al \etal\ 2014,  Fruth \etal\ 2013,
Alfonso-Garz\'on \etal\ 2012, Donati \etal\ 2012,  Messina \etal\
2010, 2011, Broeg \etal\ 2007,  Christiansen \etal\ 2008, Strassmeier
\etal\ 2005,  Batalha \etal\ 1998, Wichmann \etal\ 1998).  There are a
total of 56 such literature stars in common with us.  
The union of all unique stars in our ensemble that have any 
period in the literature is nearly 600, about 15\% of the ensemble.

The results on the whole are a mix of excellent agreement (the
majority), likely harmonics reported in the literature, and for a few,
significant disagreement with the periods derived from the
high-quality (but often subject to source confusion) {\it TESS} data.   We
recover most periods to better than 20\%; see percentages in
Fig.~\ref{fig:comparelit}. Limiting the comparison to $P<$20 days
(where we have the highest likelihood of recovering periods with {\it TESS}
data), we recover a higher fraction of periods. In a few cases, the
{\it TESS} light curve has significant variations in the LC such that even a
period comparison $>$20\% could reasonably be considered a match. In 
some other cases, the literature reports a period similar to one that we had
discarded as likely instrumental or less secure 
(see Appendix~\ref{app:lit}); 
we did not go back and resurrect our retrieved period, since we do not have
the ability to do so for all stars. 

For each mismatched period $<$20 days, we investigated the prior
period(s) in comparison to the {\it TESS} light curve, and we believe that
the period(s) we report is/are the correct periods for these stars --
or at least the LCs we have associated with them -- 
during the {\it TESS} campaign(s). Given the relatively coarse sampling rate
of ASAS-SN compared to {\it TESS}, it is perhaps not surprising that many of
the  short periods reported by ASAS-SN are longer aliases of the
periods we find.  It is encouraging that source confusion does not 
appear to be a factor in the overwhelming majority of cases.

%\clearpage

\subsection{Interpretation of Periods}
\label{sec:interpofperiods} %\textcolor{red}{sec:interpofperiods}

Periodic behavior in stars can have different origins and 
interpretations, and the shape of the LC or the periodogram  can shed
light on the mechanism. Older clusters (Pleiades  and Praesepe) have
less diversity of LC shapes than younger  clusters (USco, $\rho$ Oph,
Taurus, Taurus Foreground); this is a result of circumstellar disks
(both accretion from and occultation by disks). UCL/LCC is not as old
as the Pleiades, but it is older than USco. Thus, we find LCs covering
the full range of types identified in our earlier papers (Papers
I-VI). Here we briefly summarize the physical interpretation of the
observed LCs. Table~\ref{tab:summarystats} collects all the counts and
sample fractions.

\subsubsection{The Straightforward Rotators}

About 90\% of the members are periodic (see
Table~\ref{tab:summarystats}).  Of those, $>$60\% of the stars are
sinusoidal periods, or close to sinusoidal periods.  These are all
consistent with variations due to a star spot on the surface, rotating
into and out of view.  Some stars have $>$1 period, though we use the
first period reported in most of this analysis.  We find all the same
types of rotational variables as in Papers I-VI  (see
Table~\ref{tab:summarystats}).  The categories are: single period (one
spot or group rotating into and out of view), multi-period (more than
one spot or group rotating into and out of view, or binary (or higher-order
multiple) each with a
spot/spot group rotating into and out of view), double-dip (two
spots/spot groups rotating into and out of view), moving double-dip
(two spots/spot groups rotating into and out of view that are
moving/evolving with respect to each other and/or latitudinal
differential rotation), shape changers (spot/spot group evolution
and/or latitudinal differential rotation), beaters ($>$1 close $P$;
multiple stars or latitudinal differential rotation), complex peak
(spot/spot group evolution and/or latitudinal differential rotation),
resolved close peaks ($>$1 close P; multiple stars or latitudinal
differential rotation, or source confusion), and resolved distant
peaks ($>$1 very different P; multiple stars or source confusion). 

Separately, there are pulsators, typically manifesting as a  forest of
very short periods, but in the case of {\it TESS} can  also manifest
as just a very short period. In our {\it K2} work, we found in many
cases that the strongest peak of those pulsators is likely  closely
related to rotation, based on where such stars fall in the  $P$ vs.\
color diagrams.  We  also have five stars with phased LCs that
resemble those of RR Lyr stars, but at significantly faster periods;
we have identified  them as possible pulsators. In the present
analysis, we have left the pulsation periods in the sample, but
removed them from the plots. They are listed in
Appendix~\ref{app:weirdos}. 

The sample fractions and numbers for these LC types in UCL/LCC are in
Table~\ref{tab:summarystats}, which could be compared to Table 3 in
paper VI which includes the same analysis for all the clusters in
papers I-VI. Direct comparison is somewhat complicated, however. With
more than 3500 member stars, the UCL/LCC member sample has more than 3
times as many stars  as in our USco member sample, more than 4 times
as many as in our Pleiades member sample, and more than 20 times as
many as in Taurus or $\rho$ Oph. With the various issues affecting the
periods we can find using {\it TESS} data (see
Sec.~\ref{sec:confusioncontamination}), it may not be fair to compare,
e.g., the fraction of multi-period stars between  {\it TESS} and {\it
K2}, or the fraction of stars that are shape changers, when the noise
characteristics of {\it TESS} are so different than {\it K2}. Overall,
the fraction of periodic stars is in line with what we expect for  a
situation where some of the stars have circumstellar disks. Detailed
comparison beyond that may not be easily possible until we have more 
experience with {\it TESS} data, more specifically understanding the
noise characteristics enough to know if we are incorrectly
categorizing some of the {\it TESS} LCs using the classes developed
with {\it K2} data.

\subsubsection{The More Complex Rotators}

Stauffer \etal\ (2017) discussed stars then identified as scallop
shells, flux dips, and transient flux dips. 
These unusual LCs have sharp, angular features in the {\it phased LC},
that are too broad for planets, and too small for spots.  Most such
sources are fast-rotating, disk-free M stars. We interpreted these LCs
as due to matter entrained in coronal loops (Stauffer \etal\ 2017). 
Stauffer \etal\ (2018) found more, as did Zhan \etal\ (2019), Bouma
\etal\ (2020), and G\"unther \etal\ (2022).  

Stauffer \etal\ (2021) identified $\sim$30 such objects in UCL/LCC,
based on early explorations of the data we present here. After a  more
complete analysis, we find $\sim$100 stars total ($\sim$70 new)  in
this LC category (see Table~\ref{tab:summarystats} and
Appendix~\ref{app:weirdos}).   Due largely to lower signal-to-noise in
the LCs, some of these new candidates  may not be as unambiguously
identifiable as  the scallop shells/flux dips in Stauffer \etal\
(2021).  While most of the scallop shells/flux dip LCs can be linked
to specific stars, unfortunately, two of the stars from Stauffer
\etal\ (2021) are identified here as clearly confused with other
sources (TIC 89026133  and 89026136; see
Sec.~\ref{sec:confusioncontamination} and  Appendix~\ref{app:weirdos}).
Several other targets having these kinds of LCs are also currently
impossible to link to specific stars. Secure identification of exactly
which star is creating the patterns awaits a later investigation.

The fact that we have $\sim$100 stars in this LC category  is not
surprising. The previous literature has shown that these LCs occur 
more frequently in younger clusters.  Fractionally, we have found them
to be $\sim$1\% of the Pleiades sample, $\sim$3\% of the USco sample,
and $\sim$4\% of the Taurus sample, so it is not at all surprising to
find $\sim$3\% among our UCL/LCC sample; the fraction is consistent
with the discovery rate in other young clusters studied with {\it
K2}.

Stauffer \etal\ (2021) also identified a star described there as 
having ``icicle-like'' features in the LC, arising from a beating
between the intrinsic period (where the shape of the LC involves a
dip) and the {\it TESS} sampling rate. In that case, it was a
photometric binary with two periods. Now, with the larger member
sample, there are a few more LCs exhibiting this broad characteristic,
some of which have have only one period and few of which are obviously
binaries.  We have simply identified  them as being periodic at the
appropriate period. Some of the ``icicle" stars have dips suggestive
of eclipsing binaries.

There are many obvious eclipsing binary LCs among the UCL/LCC stars,
and some additional LCs that could be eclipsing binaries or could
be flux dips. Of the eclipsing binaries, in some cases, we can 
omit the eclipses and still derive a rotation period from the
photospheric component of the LC. We have listed the $P_{\rm rot}$
in the tables (and used those in the plots). 
In Appendix~\ref{app:weirdos} we have listed the
eclipsing binaries (and candidates) along with the $P_{\rm binary}$
where we can derive it.

We tabulate `timescales' for LCs with repeating patterns that are
probably not rotation periods in Appendix~\ref{app:timescales}.

\subsubsection{The Disk-Influenced Periodic Patterns}

Some stars in UCL/LCC still retain their disks, and they exhibit LC
types associated with circumstellar material, namely dippers and bursters 
(see, \eg, Cody \etal\ 2014;  Cody \& Hillenbrand 2018; Cody \etal\ 2022).
These sources have a ``continuum"  LC that is punctuated by dips (fading)
or bursts (brightening).  Dippers are interpreted as
occultations by disk material, while bursters are interpreted as accretion impacts.  
These stars all have large IR excesses. 

Note that the LCs identified as being bursters or dippers are identified
as such without reference to the SED; they all turn out to be disked, 
however, based on the SED.

Tajiri \etal\ (2020) also identify dippers from {\it TESS} full-frame images. 
Among their sample, we independently identified TIC 226241509,
243324939, 266079454, 334999132, \& 412308868 as dippers in this work.

\subsection{Range of Periods}
\label{sec:rangeofperiods} %\textcolor{red}{sec:rangeofperiods}

In addition to the brightness and faintness limits imposed by {\it
TESS} data availability,  we are limited by {\it TESS} cadence in the
range of $P$ to which we are sensitive. The shortest reliable period
that we have identified is 0.436 days (=1.05 hrs); the longest is 17.1
days.   Our range of identified periods is  comparable to the range we
expect for stars with ages similar to the UCL/LCC clusters. 

Our approach to finding periods in the {\it TESS} data has to
include star-by-star inspection and refinement after initial automatic
processing, despite the potential for biases introduced by this
human-based process. It enables us to easily find targets subject to
obvious source confusion (Sec.~\ref{sec:confusioncontamination}). 
There are enough very large outliers in the LCs, and a suspiciously
common $\sim$6-7 day period (which we suspect is instrumental or
introduced by LC extraction), that the investment of time per star is
well-spent.   Given the comparison to the literature, we suspect that
we may have conservatively dropped some legitimate $\sim$6-7 day
periods.   While no one else in the literature to our knowledge has
reported problems with $\sim$6-7 day periods, others have reported
difficulty with $P>$27 days (see, e.g., Avallone \etal\ 2022, and
references therein); we don't have any periods that long.  

It can be harder to obtain rotation periods for disked stars as a
result of stochastic contributions from the disk and/or accretion (see
e.g., Cody \& Hillenbrand 2018 and references therein).  Among all the
disked stars, the periodic fraction is $\sim$85\%, to be compared with
$\sim$90\% of the non-disked stars (Table~\ref{tab:summarystats}). 
Thus, our periodic member sample is likely biased against disked stars for
astrophysical reasons.

\section{UCL/LCC Color-Magnitude and Period-Color Diagrams}
\label{sec:ucllccwhole} %\textcolor{red}{sec:ucllccwhole}

In this section, we present the UCL/LCC sample in the Gaia-based
absolute color-magnitude diagram (Sec.~\ref{sec:distancescmd}), in the \ks\ 
vs.~\vmkz\ observed color-magnitude diagram (Sec.~\ref{sec:representativecmd}),
and in the period vs.~\vmkz\ diagram (Sec.~\ref{sec:periodcolor}). In 
each case, we discuss how the subsamples are similar or different
in these diagrams.

\subsection{Gaia Color-Magnitude Diagrams}
\label{sec:distancescmd} %\textcolor{red}{sec:distancescmd}

\begin{figure}[htb!]
\epsscale{1.0}
\plotone{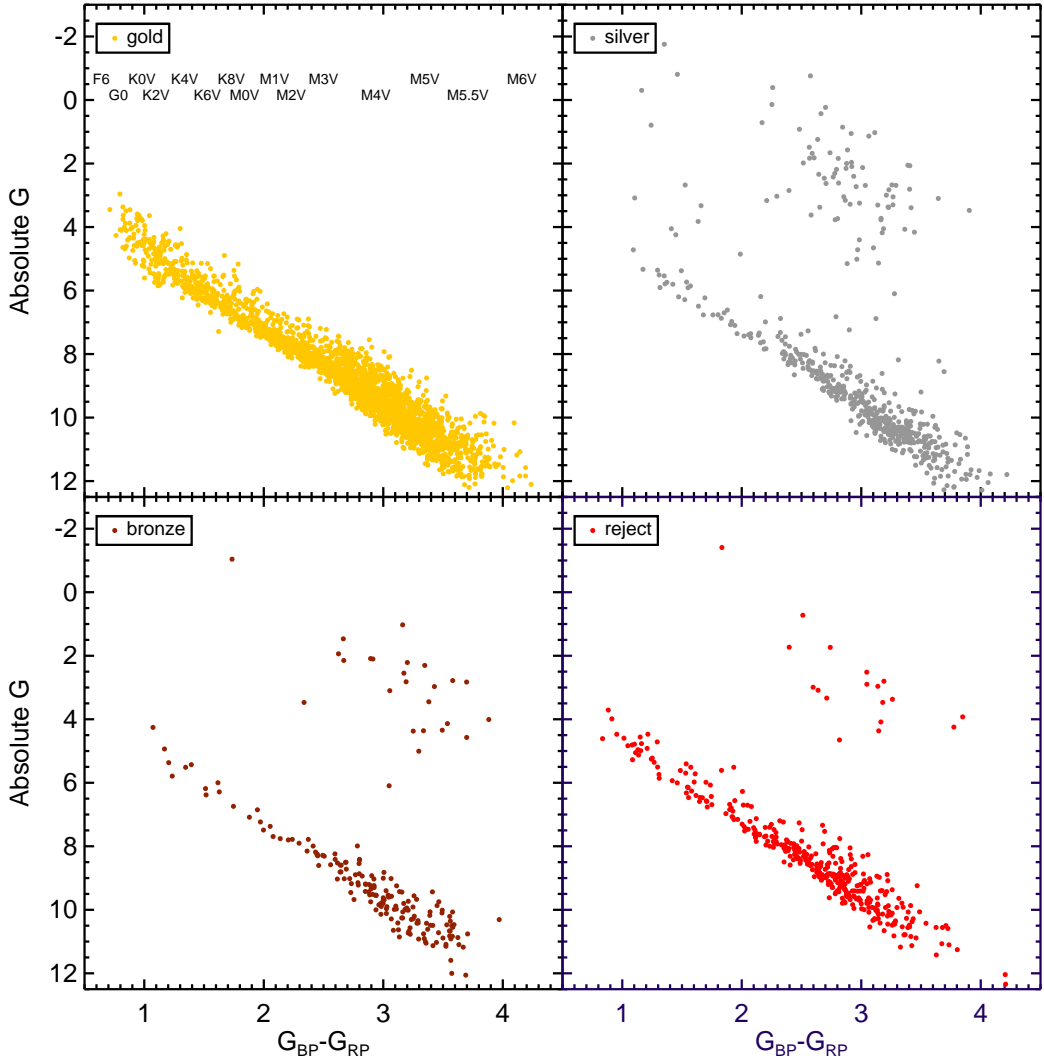}
\caption{Absolute Gaia color-magnitude diagram, using  distances from
Bailer-Jones \etal\ (2018)  for the gold, silver, and bronze members,
and the rejected targets (see Sec.~\ref{sec:membership}).  The stars
that appear to be giants (or even just significantly above the ZAMS 
defined by the rest of the sample) are immediately apparent. The gold
member sample, by definition, has no objects with Bailer-Jones
distances $>$ 300 pc. }
\label{fig:distancescmd}
\end{figure}

Figure~\ref{fig:distancescmd} presents absolute Gaia color-magnitude
diagrams for the three member samples (gold, silver, and bronze) and
the rejected sample.  The stars that appear to be  significantly above
the zero-age main sequence (ZAMS) or near-ZAMS,  as defined by the
rest of the sample, are immediately apparent; these are candidate 
giants or other non-members.  Our gold sample is defined to have none
of these stars included.  A significant fraction of the silver members
(Table~\ref{tab:summarystats} and  Fig.~\ref{fig:distancescmd}) have
as the one mark against them only a discrepant Bailer-Jones
distance.  

At first glance, the ZAMS in all of the member subsamples (gold,
silver, bronze) looks similar. Omitting all of the giants and using a
2-dimensional 2-sided KS test in addition to histograms of color or
absolute $G$, we can compare member subsamples in more detail. The
gold sample has, by far, the most well-populated early ZAMS, down to
about $G_{BP}-G_{RP}<\sim$2 (and $M_G\sim$7), such that it is still
significantly different from the other subsamples, even after omitting
the giants. This makes sense, as the brighter stars will be easier to
measure  and should therefore be easier to identify as secure members.
The silver and bronze subsamples are most similar to each  other, with
fractionally more fainter stars than the other subsamples  near
$G_{BP}-G_{RP}<\sim$3.2.

As in Fig.~\ref{fig:distancescmd}, the $JHK_s$ diagrams presented 
above in Fig.~\ref{fig:jhk} also suggested that all the member subsamples are
similar. In detail, though, the gold sample has more early-type
stars, making it different at a statistically significantly level, 
while silver and bronze are the most similar.

\subsection{$V$ and \ks\ Color-Magnitude Diagrams}
\label{sec:representativecmd} %\textcolor{red}{sec:representativecmd}

Fig.~\ref{fig:distancescmd}
shows a tightly constrained empirical ZAMS, but Gaia data are not
available for most of the stars in the other clusters in Papers I-VI.
Papers I-VI used \vmkz\ as a proxy for mass, and to facilitate the
comparisons we wish to make here, we employ \vmkz\ colors for UCL/LCC as well, 
over the readily available and reliable Gaia colors. 

\begin{figure}[htb!]
\epsscale{1}
\plotone{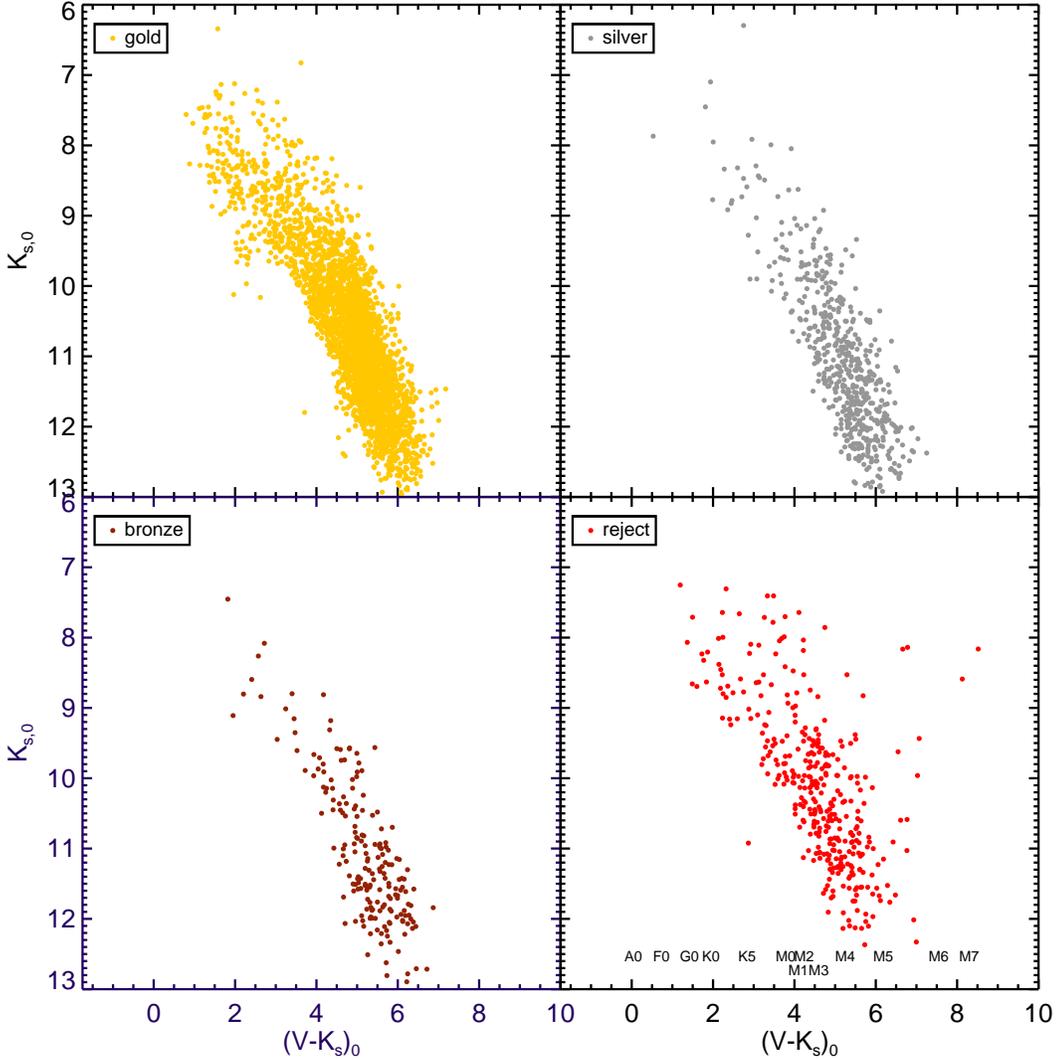}
\caption{Dereddened \ks\ vs.\ \vmkz\ for the  gold, silver, and bronze
members, and the  rejected targets.  The gold sample is the most
well-populated across the full color range. There is more scatter here
in all member subsamples than in Fig.~\ref{fig:distancescmd}, 
in part because the stellar distances vary significantly over the
cluster.  The discarded sample has the most scatter, which makes sense
given that source confusion is important for many of the rejected
stars. }
\label{fig:vmkcmd}
\end{figure}

Figure~\ref{fig:vmkcmd} shows the observed (not absolute)  \ks$_0$ vs.\
\vmkz\ diagrams for our UCL/LCC member sample, revealing a cluster locus that
is broader than expected given Fig.~\ref{fig:distancescmd}. A
significant contributor to the scatter in Fig.~\ref{fig:vmkcmd} comes
from variations in distance across UCL/LCC members. 
The gold sample spans the entire range of \vmk\ colors, and the M
stars are particularly well-populated. The silver and bronze samples
do not span the full range of \vmk, and have fractionally fewer M
stars.  Nearly all the stars are periodic
(Table~\ref{tab:bigperiods}). Fig.~\ref{fig:vmkcmd} also reveals that
there is more scatter in the rejected targets than the other
subsamples, which makes sense particularly if there is source
confusion even in 2MASS. 

\subsection{Period-Color Diagrams}
\label{sec:periodcolor} %\textcolor{red}{sec:periodcolor}

\begin{figure}[htb!]
\epsscale{1.0}
\plotone{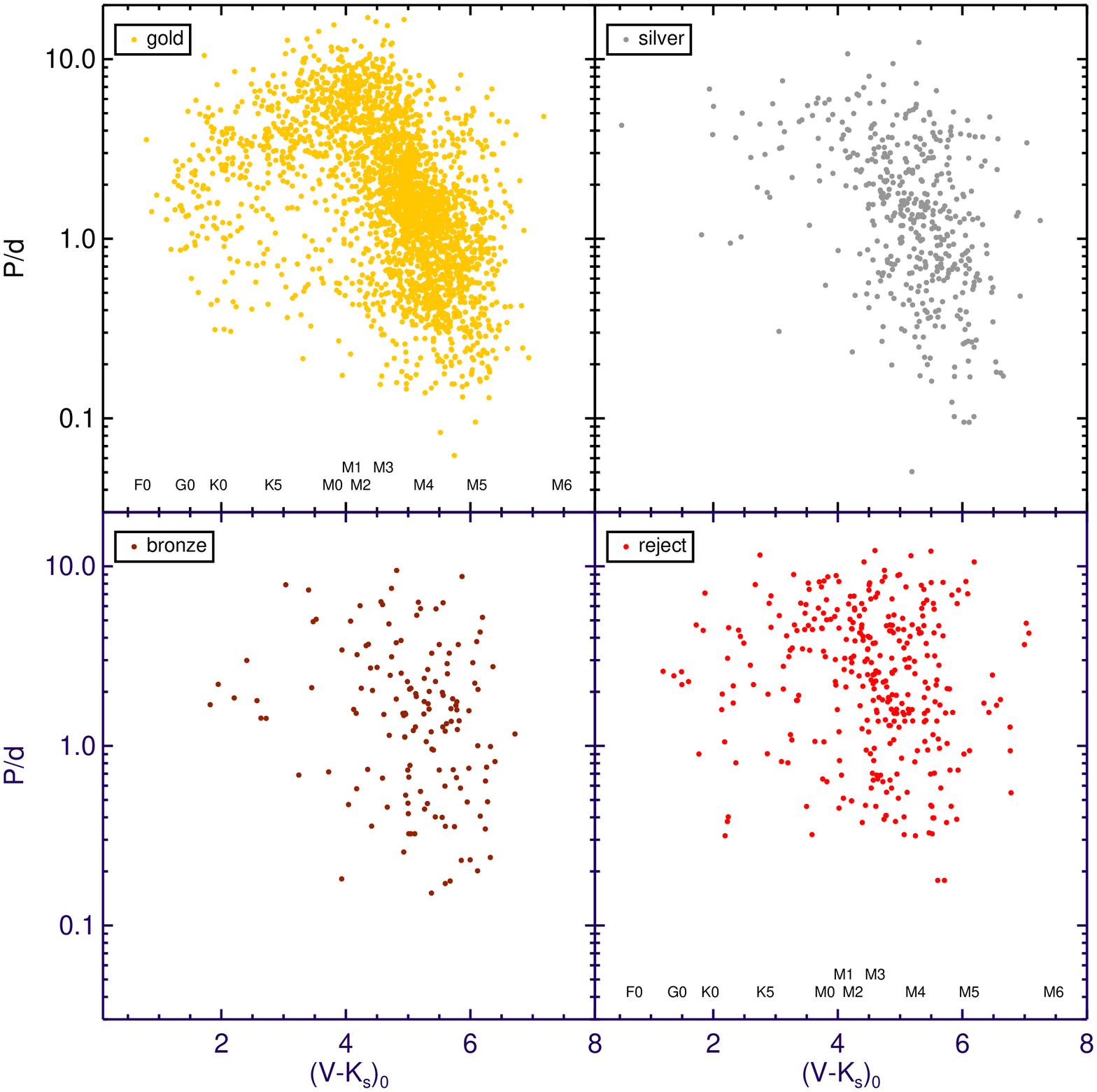}
\caption{$P$ (in days) vs.\ \vmkz\ for the gold, silver, and bronze
members, and the rejected targets. Note that there are two stars from
the discarded sample with \vmkz$>$8, beyond the plot limits.  The gold
sample is the best populated and it is easiest to see structure in
this distribution. The structure is similar to that found in other 
clusters; see the text and Fig.~\ref{fig:pvmkheatmap1}.}
\label{fig:pvmk}
\end{figure}

Figure~\ref{fig:pvmk} shows the $P$ vs. \vmkz\ plot for the members
and rejected targets in UCL/LCC.   Both the $P$ and \vmkz\ values are
listed in Table~\ref{tab:bigperiods} (or Table~\ref{tab:bignm} for the
rejected targets).  As in our {\it K2} rotation papers, for stars with
more than one period, we have taken the first period and the
identified \vmkz\ as representative of the same star (likely the
primary if it is a multiple).

The gold sample has clear structure in $P$ vs. \vmkz. 
Most of the M stars are organized into a sequence of
steeply increasing rotation rate (decreasing periods) going from early M
through at least M4/M5. The higher mass stars (G and K spectral types) have 
considerable scatter, but they seem to be, on average, more slowly rotating
than most of the M stars, with the Gs also rotating much faster than the
Ks. The bronze and silver distributions seem to have similar structure, 
but there are too few stars to define it clearly.  It is
unsurprising that the rejected stars have the most scatter, since it
is less likely here that the measured $P$ goes with the assumed
\vmkz.

Anticipating discussion later in the paper, the scatter found in 
Fig.~\ref{fig:pvmk} is substantial, even for the gold sample.  In
other clusters, any scatter had been a result of reddening 
corrections or unresolved binaries. Reddening corrections are not as
big a concern in UCL/LCC (Sec.~\ref{sec:dereddening}).  There is not
enough literature for us to constrain significantly the binary
fraction among our UCL/LCC member sample, though any such binaries
would have had to survive  the culling based on accurate Gaia
measurements, which is likely to omit binaries.  

If we look at the fraction of stars that have more than one period, as
a function of location in the $P$ vs.\ \vmkz\ plot, the multi-period
fraction rises steeply on the edges of the structure traced by the
bulk of the points. With {\it TESS}, especially given the relative
apparent difficulty of finding multiple periods
(Sec.~\ref{sec:lookingforperiods}),  without additional data, we
cannot be sure that the outliers in $P$ vs.\ \vmkz\ are photometric
binaries, or instead primarily a result of source confusion affecting
the $P$ attributed to a given star.  Given the likely bias against
binaries from Gaia, we have to suspect that source confusion is a
factor.

As the UCL/LCC sample is large, and there is some evidence for 
variation of stellar ages across the association, we explored whether
or not there is any evidence for changes in the distribution of $P$
vs.\ color across the sky in various member subsamples. We did not
find any evidence for this.

%\clearpage

\section{Period-Color Distribution in Context with Other Clusters}
\label{sec:rotationdistribclusters} %\textcolor{red}{sec:rotationdistribclusters}

In this section, we put UCL/LCC in context with the other
clusters we have studied with {\it K2} data. 
We assume that the stars in UCL/LCC represent snapshots in time of the
same population as found in the other clusters (c.f., Coker \etal\ 2016). 
We note as well that the rotation evolution is likely influenced by the 
local UV environment (see Roquette \etal\ 2021), which would matter locally if 
the high-mass stars in UCL/LCC have influenced local star evolution.
We first compare the entire distribution of UCL/LCC to well-populated
clusters (Sec.~\ref{sec:wholeclusters}), then narrow down to 
consider just the disk-free M stars in those clusters (Sec.~\ref{sec:mstars}). 
Next, we put UCL/LCC in context with rotation rates from 8 other 
clusters studied with {\it K2} (Sec.~\ref{sec:multicluster}), some very 
sparsely populated. Finally, we consider the influence of disks, 
particularly among the M stars (Sec.~\ref{sec:disksandrotation}).

\subsection{Comparing to Well-Populated Clusters}
\label{sec:wholeclusters} %\textcolor{red}{sec:wholeclusters}

\begin{figure}[htb!]
\epsscale{1.0}
\plotone{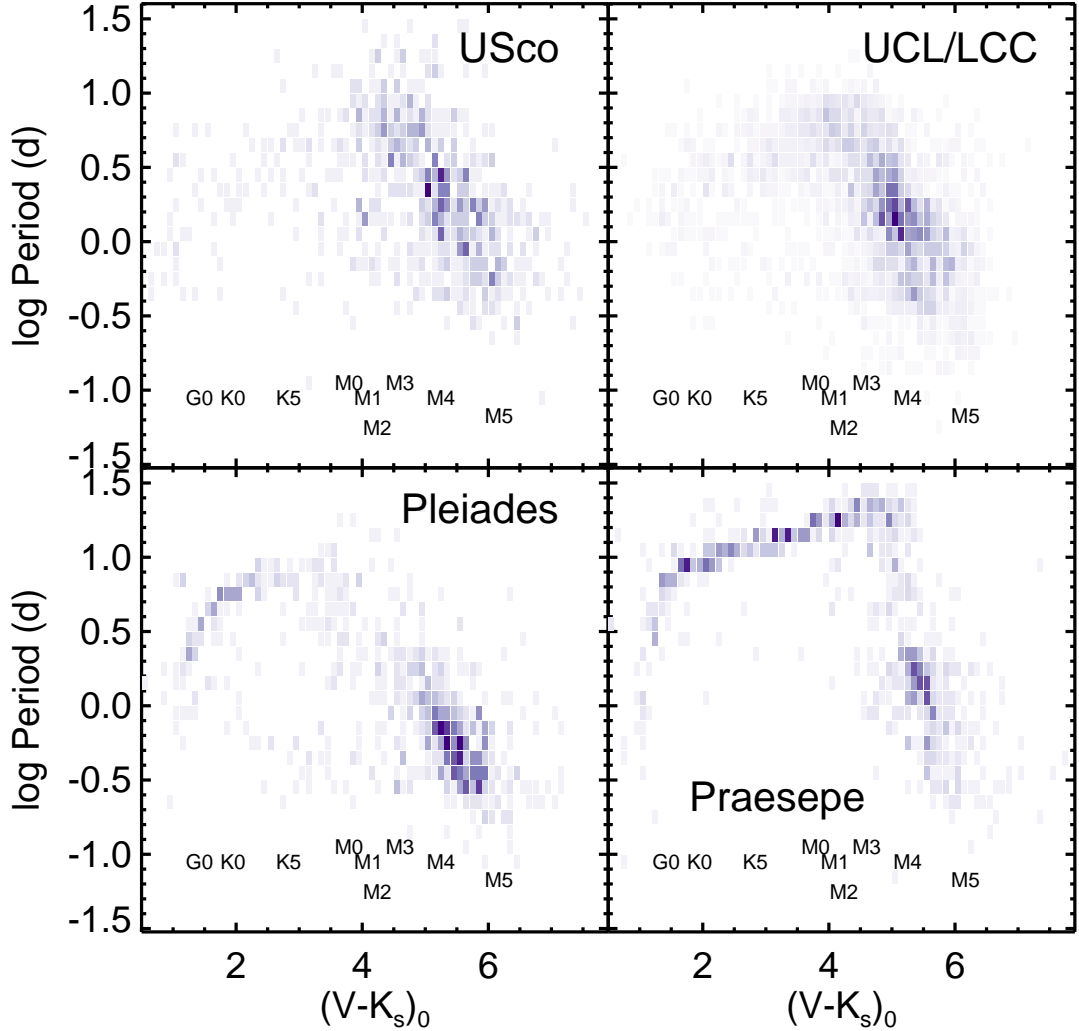}
\caption{$P$ (in days) vs.\ \vmkz, disks removed: density map, where darker
shades indicates more sources in that cell. Each cluster is self-normalized 
such that the darkest shade is the best populated within that cluster, which 
is a different absolute number (and sample fraction) in each cluster.
On the high-mass end, the distribution is well-defined in Pleiades and 
Praesepe, but far less so in USco or UCL/LCC (which is just the gold 
member sample). The low-mass end is well-defined 
in all four clusters.  See the text for much more discussion.}
\label{fig:pvmkheatmap1}
\end{figure}

The period-color structure in UCL/LCC that is most obvious in the gold
sample is suggestive of something intermediate between the
distribution in USco (Paper V) at $\sim$8 Myr and the Pleiades (Paper
I) at $\sim$125 Myr, which is consistent with the accepted age of
UCL/LCC of $\sim$16 Myr. Figure~\ref{fig:pvmkheatmap1} shows UCL/LCC
(just the gold member sample)  in context with the most well-populated
clusters that we have studied using {\it K2}: USco, Pleiades, and
Praesepe (Paper IV; 790 Myr).  In this plot, all the disks (secure and
possible) have been removed from USco and UCL/LCC, leaving just the
disk-free stars. The density map shows where sources are clustered
more tightly; there are so many stars, especially in UCL/LCC,  that it
is hard to appreciate the point density if only individual points are
plotted. 

The distribution for the G and K spectral types that is very obvious
and well-defined by the Pleiades age is far less obvious in the
younger UCL/LCC,  and even less obvious in the younger yet USco
cluster. In Paper V, we suspected that the reddening in USco, which is
patchy and sometimes large, meant the uncertainties in the reddening
correction added artificial ``smearing'' to the distribution, 
rendering the higher masses appearing to be less organized.  However,
in UCL/LCC, there is far less reddening, and less patchy reddening 
(Sec.~\ref{sec:dereddening} and Fig.~\ref{fig:jhk}), yet the early-type
``branch'' is still not anywhere near as organized as in the
Pleiades.  The most extreme outliers in UCL/LCC may be subject to
source confusion  (Sec.~\ref{sec:periodcolor}), but much of this
`disorganization' may be a real feature of this rotation distribution
for G and K spectral types in clusters $\lesssim$20 Myr old.

\subsection{Comparing the M Stars in the Well-Populated Clusters}
\label{sec:mstars} %\textcolor{red}{sec:mstars}

\begin{figure}[htb!]
\epsscale{1.0}
\plotone{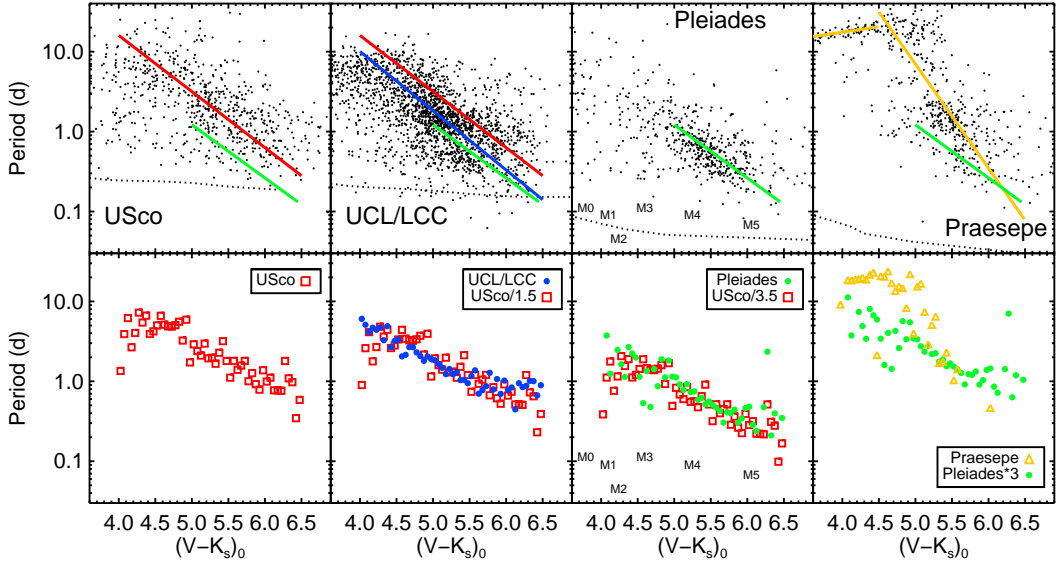}
\caption{$P$ (in days) vs.\ \vmkz\ for M stars, disks 
(high-confidence and possible) removed, and just UCL/LCC gold members. 
Top row: observations (breakup is dotted line);
Bottom row: medians of observations in \vmkz\ bins.
Column 1: USco;
Column 2: UCL/LCC;
Column 3: Pleiades;
Column 4: Praesepe.
In top row: USco fit: red line; UCL/LCC fit: blue line; 
Pleiades fit: green line; Praesepe fit: yellow line.
In bottom row: USco binned data: red symbols; UCL/LCC binned data: 
blue symbols; Pleiades binned data: green symbols; Praesepe
binned data: yellow symbols. USco binned data are shifted in second
panel to match UCL/LCC, and shifted in third panel to match Pleiades. 
Pleiades is shifted in fourth panel to attempt to match Praesepe.
UCL/LCC fits neatly in between USco and Pleiades, with a slope 
that is indistinguishably the same. }
\label{fig:pvmkmstars}
\end{figure}

Most of the stars in the present analysis are M stars, and
Figure~\ref{fig:pvmkmstars} highlights their behavior across the
clusters.  Again, disks have been removed from USco and UCL/LCC, so
these are just the disk-free, gold member M stars. The top row has a linear fit to
the M stars and  the bottom row includes running medians and scaled
running medians. Paper V pointed out that the M star slope between
USco and Pleiades is the same, just shifted as the stars contract onto
the main sequence. UCL/LCC fits neatly in between USco and Pleiades,
with a slope that is indistinguishably the same.  The slope for
USco/UCL-LCC/Pleiades is vastly different than the slope for 
Praesepe, which makes sense since the M stars are spinning 
up through the Pleiades, and spinning down
by the older age of Praesepe, likely incorporating wind braking (Paper V). 

In both Figs.~\ref{fig:pvmkheatmap1} and \ref{fig:pvmkmstars}, 
particularly for UCL/LCC, it is apparent that there is a denser 
region of points near \vmkz$\sim$5 mag, log $P\sim$0.15 ($P\sim$1.4
days).  This dense peak is very obvious in UCL/LCC and far less
prominent in the other clusters.  In USco, the distribution is more
diffuse, and the peak occurs at the same \vmkz, but slightly slower, 
log $P\sim$0.35 ($P\sim$2.2 days). For the Pleiades, the distribution
is elongated along the entire distribution of M stars, but the peak is 
much redder and faster, at \vmk$\sim$5.55, log $P\sim-$0.25
($P\sim$0.56 days).  The Praesepe M stars have an entirely different
distribution. The M star  peak is actually up on the slow branch,  at
\vmk$\sim$4.15, log $P\sim$1.25 ($P\sim$17 days); removing the slow
branch, the peak is at the same color as the Pleiades, \vmk$\sim$5.55,
and about the same period as UCL/LCC, log $P\sim$0.15 ($P\sim$1.4
days).  

There are several stars with very fast periods, apparently faster than
breakup in some cases;  they are found in all of the member
subclasses, and some appear to have IR excesses.  Those LCs, and their
periods, appear legitimate and unambiguous, leading us to conclude
that perhaps something may be wrong with the linkage between  the LC
and the stellar source.    In cases where the apparent IR excess
originates in a single band at WISE-3 or WISE-4, the possibility
exists that the relatively  low-spatial-resolution of WISE is also
subject to source confusion, so neither the $P$ nor the IR excess may
be correctly tied to the star  associated with the \vmkz;  further
exploration of these cases, with higher spatial resolution data, is
needed.

\subsection{Comparing to Other Clusters}
\label{sec:multicluster} %\textcolor{red}{sec:multicluster}

\begin{figure}[htb!]
\epsscale{1.0}
\plotone{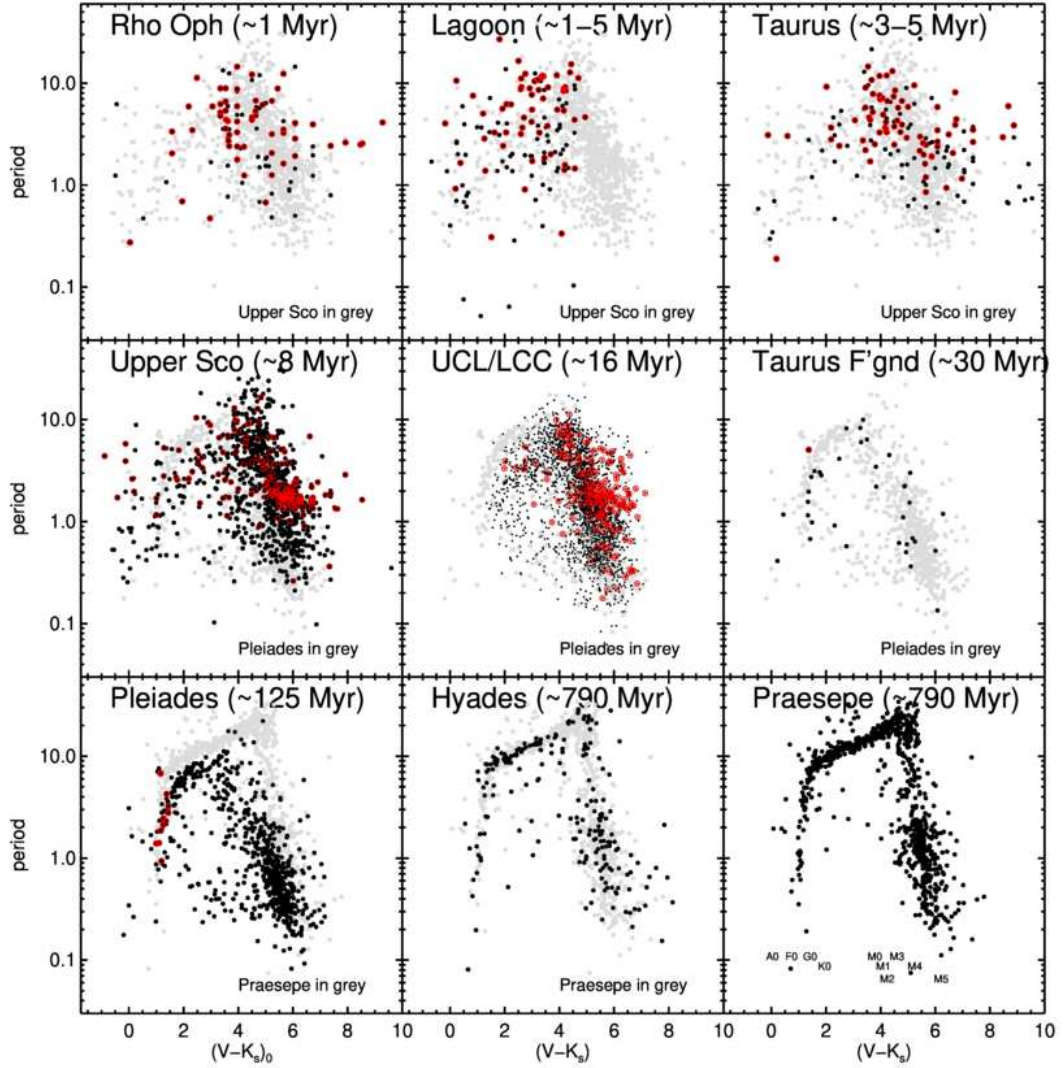}
\caption{$P$ (in days) vs.\ \vmkz\ for all stars in 9 clusters, all
except UCL/LCC from {\it K2}. Stars with IR excesses (e.g., disks) are shown
with an additional red circle. The first row has the youngest most
populous cluster, USco, underplotted in grey. The second row has the
Pleiades underplotted in grey. The third row has Praesepe underplotted
in grey. The points are smaller for UCL/LCC because there are so many
points it is hard to see the patterns otherwise. UCL/LCC points are all
from the gold member sample. It again seems to fit well ``in sequence'' with the 
other clusters around it. References for data:
Rho Oph: Paper V;  Lagoon: Rebull \etal\ (in prep); Taurus: Paper VI; 
USco: Paper V; UCL/LCC: this work;  Taurus Foreground: Paper VI;
Pleiades: Papers I \& II;  Hyades: Rebull \etal\ (in prep); Praesepe:
Paper IV.}
\label{fig:multicluster}
\end{figure}

Fig.~\ref{fig:multicluster} shows the data from UCL/LCC in context
with all the other space-based rotation rates from {\it K2} from our
prior work. For each cluster, an older well-populated cluster is
underplotted to guide the eye regarding the evolutionary patterns. 
The UCL/LCC distribution is in the middle of the evolutionary age
sequence.  The structure in the Hyades is indistinguishable from that
in Praesepe. In the case of the Taurus Foreground, there are so few
stars that it is hard to identify any structure unique to that
cluster.  At the youngest ages, determining the structure is greatly
complicated by the uncertainties added by the ``smearing'' imposed by
reddening corrections and contributions from disk excesses.   The
structure seen in USco is vaguely apparent in Rho Oph, Lagoon, and
Taurus. In the case of the Lagoon, limited sensitivity complicates
interpretation at the lower masses. 

In all of the three youngest clusters, there are relatively few stars,
there is additional smearing due to reddening corrections, and  there
is a substantial disk fraction. The disks appear to affect the
rotation distribution. Most notably there is an obvious $\sim$2 day
pileup of disks in USco as well as in UCL/LCC. We investigate the
influence of disks in the next subsection.

\subsection{Disks and Rotation}
\label{sec:disksandrotation} %\textcolor{red}{sec:disksandrotation}

\begin{figure}[htb!]
\epsscale{1}
\plotone{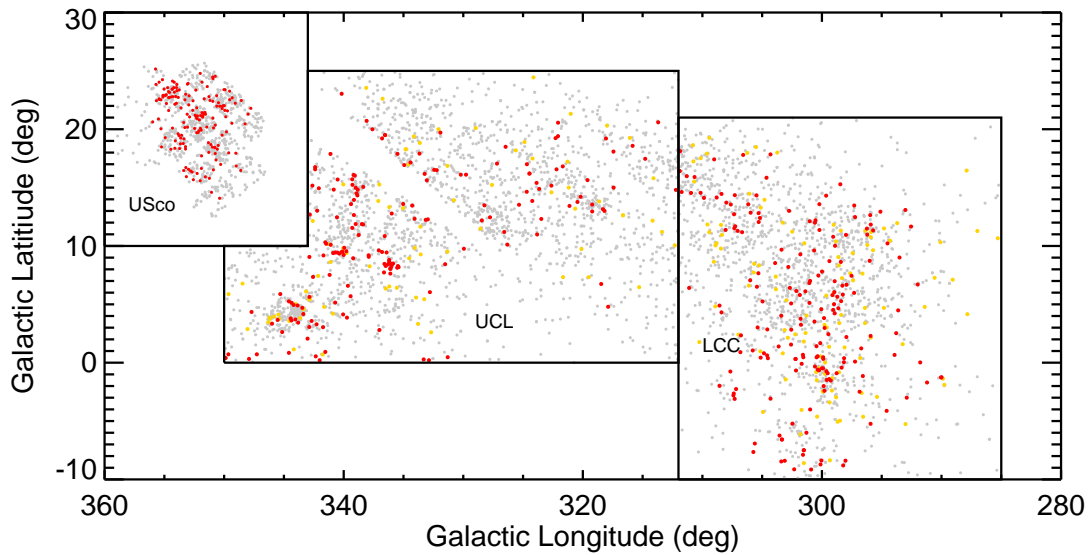}
\caption{Location of targets in Galactic coordinates, as in Fig.~\ref{fig:where}.
Stars with high-confidence disks from UCL/LCC (or USco; paper IV) are red points;
stars with lower-confidence disks from UCL/LCC are yellow points. Grey points
are simply stars with LCs. The high-confidence disks are clustered, but not 
nearly as much as they were in USco. The high-confidence
disk fraction in both UCL and LCC is $\sim$7-8\%, and there is no
obvious large-scale gradient in disk fraction with position among UCL/LCC.}
\label{fig:wheredisks}
\end{figure}

Figure~\ref{fig:wheredisks} shows where the disks are on the sky,
using the same orientation and units as Fig.~\ref{fig:where}, but with
different interpretation of the point color. Disks in USco
are as in Paper VI; high-confidence and lower-confidence disks in
UCL/LCC are identified here and comprise  $\sim$7-8\% of the sample. 
There are obvious clumps of stars with disks, 
but little large-scale systematic structure. There is also no obvious systematic 
difference (in distribution or in disk fraction) between UCL and LCC; 
given their similarity in ages, this is not surprising. 

There is a theoretical expectation that primordial disks lock the 
rotation rate of the star to that of the inner disk (e.g., Ghosh \& 
Lamb 1977; K\"onigl 1991). When the disk disperses, which is thought to happen
at about the age of UCL/LCC for M stars, the star is free to spin up. 
We thus now explore the relationship between disks and rotation 
for the M stars in UCL/LCC.

\begin{figure}[htb!]
\epsscale{1.0}
\plotone{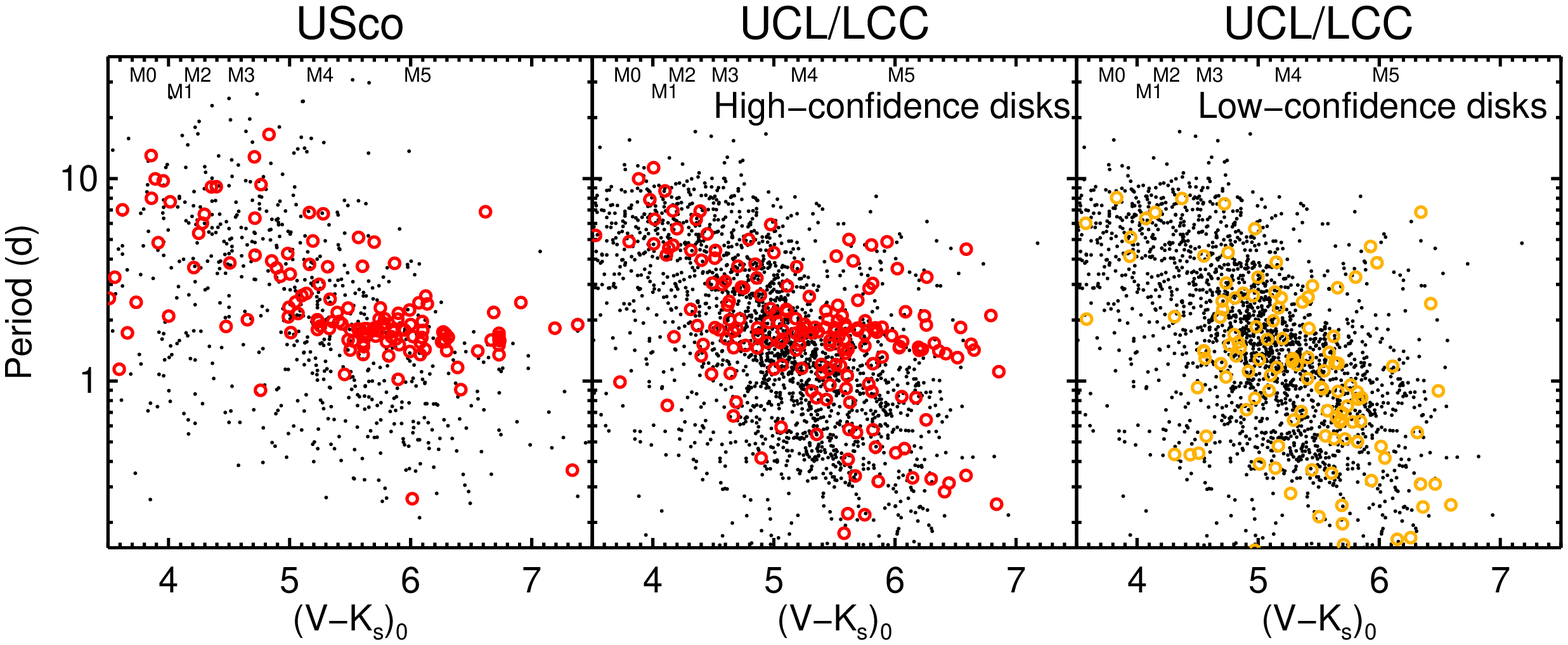}
\caption{$P$ (in days) vs.\ \vmkz\ for M stars from the gold member sample, 
where a black point is 
a disk-free M star, a red circle is a high-confidence disked M star,
and a yellow circle is a lower-confidence disked M star. The left panel
is USco, and the other two panels are UCL/LCC (the black dots are the same 
in the middle and right panels). The disked stars have an obvious
``pile-up'' near 2 days among the M stars, and there may be a slope
to that 2-day pile-up; see the text. The lack of structure in the last 
panel reinforces our lack of confidence in those disks in general.}
\label{fig:pvmkmstardisks}
\end{figure}

Figure~\ref{fig:pvmkmstardisks} zooms in on the M stars in the $P$
vs.\ \vmkz\ plot for USco, the confident UCL/LCC disks, and the
possible UCL/LCC disks.  As discussed in detail in Paper V, the USco
plot has a clear pileup of disked stars at $\sim$2 days, which we
believe to be a signature of disk locking.  For the confident disks in
UCL/LCC (middle panel), there is a similar very obvious $\sim$2 day
pileup in the mid-M stars.  Using a 2-dimensional 2-sided KS test, the disked and
non-disked samples are statistically significantly different from each
other in the first two panels of Figure~\ref{fig:pvmkmstardisks}.

Few disked M stars are found to rotate faster than $\sim$2 days in
USco (first panel; only 3\% of the disked stars have $P<$1 day,
compared to 26\% of the non-disked). However, there are more disked
stars rotating faster than $\sim$2 days in UCL/LCC (middle panel; 
20\% of the disked stars have $P<$1 day, compared to 32\% of the
non-disked). It is not immediately obvious why this would be the
case. Since source confusion is an important concern, this is 
perhaps the most likely reason, meaning largely TESS confusion, but also
possibly confusion in WISE. These stars could be fast rotating
because they are binaries, but they do not appear to be photometric 
binaries in an optical color-magnitude diagram, nor are they in
general those stars with multiple periods (in {\it K2}, we found that 
in particular, M stars with multiple periods were also often 
photometric binaries). 

\begin{figure}[htb!]
\epsscale{1.0}
\plotone{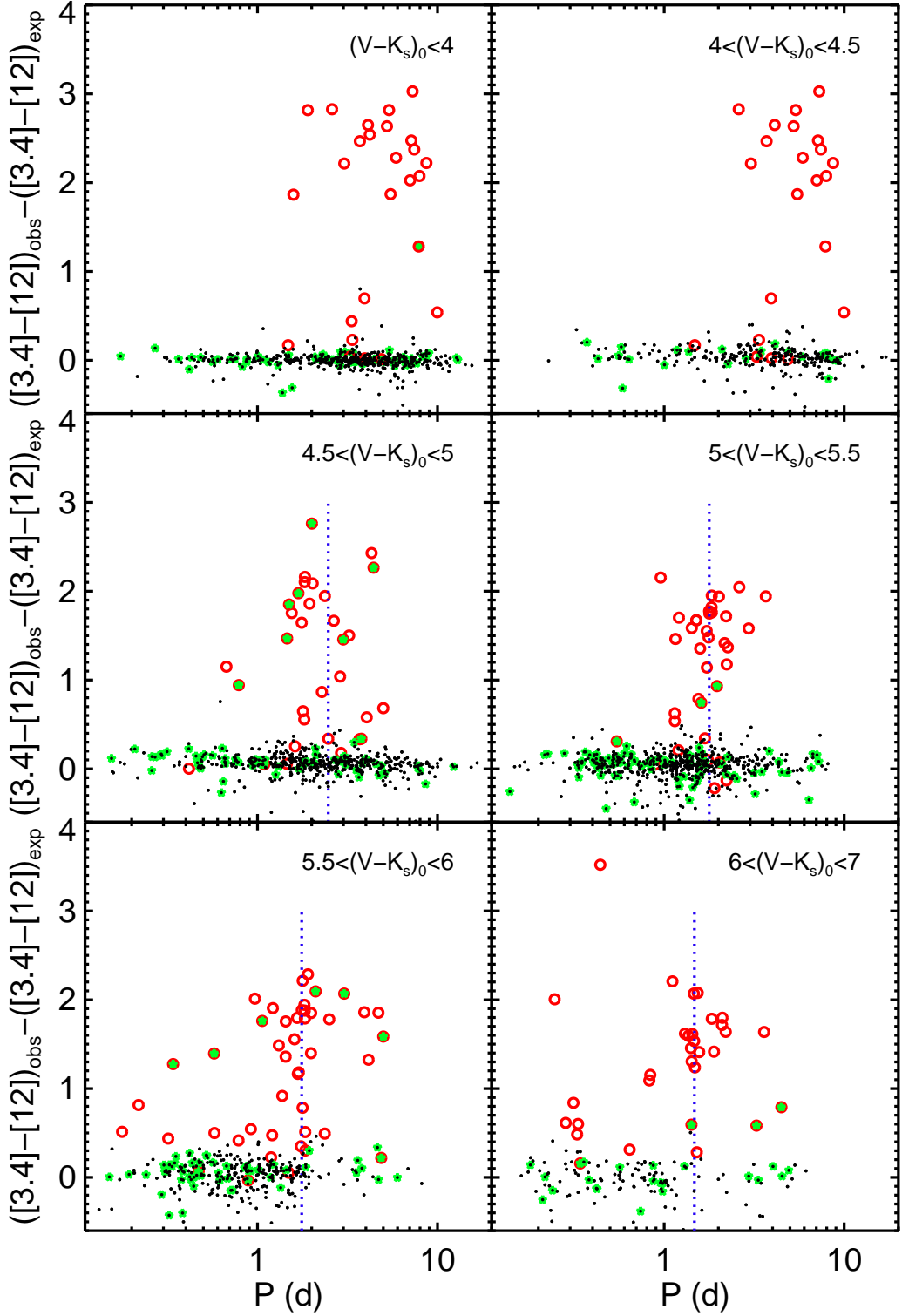}
\caption{WISE IR excess (observed $-$ expected [3.4]$-$[12]) for gold
member stars unlikely to be pulsators,  in bins of \vmkz, as shown in
each panel. Black dots are stars without disks; red circles are stars
with high-confidence disks. An additional  green symbol means that
there are multiple periods associated with the object, e.g., it could
be a multiple or subject to source confusion.  UCL/LCC stars with
disks tend to be rotating slower than the ensemble,  or, rather, fast
rotating stars tend not to have disks.  The vertical blue dotted lines
in the bottom four panels is located at the mean period for disked
stars with  \vmkz$>$4.5 and  $([3.4]-[12])_{\rm observed} -
([3.4]-[12])_{\rm expected} >0.3$ in each panel  (2.5, 1.8, 1.8, \&
1.5 days). Among the disked mid-M stars with large IR excesses, M5
stars are rotating faster than M3 stars.}
\label{fig:irx}
\end{figure}

Figure~\ref{fig:irx} shows the WISE IR excess $([3.4]-[12])_{\rm
observed} - ([3.4]-[12])_{\rm expected})$ as a function of $P$ for
bins in \vmkz. It is straightforward to see that the fast rotating
disked stars do not, in general, have multiple periods, nor are they
the stars with the larger 12 \mum\ IR excesses. 
A handful of these stars do have secondary
periods that are $>$1 day, or have discarded periods that are $>$1 day. 
Follow-up observations of these targets are warranted.
It is still the case, as
it is for USco, that UCL/LCC stars with disks tend to rotate slower
than the ensemble. Alternately, fast rotating stars tend not to have disks. 

There are enough stars in this UCL/LCC member sample that we can break down
the distribution into relatively fine bins of \vmkz, and 
investigate a relationship that was hinted at in USco. In
Fig.~\ref{fig:pvmkmstardisks}, in both USco and UCL/LCC, the pileup at
\vmkz$\sim$5-6 and $\sim$2 days seems  to have a downward slope.
Fig.~\ref{fig:irx} attempts to explore this further; for stars with
\vmkz$>$4.5, it shows the mean $P$  calculated for disked stars with 
$([3.4]-[12])_{\rm observed} - ([3.4]-[12])_{\rm expected} >0.3$  in
each panel. The M3 stars have a mean P near  2.5 days, and the M5
stars are near 1.5 days, suggesting that among the disked stars with
large IR excesses, less massive stars  are rotating faster.

The discussion with Fig.~\ref{fig:pvmkheatmap1} above noted that the
distribution of disk-free UCL/LCC M stars was very sharply peaked at
$P\sim$1.5d, which is quite close to the  pileup of the M star disked
stars at $\sim$1.5-1.8 days.  We speculate whether this clump consists
of stars that have relatively recently shed their dust disks and
therefore unlocked from their disk, but have not yet had the time to
spin up in response, a process that may take a few million years
(e.g., Roquette \etal\ 2021).

We note that the last panel in Fig.~\ref{fig:pvmkmstardisks}
has very little structure reminiscent of any of the structure seen
in the other panels, reinforcing our lack of confidence in those 
stars having believable disks.

\section{Summary and Conclusions}
\label{sec:concl} %\textcolor{red}{sec:concl}

We have presented space-based spot-modulated rotation rates from {\it
TESS} for $\sim$90\% of $\sim$3700 UCL/LCC members ($\sim$91\% of the
$\sim$3000 most likely members).  

We started from a set of members assembled in the literature  largely
from Gaia DR2 analyses. We winnowed down the sample based on what
supporting data we could find, whether or not we could obtain a {\it
TESS} LC, and whether or not that {\it TESS} LC could reliably be tied
to that source. Based on the available information, we placed the
stars into three member bins: gold (the overwhelming majority of the
members), silver, and bronze. We also kept track of the sources
rejected as members. We identified IR excesses from any available IR
band, but WISE-3 (12 \mum) was the most widely available band, which
we used for plotting purposes. 

We identified periods in the {\it TESS} LCs  that compare well with
those in the literature, up to $\sim$15 days of half a {\it TESS}
campaign duration.   We find periods for $\sim$90\% of the UCL/LCC
members,  with $>$60\% of those sinusoidal periods, or close to it,
and interpreted as arising to spots or spot groups that rotate into
and out of view. We find a similar variety of LC and periodogram types
as we found in our {\it K2} work, but we found a lower fraction of
stars with multiple periods. This could be a result of the noise
characteristics of {\it TESS} LCs, but it could also be a result of
the target selection. Most of the targets were identified based on
Gaia selection, which can select against multiplicity.  In our {\it
K2} work, we found that multi-period stars, especially among  the Ms,
were often photometric binaries. Thus, if our input target list was
biased against multiples, it makes sense that our  sample is less
likely to find multi-period stars. However, the  noise characteristics
of the {\it TESS} LCs are indeed very different than {\it K2}, so that
cannot be entirely dismissed as a cause for  finding fewer
multi-period stars.

We find more of the scallop shells, flux dips, and transient flux dips
initially identified in (Stauffer \etal\ 2017, 2018, 2021). 
Fractionally, we find them at a rate of 3\% in the UCL/LCC sample, 
which is quite consistent, given relative ages, with  1\% of the
Pleiades sample, 3\% of the USco sample, and 4\% of the Taurus sample.

The distribution of $P$ vs.\ \vmkz\ color as a proxy for mass, reveals
that UCL/LCC fits in well with evolutionary expectations, given its
age. Figs.~\ref{fig:pvmkheatmap1} and \ref{fig:multicluster} both 
show that the highest and lowest masses probed here are among the
fastest rotators; the slowest rotators are in the middle masses. The
higher masses are not as `organized' as they presumably will be  by
the Pleiades age, but at the same time, they are not as `disorganized'
as they  presumably were at the USco age. The M stars are already
well-organized, and follow the same slope as is found in both USco and
the Pleiades. 

The disk fraction in UCL/LCC is a few percent, and as a result, there
are some disk-influenced  LCs -- we find both bursters and dippers
among the disked stars. The disked M stars form a very obvious pile-up
at about 2 days in the $P$ vs.\ \vmkz\ plot
(Figs.~\ref{fig:pvmkmstardisks} and \ref{fig:irx}), a pile-up that
seems to drift slightly  longer as the mass decreases. Interestingly,
the distribution of disk-free M stars is also sharply peaked at
$\sim$2 days,  at the same location as the disked stars, suggesting
that perhaps these stars have just freed themselves from their disks
but have not yet had the time to react yet.

\begin{acknowledgments}

Some of the data presented in this paper were obtained from the
Mikulski Archive for Space Telescopes (MAST). Support for MAST for
non-HST data is provided by the NASA Office of Space Science via grant
NNX09AF08G and by other grants and contracts. This paper includes data 
collected by the {\it TESS} mission. Funding for the {\it TESS} mission is provided 
by the NASA Explorer Program. 
This research has made use of the NASA/IPAC Infrared Science Archive
(IRSA), which is operated by the Jet Propulsion Laboratory, California
Institute of Technology, under contract with the National Aeronautics
and Space Administration.    This research has made use of NASA's
Astrophysics Data System (ADS) Abstract Service, and of the SIMBAD
database, operated at CDS, Strasbourg, France. 
This research has made use of the VizieR catalogue access tool, CDS,
Strasbourg, France (DOI : 10.26093/cds/vizier). The original description 
of the VizieR service was published in 2000, A\&AS 143, 23.  
This research has made
use of data products from the Two Micron All-Sky Survey (2MASS), which
is a joint project of the University of Massachusetts and the Infrared
Processing and Analysis Center, funded by the National Aeronautics and
Space Administration and the National Science Foundation. The 2MASS
data are served by the NASA/IPAC Infrared Science Archive, which is
operated by the Jet Propulsion Laboratory, California Institute of
Technology, under contract with the National Aeronautics and Space
Administration. This publication makes use of data products from the
Wide-field Infrared Survey Explorer, which is a joint project of the
University of California, Los Angeles, and the Jet Propulsion
Laboratory/California Institute of Technology, funded by the National
Aeronautics and Space Administration. 
\end{acknowledgments}

\facility{K2, 2MASS, WISE, IRSA, Exoplanet Archive}

%\clearpage

\appendix

\section{Rejected/Discarded Sources}
\label{app:nm}

As mentioned in detail in the main body of the paper, there are sources that we
have discarded from the analysis above for any of a number of reasons, ranging from
not having a {\it TESS} LC at all or not having a {\it TESS} LC that was uncontaminated, 
to not having a $V$ or a \ks, or other reasons. This appendix largely consists of
the table of values for these discarded sources, should later investigators wish
to use any of this information. Note that the periods are right for the LCs that we
have, but the LCs {\bf often do not correspond to the stars listed here}; the most common
reason we have for rejecting a source is that there is obvious source confusion
in the {\it TESS} LCs.

\floattable
\begin{deluxetable}{ccp{13cm}}
\tabletypesize{\scriptsize}
%\rotate
\tablecaption{Contents of Table: Periods and Supporting Data for
Discarded/Rejected Sources\label{tab:bignm}}
\tablewidth{0pt}
\tablehead{\colhead{Number} & \colhead{Label} & \colhead{Contents}}
\startdata
1 & TIC & Number in the {\it TESS} Input Catalog (TIC)\\
2 & coord & Coordinate-based (right ascension and declination, J2000) name for target \\
3 & othername & Alternate name for target \\
4 & gaiaid & Gaia DR2 ID \\
5 & distance & Distance from Bailer-Jones \etal\ (2018) in parsecs\\
6 & Kmag & \ks\ magnitude (in Vega mags), if observed\\
7 & vmk-used & \vmk\ used, in Vega mags (observed or inferred; see text)\\
8 & evmk & $E(V-K_s)$ adopted for this star (in mags; see \S~\ref{sec:dereddening}) \\
9 & Kmag0 & dereddened $K_{s,0}$ magnitude (in Vega mags), as inferred (see \S\ref{sec:dereddening})\\
10 & vmk0 & $(V-K_s)_0$, dereddened $V-K_s$ (in Vega mags), as inferred (see \S~\ref{sec:dereddening}; rounded to nearest 0.1 to emphasize the relatively low accuracy)\\
11 & color\_uncertcode & two digit uncertainty code denoting origin of \vmk\ and \vmkz\
(see \S\ref{sec:litphotom} and \ref{sec:dereddening}):
First digit (origin of \vmk): 
1=$V$ measured directly from the literature (including SIMBAD) and $K_s$ from 2MASS; 
2=$V$ from the literature (see \S\ref{sec:litphotom}) and $K_s$ from 2MASS;
3=\vmk\ inferred from Gaia DR1 $G$ and $K_s$ from 2MASS (see \S\ref{sec:litphotom});
4=\vmk\ inferred from Pan-STARRS1 $g$ and $K_s$ from 2MASS (see \S\ref{sec:litphotom});
6=$V$ inferred from well-populated optical SED and $K_s$ from 2MASS (see \S\ref{sec:litphotom});
7=\vmk\ inferred from Gaia DR2 $G$ and $K_s$ from 2MASS (see \S\ref{sec:litphotom});
-9= no measure of \vmk.
Second digit (origin of $E(V-K_s)$ leading to \vmkz): 
1=dereddening from $JHK_s$ diagram (see \S\ref{sec:dereddening});
2=dereddening back to \vmkz\ expected for spectral type;
3=dereddening from SED fits; 
4=used median $E(V-K_s)$=0 (see \S\ref{sec:dereddening});
-9= no measure of  $E(V-K_s)$ \\
12 & P1 & Primary period, in days (taken to be rotation period in cases where there is $>$ 1 period)\\
13 & P2 & Secondary period, in days\\
14 & P3 & Tertiary period, in days\\
15 & P4 & Quaternary period, in days\\
16 & p\_uncertcode & uncertainty code for period -- is there any reason to worry about 
this period? Values are `n' (no worry, full confidence; by far the most common
value), `(n)' (no period), 
and then `n?' and `y?' are progressively less confident periods.\\
17 & IRexcess & Whether an IR excess is present or not (see \S\ref{sec:disks})\\
18 & IRexcessStart & Minimum wavelength at which the IR excess is detected or 
the limit of our knowledge of where there is no excess (see \S\ref{sec:disks}) \\
19 & SEDslope & best-fit slope to all detections between 2 and 25 microns \\
20 & SEDclass & SED class (I, flat, II, or III) based on the SED slope between 2 and 25 microns \\
21 & dipper & LC matches dipper characteristics (see \S\ref{sec:interpofperiods})\\
22 & burster & LC matches burster characteristics (see \S\ref{sec:interpofperiods})\\
23 & single/multi-P &  single or multi-period star \\
24 & dd &  LC and power spectrum matches double-dip characteristics (see \S\ref{sec:interpofperiods})\\
25 & ddmoving & LC and power spectrum matches moving double-dip characteristics (see \S\ref{sec:interpofperiods})\\
26 & shapechanger & LC matches shape changer characteristics (see \S\ref{sec:interpofperiods})\\
27 & beater &  LC has beating visible (see \S\ref{sec:interpofperiods})\\
28 & complexpeak & power spectrum has a complex, structured peak and/or has a wide peak (see \S\ref{sec:interpofperiods})\\
29 & resolvedclose & power spectrum has resolved close peaks (see \S\ref{sec:interpofperiods})\\
30 & resolveddist & power spectrum has resolved distant peaks (see \S\ref{sec:interpofperiods})\\
31 & pulsator & power spectrum and LC match pulsator characteristics (see \S\ref{sec:interpofperiods})\\
32 & scallop &  LC matches scallop or flux dip characteristics (see \S\ref{sec:interpofperiods} and App.~\ref{app:weirdos}) \\
33 & EB &  LC has characteristics of eclipsing binary (see \S\ref{sec:interpofperiods} and App.~\ref{app:weirdos}) \\
\enddata
\end{deluxetable}

\section{Comparison to Literature Periods}
\label{app:lit}

Section~\ref{sec:comparelit} discussed in broad terms the comparison
of our period finding to that from the literature.  In this appendix,
we list periods for $\sim$600 specific stars in common between our study and
several literature studies, regardless of whether or not the stars are
currently thought to be UCL/LCC members or not. The  stars are listed
in Table~\ref{tab:comparelit}. If the {\it TESS} period  matches the
literature period within 15\%, we take it as a match.  If the {\it
TESS} period is a match between 15 and 20\%, we take it as  close but
not quite  a match. Several periods are too long for {\it TESS} to
have recovered.  In a few cases, the literature period matches a
provisional period identified in {\it TESS} or in the literature; in
some additional  cases, there is a peak in {\it TESS} at or near the
literature $P$, but we had discarded the {\it TESS} $P$ as  unlikely
to be astrophysical and/or more likely to be a timescale than a 
rotation period. See the notes in the table.

\floattable
\begin{deluxetable}{ccp{8cm}}
\tablecaption{Contents of Table: Comparison to Literature Periods\label{tab:comparelit}}
\tablewidth{0pt}
\tablewidth{0pt}
\tablehead{\colhead{Number} & \colhead{Label} & \colhead{Contents}}
\startdata
1 & tic & TIC number \\
2 & catnum & Position-based catalog number \\
3 & pkiraga & $P$ in days from Kiraga (2012) \\
4 & pmellon & $P$ in days from Mellon \etal\ (2017) \\
5 & pgdr2 & $P$ in days from Gaia DR2 \\
6 & pasassn & $P$ in days from ASAS-SN \\
7 & plit & $P$ in days from elsewhere in the literature \\
8 & pcite & citation for literature source of $P$ in col 7 \\
9 & p1 & first period reported here \\
10 & p2 & second period reported here \\
11 & p3 & third period reported here \\
12 & p4 & fourth period reported here \\
13 & match? & is this a match?\\
14 & notes & notes \\
\enddata
\end{deluxetable}

\section{Unusual LCs}
\label{app:weirdos}

As discussed in Sec.~\ref{sec:interpofperiods},  most of the LCs
considered here are well-behaved, e.g., sinusoidal and  clearly
periodic (or clearly not), and the periodic signal is easily
interpreted as the rotation of the star. However, there are $\sim$150
stars ($<$4\% of the entire sample) that we wanted to call out here
because they are unusual in some way. They are all listed in
Table~\ref{tab:weirdos} with notes as to why they merited listing
here. Parameters for all of them (including periods) are listed in the
big data tables above; this table here is largely notes on these
stars. The table includes TIC number, coordinate-based name, what
member sample they are in (gold, silver, bronze, or reject), how
confident we are that they should be in the category we have placed
them (e.g.,  scallop or EB, etc.) and whether or not the detected
period is due to rotation, and any additional notes about them.  In
the context of the entire sample (or the entire member sample), there
are very few for which we are not confident about the bin in which we
have placed them, and errors in their categorization doesn't affect
any of our conclusions. More detailed, statistical characterization of
any of these these stars or categories is beyond the scope of the
present paper,  so they are simply listed here. 

This list includes scallops and flux dips published in Stauffer \etal\
(2021), two of which have had to be rejected here due to source
confusion; it also includes new scallops and flux dips identified
here, as well as stars that may be scallops or flux dips but that are
lower signal-to-noise. 

Also in the table are candidate pulsators, including some that have
phased waveforms that resemble a `ski jump' -- e.g., they look like
those  of RR Lyr, but these are, in general, very fast (hours).  For
the pulsators, note that their periods appear in the tables above, 
but they are omitted from the plots involving $P_{\rm rot}$, since
they are not  rotation. 

Many obvious eclipsing binaries (EBs) appear in the LCs; some phased
LCs could be EBs or could be flux dips, and it's very hard to
determine from the available (often low signal-to-noise) LC which one
it is. In cases of obvious EBs, we have omitted $P_{\rm EB}$ from
plots, but have retained $P_{\rm rot}$ where it was possible to
derive a plausible rotation period estimate from the LC. Of the EBs,
about half of them appear above the MS, e.g., consistent with being a
binary, in Fig.~\ref{fig:distancescmd}.

\startlongtable
\begin{deluxetable}{llllp{5cm}}
\tabletypesize{\scriptsize}
\tablecaption{List of Unusual Objects\label{tab:weirdos}}
\tablewidth{0pt}
\tablehead{\colhead{TIC} & \colhead{Name} 
& \colhead{Member bin} & \colhead{sample confidence\tablenotemark{a}} & \colhead{notes}  }
\startdata
405235910	&	102115.42-622604.3	&	gold	&	high but not rotation	&	pulsator			\\
93952051	&	102646.51-595526.8	&	silver	&	low but not rotation	&	pulsator?			\\
273703001	&	103137.10-690158.7	&	bronze	&	high but not rotation	&	pulsator			\\
412006082	&	103341.80-641345.7	&	gold	&	high but not rotation	&	pulsator			\\
242558208	&	103422.70-641809.3	&	silver	&	high but not rotation	&	pulsator			\\
351521478	&	103819.04-612100.7	&	silver	&	high but not rotation	&	pulsator			\\
390470134	&	103844.36-595443.7	&	silver	&	high but not rotation	&	pulsator			\\
419686932	&	104122.99-694043.1	&	silver	&	high but not rotation	&	pulsator			\\
92574021	&	113249.19-494907.8	&	gold	&	low	&	scallop?			\\
451984338	&	113319.04-594822.0	&	gold	&	high but not rotation	&	EB, only $P$ is that of binary		; $P_{\rm EB}$=1.132	\\
290889135	&	113556.21-653012.1	&	gold	&	high	&	scallop in p2			\\
280945693	&	113616.71-692751.8	&	gold	&	high	&	Published as flux dip in Stauffer+21			\\
452345586	&	113658.55-582237.5	&	gold	&	high	&	flux dip			\\
93763678	&	114322.41-532711.4	&	gold	&	high	&	flux dip			\\
323478101	&	114557.95-635246.3	&	gold	&	high but not rotation	&	EB, only $P$ is that of binary		; $P_{\rm EB}$=0.477	\\
296790810	&	114625.66-664135.6	&	gold	&	high	&	Published as flux dip in Stauffer+21; possible small IR excess			\\
268665785	&	114902.67-570014.1	&	gold	&	low	&	possible EB; extracted $P_{\rm rot}$			\\
301432612	&	115404.70-580241.3	&	gold	&	high	&	Published as scallop in Stauffer+21			\\
307686978	&	120017.79-635549.0	&	bronze	&	high	&	EB, but can pull out $P_{\rm rot}$		; $P_{\rm EB}$=2.778	\\
398768350	&	120119.80-564902.7	&	gold	&	high	&	Published as scallop in Stauffer+21			\\
379774242	&	120504.18-644721.6	&	gold	&	low	&	scallop?			\\
994964114	&	120858.09-513019.6	&	reject	&	(reject)	&	source confusion so rejected, but p1 could be a flux dip			\\
994964138	&	120858.19-513017.2	&	reject	&	(reject)	&	source confusion so rejected, but p1 could be a flux dip			\\
334409011	&	121318.43-515641.6	&	gold	&	high	&	EB, but can pull out $P_{\rm rot}$		; $P_{\rm EB}$=3.361	\\
288093002	&	122028.00-543537.6	&	gold	&	high	&	Published as flux dip in Stauffer+21			\\
310412874	&	122048.65-640904.1	&	silver	&	low	&	flux dip? Small IR excess			\\
448002486	&	122138.04-690838.4	&	gold	&	high	&	Published as flux dip in Stauffer+21			\\
310720311	&	122153.09-634733.9	&	gold	&	high	&	Published as scallop in Stauffer+21			\\
411614400	&	122201.40-573757.1	&	gold	&	high but not rotation	&	EB, only $P$ is that of binary		; $P_{\rm EB}$=1.535	\\
135162879	&	122213.16-414802.7	&	gold	&	high	&	Published as scallop in Stauffer+21			\\
311447879	&	122220.20-650950.6	&	bronze	&	low	&	flux dip?			\\
311333943	&	122247.22-633757.6	&	silver	&	low	&	flux dip, possible IR excess, possible icicles?			\\
261573174	&	122303.08-542526.9	&	silver	&	low	&	possible EB; extracted $P_{\rm rot}$		; $P_{\rm EB}$=1.084	\\
311592558	&	122348.30-633248.7	&	silver	&	low	&	possible EB; extracted $P_{\rm rot}$			\\
311585720	&	122354.75-641730.0	&	silver	&	high but not rotation	&	EB, only $P$ is that of binary		; $P_{\rm EB}$=0.612	\\
450386147	&	122424.02-645344.0	&	gold	&	low	&	flux dip?			\\
273460357	&	122500.61-521627.1	&	reject	&	(reject)	&	source confusion so rejected, but an interesting LC shape.			\\
273460338	&	122501.42-521614.6	&	reject	&	(reject)	&	source confusion so rejected, but an interesting LC shape.			\\
281742840	&	122504.80-655942.1	&	bronze	&	high but not rotation	&	EB, only $P$ is that of binary		; $P_{\rm EB}$=4.760	\\
271221172	&	122657.75-554620.0	&	gold	&	high but not rotation	&	EB, only $P$ is that of binary		; $P_{\rm EB}$=0.266	\\
450957950	&	122902.23-645500.6	&	gold	&	high but not rotation	&	EB, only $P$ is that of binary		; $P_{\rm EB}$=2.360	\\
272407484	&	123108.44-545644.5	&	gold	&	low	&	flux dip in p2; clear IR excess.			\\
179968331	&	123410.56-635241.6	&	gold	&	high but not rotation	&	EB, only $P$ is that of binary		; $P_{\rm EB}$=2.825	\\
273821594	&	123624.97-550710.6	&	gold	&	high	&	scallop and flux dip?			\\
411681763	&	123742.99-554851.2	&	gold	&	high but not rotation	&	EB, only $P$ is that of binary		; $P_{\rm EB}$=1.415	\\
73183013	&	123941.30-464444.6	&	gold	&	high	&	three dips per cycle, still sinusoidal			\\
161734785	&	124147.22-511006.7	&	gold	&	high	&	Published as scallop in Stauffer+21			\\
412182239	&	124231.62-590512.6	&	silver	&	low	&	Two very clear v-shaped dips per cycle, not evenly spaced. Substantial IR excess \& long $P$ (1.3 d), both of which are inconsistent with scallop category. Counted as scallop but..?			\\
328468855	&	124301.62-675620.8	&	gold	&	high	&	flux dip			\\
419502814	&	124434.42-551246.5	&	gold	&	high	&	EB, but can pull out $P_{\rm rot}$		; $P_{\rm EB}$=1.059	\\
419779703	&	124631.20-540431.4	&	gold	&	high	&	scallop			\\
405695928	&	124826.12-553816.1	&	silver	&	high	&	flux dip			\\
405754448	&	124831.44-594449.8	&	gold	&	low	&	scallop shape but at \vmkz$\sim$3, this is too blue for a scallop, and at $P\sim$0.5, it is much faster than other stars of its color. Source confusion?			\\
165904363	&	124942.09-495316.3	&	gold	&	high	&	flux dip			\\
165912675	&	125033.96-482656.5	&	gold	&	low	&	flux dip?			\\
405970436	&	125058.63-593400.6	&	gold	&	high	&	scallop			\\
405910532	&	125100.26-564313.1	&	reject	&	(reject)	&	source confusion so rejected, but could be EB			\\
405910546	&	125100.61-564321.0	&	reject	&	(reject)	&	source confusion so rejected, but could be EB			\\
412376751	&	125129.47-592421.7	&	gold	&	low	&	flux dip?			\\
412376096	&	125137.68-592945.0	&	bronze	&	high	&	flux dip			\\
435899024	&	125222.75-641839.4	&	gold	&	high	&	Published as scallop in Stauffer+21; possible IR excess			\\
406040223	&	125521.95-584641.8	&	gold	&	high	&	Published as flux dip in Stauffer+21			\\
248145126	&	125545.32-445151.7	&	gold	&	high	&	Published as scallop in Stauffer+21; possible IR excess			\\
335598085	&	125935.63-680801.6	&	gold	&	high	&	Published as scallop in Stauffer+21			\\
253239639	&	130159.57-592023.8	&	gold	&	high but not rotation	&	`ski jump'; shaped like RR Lyr but very fast			\\
404335106	&	130317.73-515013.0	&	gold	&	low	&	flux dip, though at $\sim$3 days, the period is very long compared to other flux dips.			\\
404387832	&	130354.33-501502.3	&	gold	&	high	&	scallop			\\
439883940	&	130416.15-640557.0	&	silver	&	high	&	EB, but can pull out $P_{\rm rot}$		; $P_{\rm EB}$=6.176	\\
258583707	&	130730.60-454919.7	&	gold	&	low	&	flux dip? EB? Counted as flux dip			\\
441263248	&	130927.70-653325.5	&	gold	&	high	&	EB, but can pull out $P_{\rm rot}$		; $P_{\rm EB}$=6.877	\\
245002119	&	130949.27-513546.3	&	gold	&	high but not rotation	&	EB, only $P$ is that of binary		; $P_{\rm EB}$=0.581	\\
363656704	&	131920.44-450800.7	&	gold	&	high	&	EB, but can pull out $P_{\rm rot}$			\\
973449111	&	131935.91-623436.9	&	gold	&	high	&	Published as flux dip in Stauffer+21			\\
438709950	&	132424.75-475730.1	&	gold	&	low	&	flux dip?			\\
243192504	&	133157.69-470614.9	&	gold	&	high	&	EB, but can pull out $P_{\rm rot}$			\\
261272259	&	133521.22-493921.0	&	bronze	&	high	&	EB? Dominant $P$ is $P_{\rm rot}$, with eclipses			\\
241375625	&	133852.06-510507.0	&	gold	&	high	&	flux dip in p2			\\
243381460	&	134001.65-434857.3	&	gold	&	high	&	Published as scallop in Stauffer+21			\\
243449997	&	134233.80-460826.4	&	gold	&	high	&	flux dip			\\
243499565	&	134435.82-470613.8	&	gold	&	high	&	Published as flux dip in Stauffer+21			\\
166302995	&	134518.06-410202.2	&	gold	&	low	&	possible EB; extracted $P_{\rm rot}$		; $P_{\rm EB}$=2.881	\\
207621404	&	134732.79-553301.9	&	gold	&	high	&	Published as scallop in Stauffer+21			\\
243611773	&	134816.29-440238.8	&	gold	&	high	&	Published as flux dip in Stauffer+21; flux dip in p2			\\
208351772	&	135345.11-554420.0	&	silver	&	high	&	EB, but can pull out $P_{\rm rot}$		; $P_{\rm EB}$=5.003	\\
448165364	&	135407.42-673344.9	&	gold	&	high	&	three dips per cycle, still sinusoidal			\\
312410638	&	140146.25-501537.3	&	gold	&	high	&	flux dip			\\
328906141	&	140554.60-522600.5	&	gold	&	high	&	Published as scallop in Stauffer+21			\\
179368022	&	140729.83-385427.0	&	gold	&	high	&	scallop in p2			\\
179367270	&	140731.25-393216.8	&	gold	&	high	&	flux dip, though at $\sim$3 days, the period is very long compared to other flux dips.			\\
329694185	&	140953.34-521717.7	&	silver	&	high	&	scallop			\\
330560000	&	141424.51-510319.3	&	gold	&	high	&	Published as flux dip in Stauffer+21			\\
242407571	&	141424.86-455643.6	&	gold	&	high	&	Published as scallop in Stauffer+21; noted has having second period that manifests as ``icicles''			\\
330791148	&	141529.27-495747.5	&	gold	&	high	&	EB? Can pull out $P_{\rm rot}$		; $P_{\rm EB}$=3.430	\\
448852739	&	141605.67-691735.8	&	gold	&	high	&	EB, but can pull out $P_{\rm rot}$			\\
167448346	&	142229.54-385517.1	&	gold	&	high but not rotation	&	EB, only $P$ is that of binary		; $P_{\rm EB}$=0.548	\\
242594123	&	142347.41-432457.3	&	gold	&	high but not rotation	&	EB, only $P$ is that of binary		; $P_{\rm EB}$=2.548	\\
241841997	&	142758.40-392328.7	&	gold	&	low	&	scallop in p3, or possible EB. P3 is $<$0.5 days, so counted as scallop.			\\
127309526	&	142759.62-432629.2	&	gold	&	high	&	Published as flux dip in Stauffer+21; flux dip in p2			\\
211513644	&	142809.56-491545.9	&	gold	&	high	&	Published as flux dip in Stauffer+21			\\
127246012	&	142829.78-455715.4	&	gold	&	high	&	scallop			\\
241884143	&	143112.81-410358.6	&	gold	&	low	&	Could be flux dip in p2, could be EB. High enough in CMD to be photometric binary. Counted as flux dip.			\\
127866051	&	143333.84-474354.5	&	silver	&	high	&	scallop			\\
159427926	&	143958.40-402809.3	&	reject	&	(reject)	&	source confusion so rejected, but p1 is a flux dip			\\
159427927	&	143958.71-402809.4	&	reject	&	(reject)	&	source confusion so rejected, but p1 is a flux dip			\\
129116176	&	144134.97-470029.3	&	gold	&	high but not rotation	&	EB, only $P$ is that of binary		; $P_{\rm EB}$=2.023	\\
129116164	&	144139.29-470015.1	&	silver	&	high but not rotation	&	EB, only $P$ is that of binary		; $P_{\rm EB}$=2.019	\\
129309458	&	144343.05-433756.0	&	gold	&	high	&	flux dip			\\
309321971	&	145537.81-493427.7	&	gold	&	high	&	scallop			\\
461643692	&	145823.13-334415.4	&	gold	&	high	&	Published as flux dip in Stauffer+21			\\
334325329	&	150031.41-462131.4	&	gold	&	high	&	flux dip			\\
75489110	&	150033.42-342954.6	&	gold	&	high but not rotation	&	`ski jump' in P2; shaped like RR Lyr but very fast			\\
334838280	&	150234.31-441933.2	&	gold	&	low	&	flux dip?			\\
121727134	&	150455.33-375746.9	&	gold	&	high but not rotation	&	EB, contact binary, only $P$ is that of binary		; $P_{\rm EB}$=0.374	\\
121840452	&	150516.87-381412.4	&	gold	&	high	&	Published as scallop in Stauffer+21			\\
366178112	&	150638.97-435728.9	&	gold	&	low	&	possible `ski jump'? Or flux dip? Unclear.			\\
160539036	&	150757.83-440458.7	&	gold	&	low	&	flux dip in p2			\\
76048114	&	150804.78-314231.8	&	reject	&	(reject)	&	source confusion so rejected, but p2 is a scallop			\\
76048113	&	150805.11-314231.4	&	reject	&	(reject)	&	source confusion so rejected, but p2 is a scallop			\\
140765939	&	151339.89-434020.8	&	gold	&	high	&	flux dip			\\
276502773	&	151433.88-405419.8	&	silver	&	high	&	scallop			\\
76698847	&	151616.51-322309.6	&	gold	&	high	&	flux dip			\\
148218929	&	151858.60-374518.5	&	gold	&	high but not rotation	&	EB, only $P$ is that of binary		; $P_{\rm EB}$=1.768	\\
272398365	&	151903.08-321453.9	&	gold	&	high	&	flux dip			\\
89026133	&	152234.45-350414.6	&	reject	&	(reject)	&	Published as flux dip in Stauffer+21; possible IR excess; now have to reject for source confusion			\\
89026136	&	152235.51-350357.7	&	reject	&	(reject)	&	Published as flux dip in Stauffer+21; possible IR excess; now have to reject for source confusion			\\
147828926	&	152257.19-455619.2	&	gold	&	high	&	scallop			\\
173267960	&	153758.40-403805.3	&	gold	&	high	&	flux dip in p3			\\
176205621	&	154212.72-410955.5	&	gold	&	high	&	flux dip			\\
177906792	&	154426.96-362543.2	&	gold	&	low	&	scallop in p2			\\
254612758	&	154508.33-442607.9	&	gold	&	high	&	Published as flux dip in Stauffer+21			\\
99207324	&	154606.01-352510.7	&	gold	&	high	&	Published as flux dip in Stauffer+21			\\
179357494	&	154713.17-411451.1	&	gold	&	low	&	flux dip?			\\
442575691	&	154802.96-305431.8	&	gold	&	low	&	scallop?			\\
364291454	&	155607.07-400036.2	&	gold	&	low	&	p1 is `ski jump' (shaped like RR Lyr but very fast); p2 is flux dip.			\\
1172201258	&	155709.59-320434.2	&	reject	&	(reject)	&	source confusion so rejected, but p1 is a flux dip			\\
1172201260	&	155709.76-320434.3	&	reject	&	(reject)	&	source confusion so rejected, but p1 is a flux dip			\\
58753591	&	155758.97-415846.6	&	gold	&	high but not rotation	&	`ski jump'; shaped like RR Lyr but very fast			\\
279691401	&	155915.83-365712.3	&	gold	&	high but not rotation	&	`ski jump'; shaped like RR Lyr but very fast			\\
69874547	&	160556.03-373748.2	&	silver	&	high but not rotation	&	EB, only $P$ is that of binary		; $P_{\rm EB}$=0.795	\\
4231194	&	162231.10-410527.8	&	gold	&	low	&	Possible EB	; $P_{\rm EB}$= 0.284		\\
223461868	&	162508.50-415446.5	&	silver	&	high	&	EB, but can pull out $P_{\rm rot}$		; $P_{\rm EB}$=1.325	\\
84187344	&	163854.75-405519.8	&	silver	&	low	&	flux dip? `icicle'? $\sim$4d period too long for flux dip but does not look like EB.			\\
84756561	&	163952.56-394431.1	&	gold	&	high but not rotation	&	EB, contact binary, only $P$ is that of binary		; $P_{\rm EB}$=0.449	\\
85125492	&	164033.61-390722.1	&	gold	&	low	&	flux dip? Blue (\vmkz$\sim$4) for a flux dip, and has a significant IR excess. 			\\
85356121	&	164116.54-382322.8	&	gold	&	low	&	flux dip? 			\\
77919426	&	164400.33-395943.8	&	gold	&	high but not rotation	&	EB, only $P$ is that of binary		; $P_{\rm EB}$=1.333	\\
96140771	&	164455.86-341043.3	&	gold	&	high	&	EB, but can pull out $P_{\rm rot}$		; $P_{\rm EB}$=1.290	\\
97239026	&	164717.49-324535.8	&	gold	&	low	&	scallop?			\\
337253450	&	165043.59-371212.9	&	gold	&	high	&	EB, but can pull out $P_{\rm rot}$		; $P_{\rm EB}$=1.927	\\
337243787	&	165050.98-381226.5	&	gold	&	low	&	flux dip			\\
191694822	&	165108.97-330059.4	&	gold	&	high	&	EB? Can pull out $P_{\rm rot}$			\\
\enddata
\tablenotetext{a}{``Sample confidence'' means how confident we are in both
the placement in the category (e.g., `scallop' or `EB') and in the measured
period being rotation.}
\end{deluxetable}

\section{Timescales}
\label{app:timescales}

Some LCs have some repeated patterns that we cannot
identify with certainty as a rotation period. Table~\ref{tab:timescales}
summarizes these timescales for the stars out of the entire ensemble. 
Note the frequent appearance of 6-8 d timescales (or multiples thereof) 
-- this may be a timescale introduced by {\it TESS} or the data reduction 
(see Sec.~\ref{sec:lookingforperiods}).

\startlongtable
%\floattable
\begin{deluxetable}{ccl}
\tabletypesize{\scriptsize}
%\rotate
\tablecaption{List of Objects with Timescales\label{tab:timescales}}
\tablewidth{0pt}
\tablehead{\colhead{TIC} & \colhead{Name} 
& \colhead{Timescale (d) and/or Notes}  }
\startdata
0311456083 & 122327.67-653614.1 &   2.4? 6.5? \\
0425319705 & 124406.63-555418.2 &   7.5? \\
0253309679 & 130350.87-582054.7 &   15?   \\
0253451463 & 130713.28-542732.7 &   6?  \\
0253455154 & 130807.49-543230.9 &   8? \\
0406246159 & 131022.15-591740.0 &   5? \\
0253957851 & 131600.11-570415.6 &   6? \\
0245266758 & 131723.19-512217.4 &  2? 5? 14? has IR excess  \\
1048575980 & 132507.42-430409.8 &  4? 8? very close to NGC 5128 (Cen A)  \\
0243222122 & 133408.54-472700.9 &  15?  \\
0243249579 & 133419.83-433250.1 &  4?  \\
0413084801 & 133908.59-585128.6 &  7?  \\
1052983805 & 134132.37-443044.0 &  20?  \\
0275884492 & 135209.38-433223.6 &  4?  \\
0241684780 & 135230.72-515337.2 &  5? 6?  \\
0208317834 & 135324.21-574630.0 &  5?  done on portion of LC  \\
0242153310 & 140108.61-443029.4 &  8?  \\
0329007676 & 140639.03-502509.3 &  5?  \\
0179410344 & 140809.36-391251.5 &  6?  \\
0242548815 & 142119.68-474532.7 &  7?  \\
0180028859 & 142123.18-390714.4 &  7?  \\
0392752922 & 143120.10-350835.6 &  4?  \\
0289425666 & 143749.01-492827.4 &  7.2?  \\
0159656935 & 144223.07-404149.7 &  2? 12?  \\
0159711236 & 144346.08-371908.9 &  3? 0.4? done on portion of LC  \\
0159751326 & 144426.97-415323.0 &  2? 4? has IR excess \\
0451276379 & 144431.39-485707.1 &  451276379 \& 451276381 are confused; 4?  \\
0451276381 & 144431.74-485710.3 &  451276379 \& 451276381 are confused; 4?  \\
0460218478 & 144911.31-500114.0 &  6?  \\
0335145136 & 150425.51-453050.5 &  7? 11?  \\
0121932858 & 150618.61-412016.9 &  1.3?  \\
0160476756 & 150708.81-445555.0 &  6.4?  \\
0160818175 & 150947.71-461948.4 &  4?  \\
0185053184 & 152803.22-260003.4 &  7?  \\
0054516263 & 152812.46-343606.0 &  5?  \\
0153781950 & 153141.33-472926.2 &  9? 5?   \\
0171609844 & 153538.29-410709.2 &  8?  \\
0172608416 & 153650.92-384138.6 &  5?   \\
0290559990 & 153745.91-423656.8 &  5? 15?    \\
0254394156 & 154034.80-445322.3 &  15?  \\
0178767016 & 154659.81-395344.3 &  7? 12?  \\
0179347980 & 154711.40-392414.4 &  8?   \\
0179058102 & 154723.35-364411.6 &  6? 5?  \\
0442586515 & 154854.12-351319.2 &  7?   \\
0442623945 & 154912.10-353905.5 &  10? disky LC, and has IR excess   \\
0058245194 & 155302.65-363305.7 &  7?   \\
0058170116 & 155307.05-385500.4 &  7?  \\
0255260454 & 155721.11-433342.9 &  5?  \\
0279883855 & 155916.48-415710.7 &  7? disky LC, and has IR excess; burster  \\
0059016771 & 155958.21-360255.6 &  12? matches literature!  \\
0059501088 & 160233.48-362923.4 &  59501088 \& 59501092 are confused; 5?  \\
0059501092 & 160233.48-362928.3 &  59501088 \& 59501092 are confused; 5?  \\
0257368371 & 160317.74-421502.5 &  5? has IR excess \\
0256901758 & 160319.73-442214.2 &  4?  \\
0069514999 & 160445.05-380512.0 &  3.5? 7?   \\
0457995946 & 160854.69-393743.6 &  disky, 2.4? has IR excess  \\
0382700580 & 160928.01-384854.2 &  7?  \\
0382714814 & 160942.04-405225.4 &  7?  \\
0164027526 & 161152.32-420436.2 &  2?  \\
0004231380 & 162250.15-410301.2 &  6?  \\
0225431084 & 162828.83-444903.2 &  2? short-P thing seen by eye, cannot recover   \\
0029459205 & 163121.55-394350.8 &  7?  \\
0291256981 & 163308.67-382215.2 &  8?  \\
0083468373 & 163741.14-370654.4 &  8?  \\
0458060193 & 164156.02-395300.7 &  6?  \\
0458203191 & 164212.49-392611.5 &  7?  \\
0077918101 & 164426.14-400803.6 &  7?  \\
0078874478 & 164619.08-392723.8 &  8?  \\
0337243291 & 165046.65-381526.2 &  2?  \\
\enddata
\end{deluxetable}


\begin{thebibliography}{}
\bibitem[AKARI team(2010a)]{akarifis} AKARI team, 2010a, AKARI/FIS All-Sky Survey Bright Source Catalogue, IRSA, doi:10.26131/IRSA180 
\bibitem[AKARI team(2010b)]{akariirc} AKARI team, 2010b, AKARI/IRC Point Source Catalogue, IRSA, doi:10.26131/IRSA181
\bibitem[Akeson et al.(2013)]{2013PASP..125..989A} Akeson, R.~L., Chen, X., Ciardi, D., et al.\ 2013, \pasp, 125, 989. doi:10.1086/672273
\bibitem[Alfonso-Garz{\'o}n et al.(2012)]{2012A&A...548A..79A} Alfonso-Garz{\'o}n, J., Domingo, A., Mas-Hesse, J.~M., et al.\ 2012, \aap, 548, A79. doi:10.1051/0004-6361/201220095
\bibitem[Avallone et al.(2022)]{2022arXiv220315116A} Avallone, E.~A., Tayar, J.~N., van Saders, J.~L., et al.\ 2022, arXiv:2203.15116
\bibitem[Bailer-Jones et al.(2018)]{bj18} Bailer-Jones, C.~A.~L., Rybizki, J., Fouesneau, M., et al.\ 2018, \aj, 156, 58. doi:10.3847/1538-3881/aacb21
\bibitem[Bailer-Jones et al.(2021)]{2021AJ....161..147B} Bailer-Jones, C.~A.~L., Rybizki, J., Fouesneau, M., et al.\ 2021, \aj, 161, 147. doi:10.3847/1538-3881/abd806
\bibitem[Batalha et al.(1998)]{1998A&AS..128..561B} Batalha, C.~C., Quast, G.~R., Torres, C.~A.~O., et al.\ 1998, \aaps, 128, 561. doi:10.1051/aas:1998163
\bibitem[Blaauw(1964)]{1964ARA&A...2..213B} Blaauw, A.\ 1964, \araa, 2, 213. doi:10.1146/annurev.aa.02.090164.001241
\bibitem[Bouma et al.(2019)]{cdips}Bouma, L., Hartman, J., Bhatti, W,., Winn, J., Bakos, G., 2019, \apjs, 245, 13. doi:10.17909/t9-ayd0-k727
\bibitem[Bowler et al.(2019)]{2019ApJ...877...60B} Bowler, B.~P., Hinkley, S., Ziegler, C., et al.\ 2019, \apj, 877, 60. doi:10.3847/1538-4357/ab1018
\bibitem[Broeg et al.(2007)]{2007A&A...468.1039B} Broeg, C., Schmidt, T.~O.~B., Guenther, E., et al.\ 2007, \aap, 468, 1039. doi:10.1051/0004-6361:20066793
\bibitem[Burke et al.(2020)]{tesspoint}Burke, C. J., Levine, A., Fausnaugh, M., Vanderspek, R., Barclay, T., Libby-Roberts, J. E., Morris, B., Sipocz, B., Owens, M., Feinstein, A. D., Camacho, J., 2020, 0.4.1, Astrophysics Source Code Library, record ascl:2003:001
\bibitem[Capak et al.(2013)]{seip}Capak, P., 2013, Spitzer Enhanced Imaging Products (SEIP) Source List, IRSA, doi:10.26131/IRSA3
\bibitem[Cardelli et al.(1989)]{1989ApJ...345..245C} Cardelli, J.~A., Clayton, G.~C., \& Mathis, J.~S.\ 1989, \apj, 345, 245. doi:10.1086/167900
\bibitem[Carpenter et al.(2009)]{2009ApJS..181..197C} Carpenter, J.~M., Bouwman, J., Mamajek, E.~E., et al.\ 2009, \apjs, 181, 197. doi:10.1088/0067-0049/181/1/197
\bibitem[CatWISE team(2020)]{catwisedata} CatWISE team, 2020, CatWISE Preliminary Catalog, IPAC, doi: 10.26131/IRSA126
\bibitem[Chambers et al.(2016)]{panstarrs}Chambers, K., Magnier, E., Metcalf, N., \etal, 2016, arXiv:1612.05560
\bibitem[Chen et al.(2014)]{2014ApJS..211...25C} Chen, C.~H., Mittal, T., Kuchner, M., et al.\ 2014, \apjs, 211, 25. doi:10.1088/0067-0049/211/2/25
\bibitem[Christiansen et al.(2008)]{2008MNRAS.385.1749C} Christiansen, J.~L., Derekas, A., Kiss, L.~L., et al.\ 2008, \mnras, 385, 1749. doi:10.1111/j.1365-2966.2008.13013.x
\bibitem[Cody \& Hillenbrand(2018)]{uscodisks} Cody, A.~M. \& Hillenbrand, L.~A.\ 2018, \aj, 156, 71. doi:10.3847/1538-3881/aacead
\bibitem[Cody et al.(2014)]{csi} Cody, A.~M., Stauffer, J., Baglin, A., et al.\ 2014, \aj, 147, 82. doi:10.1088/0004-6256/147/4/82
\bibitem[Cody et al.(2017)]{uscoburster} Cody, A.~M., Hillenbrand, L.~A., David, T.~J., et al.\ 2017, \apj, 836, 41. doi:10.3847/1538-4357/836/1/41
\bibitem[Coker et al.(2016)]{2016ApJ...833..122C} Coker, C.~T., Pinsonneault, M., \& Terndrup, D.~M.\ 2016, \apj, 833, 122. doi:10.3847/1538-4357/833/1/122
\bibitem[Comer{\'o}n et al.(2009)]{2009A&A...500.1045C} Comer{\'o}n, F., Spezzi, L., \& L{\'o}pez Mart{\'\i}, B.\ 2009, \aap, 500, 1045. doi:10.1051/0004-6361/200911771
\bibitem[Cotten \& Song(2016)]{2016ApJS..225...15C} Cotten, T.~H. \& Song, I.\ 2016, \apjs, 225, 15. doi:10.3847/0067-0049/225/1/15
\bibitem[Cruzal{\`e}bes et al.(2019)]{2019MNRAS.490.3158C} Cruzal{\`e}bes, P., Petrov, R.~G., Robbe-Dubois, S., et al.\ 2019, \mnras, 490, 3158. doi:10.1093/mnras/stz2803
\bibitem[Curtis et al.(2020)]{2020ApJ...904..140C} Curtis, J.~L., Ag{\"u}eros, M.~A., Matt, S.~P., et al.\ 2020, \apj, 904, 140. doi:10.3847/1538-4357/abbf58
\bibitem[Damiani et al.(2019)]{2019A&A...623A.112D} Damiani, F., Prisinzano, L., Pillitteri, I., et al.\ 2019, \aap, 623, A112. doi:10.1051/0004-6361/201833994
\bibitem[David et al.(2019)]{2019ApJ...872..161D} David, T.~J., Hillenbrand, L.~A., Gillen, E., et al.\ 2019, \apj, 872, 161. doi:10.3847/1538-4357/aafe09
\bibitem[de Zeeuw et al.(1999)]{1999AJ....117..354D} de Zeeuw, P.~T., Hoogerwerf, R., de Bruijne, J.~H.~J., et al.\ 1999, \aj, 117, 354. doi:10.1086/300682
\bibitem[DENIS team(1999)]{deniscat} DENIS team, 1999, DENIS catalog, IPAC, doi:10.26131/IRSA478
\bibitem[Desidera et al.(2015)]{2015A&A...573A.126D} Desidera, S., Covino, E., Messina, S., et al.\ 2015, \aap, 573, A126. doi:10.1051/0004-6361/201323168
\bibitem[Distefano et al.(2016)]{2016A&A...591A..43D} Distefano, E., Lanzafame, A.~C., Lanza, A.~F., et al.\ 2016, \aap, 591, A43. doi:10.1051/0004-6361/201527698
\bibitem[Donati et al.(2012)]{2012MNRAS.425.2948D} Donati, J.-F., Gregory, S.~G., Alencar, S.~H.~P., et al.\ 2012, \mnras, 425, 2948. doi:10.1111/j.1365-2966.2012.21482.x
\bibitem[Drake et al.(2017)]{catalina} Drake, A.~J., Djorgovski, S.~G., Catelan, M., et al.\ 2017, \mnras, 469, 3688. doi:10.1093/mnras/stx1085
\bibitem[Eisenhardt et al.(2020)]{catwise} Eisenhardt, P.~R.~M., Marocco, F., Fowler, J.~W., et al.\ 2020, \apjs, 247, 69. doi:10.3847/1538-4365/ab7f2a
\bibitem[Epchtein et al.(1999)]{denis} Epchtein, N., Deul, E., Derriere, S., et al.\ 1999a, \aap, 349, 236
\bibitem[Faherty et al.(2018)]{2018ApJ...863...91F} Faherty, J.~K., Bochanski, J.~J., Gagn{\'e}, J., et al.\ 2018, \apj, 863, 91. doi:10.3847/1538-4357/aac76e
\bibitem[Feinstein et al.(2019)]{eleanor} Feinstein, A.~D., Montet, B.~T., Foreman-Mackey, D., et al.\ 2019, \pasp, 131, 094502. doi:10.1088/1538-3873/ab291c
\bibitem[Fruth et al.(2013)]{2013AJ....146..136F} Fruth, T., Cabrera, J., Chini, R., et al.\ 2013, \aj, 146, 136. doi:10.1088/0004-6256/146/5/136
\bibitem[Gaia Collaboration et al.(2016a)]{dr1} Gaia Collaboration, Brown, A.~G.~A., Vallenari, A., et al.\ 2016a, \aap, 595, A2. doi:10.1051/0004-6361/201629512
\bibitem[Gaia Collaboration et al.(2016b)]{dr1data} Gaia Collaboration, Brown, A.~G.~A., Vallenari, A., et al.\ 2016b, Gaia Catalog DR1, IRSA, doi:10.26131/IRSA16
\bibitem[Gaia Collaboration et al.(2018a)]{dr2} Gaia Collaboration, Brown, A.~G.~A., Vallenari, A., et al.\ 2018a, \aap, 616, A1. doi:10.1051/0004-6361/201833051
\bibitem[Gaia Collaboration et al.(2018b)]{dr2data} Gaia Collaboration, Brown, A.~G.~A., Vallenari, A., et al.\ 2018b, Gaia Source Catalog DR2. doi:10.26131/IRSA12
\bibitem[Gaia Collaboration et al.(2021)]{edr3} Gaia Collaboration, Brown, A.~G.~A., Vallenari, A., et al.\ 2021, \aap, 649, A1. doi:10.1051/0004-6361/202039657
\bibitem[Galli et al.(2015)]{2015A&A...580A..26G} Galli, P.~A.~B., Bertout, C., Teixeira, R., et al.\ 2015, \aap, 580, A26. doi:10.1051/0004-6361/201525804
\bibitem[Galli et al.(2013)]{2013A&A...558A..77G} Galli, P.~A.~B., Bertout, C., Teixeira, R., et al.\ 2013, \aap, 558, A77. doi:10.1051/0004-6361/201220704
\bibitem[Ghosh et al.(1977)]{1977ApJ...217..578G} Ghosh, P., Lamb, F.~K., \& Pethick, C.~J.\ 1977, \apj, 217, 578. doi:10.1086/155606
\bibitem[Girard et al.(2011)]{spm} Girard, T.~M., van Altena, W.~F., Zacharias, N., et al.\ 2011, \aj, 142, 15. doi:10.1088/0004-6256/142/1/15
\bibitem[Godoy-Rivera et al.(2021)]{2021ApJS..257...46G} Godoy-Rivera, D., Pinsonneault, M.~H., \& Rebull, L.~M.\ 2021, \apjs, 257, 46. doi:10.3847/1538-4365/ac2058
\bibitem[Goldman et al.(2018)]{2018ApJ...868...32G} Goldman, B., R{\"o}ser, S., Schilbach, E., et al.\ 2018, \apj, 868, 32. doi:10.3847/1538-4357/aae64c
\bibitem[G{\"u}nther et al.(2022)]{2022AJ....163..144G} G{\"u}nther, M.~N., Berardo, D.~A., Ducrot, E., et al.\ 2022, \aj, 163, 144. doi:10.3847/1538-3881/ac503c
\bibitem[Henden et al.(2016)]{apass} Henden, A.~A., Templeton, M., Terrell, D., et al.\ 2016, VizieR Online Data Catalog, II/336 %http://adsabs.harvard.edu/abs/2016yCat.2336....0H
\bibitem[Howell et al.(2014)]{k2} Howell, S.~B., Sobeck, C., Haas, M., et al.\ 2014, \pasp, 126, 398. doi:10.1086/676406
\bibitem[Huang et al.(2020a)]{qlp1}Huang, C.~X., Vanderburg, A., P{\'a}l, A., et al.\ 2020, Research Notes of the American Astronomical Society, 4, 204. doi:10.3847/2515-5172/abca2e
\bibitem[Huang et al.(2020b)]{qlp2}Huang, C.~X., Vanderburg, A., P{\'a}l, A., et al.\ 2020, Research Notes of the American Astronomical Society, 4, 206. doi:10.3847/2515-5172/abca2d
\bibitem[Indebetouw et al.(2005)]{red} Indebetouw, R., Mathis, J.~S., Babler, B.~L., et al.\ 2005, \apj, 619, 931. doi:10.1086/426679
\bibitem[Jang-Condell et al.(2015)]{2015ApJ...808..167J} Jang-Condell, H., Chen, C.~H., Mittal, T., et al.\ 2015, \apj, 808, 167. doi:10.1088/0004-637X/808/2/167
\bibitem[Jayasinghe et al.(2018)]{asassn} Jayasinghe, T., Kochanek, C.~S., Stanek, K.~Z., et al.\ 2018, \mnras, 477, 3145. doi:10.1093/mnras/sty838
\bibitem[Joy(1945)]{1945ApJ...102..168J} Joy, A.~H.\ 1945, \apj, 102, 168. doi:10.1086/144749
\bibitem[Kapteyn(1914)]{1914ApJ....40...43K} Kapteyn, J.~C.\ 1914, \apj, 40, 43. doi:10.1086/142098
\bibitem[Kerr et al.(2021)]{2021ApJ...917...23K} Kerr, R.~M.~P., Rizzuto, A.~C., Kraus, A.~L., et al.\ 2021, \apj, 917, 23. doi:10.3847/1538-4357/ac0251
\bibitem[Kiraga(2012)]{2012AcA....62...67K} Kiraga, M.\ 2012, \actaa, 62, 67 %https://ui.adsabs.harvard.edu/abs/2012AcA....62...67K/abstract
\bibitem[Koenigl(1991)]{1991ApJ...370L..39K} Koenigl, A.\ 1991, \apjl, 370, L39. doi:10.1086/185972
\bibitem[K{\'o}sp{\'a}l et al.(2014)]{2014A&A...561A..61K} K{\'o}sp{\'a}l, {\'A}., Mohler-Fischer, M., Sicilia-Aguilar, A., et al.\ 2014, \aap, 561, A61. doi:10.1051/0004-6361/201322428
\bibitem[Kounkel \& Covey(2019)]{2019AJ....158..122K} Kounkel, M. \& Covey, K.\ 2019, \aj, 158, 122. doi:10.3847/1538-3881/ab339a
\bibitem[Kraus \& Hillenbrand(2009)]{2009ApJ...704..531K} Kraus, A.~L. \& Hillenbrand, L.~A.\ 2009, \apj, 704, 531. doi:10.1088/0004-637X/704/1/531
\bibitem[Kraus et al.(2017)]{2017ApJ...838..150K} Kraus, A.~L., Herczeg, G.~J., Rizzuto, A.~C., et al.\ 2017, \apj, 838, 150. doi:10.3847/1538-4357/aa62a0
\bibitem[Lasker et al.(2008)]{gsc?} Lasker, B.~M., Lattanzi, M.~G., McLean, B.~J., et al.\ 2008, \aj, 136, 735. doi:10.1088/0004-6256/136/2/735
\bibitem[Liu et al.(2020)]{2020AJ....159..105L} Liu, J., Fang, M., \& Liu, C.\ 2020, \aj, 159, 105. doi:10.3847/1538-3881/ab6b22
\bibitem[Luhman(2022)]{2022AJ....163...25L} Luhman, K.~L.\ 2022, \aj, 163, 25. doi:10.3847/1538-3881/ac35e3
\bibitem[Luhman(2022)]{2022AJ....163...24L} Luhman, K.~L.\ 2022, \aj, 163, 24. doi:10.3847/1538-3881/ac35e2
\bibitem[Luhman \& Mamajek(2012)]{2012ApJ...758...31L} Luhman, K.~L. \& Mamajek, E.~E.\ 2012, \apj, 758, 31. doi:10.1088/0004-637X/758/1/31
\bibitem[Marton et al.(2017)]{hpdppacs} Marton, G., Calzoletti, L., Perez Garcia, A.~M., et al.\ 2017, arXiv:1705.05693 %http://irsa.ipac.caltech.edu/data/Herschel/PPSC/docs/HPPSC\_Explanatory\_Supplement.pdf
\bibitem[Mathis(1990)]{1990ARA&A..28...37M} Mathis, J.~S.\ 1990, \araa, 28, 37. doi:10.1146/annurev.aa.28.090190.000345
\bibitem[Meisner et al.(2019)]{unwise} Meisner, A.~M., Lang, D., Schlafly, E.~F., et al.\ 2019, \pasp, 131, 124504. doi:10.1088/1538-3873/ab3df4
\bibitem[Mellon et al.(2017)]{wasp} Mellon, S.~N., Mamajek, E.~E., Oberst, T.~E., et al.\ 2017, \apj, 844, 66. doi:10.3847/1538-4357/aa77fb
\bibitem[Messina et al.(2010)]{2010A&A...520A..15M} Messina, S., Desidera, S., Turatto, M., et al.\ 2010, \aap, 520, A15. doi:10.1051/0004-6361/200913644
\bibitem[Messina et al.(2011)]{2011A&A...532A..10M} Messina, S., Desidera, S., Lanzafame, A.~C., et al.\ 2011, \aap, 532, A10. doi:10.1051/0004-6361/201016116
\bibitem[Meyer et al.(1997)]{ttlocus} Meyer, M.~R., Calvet, N., \& Hillenbrand, L.~A.\ 1997, \aj, 114, 288. doi:10.1086/118474
\bibitem[Meyer et al.(2006)]{feps} Meyer, M.~R., Hillenbrand, L.~A., Backman, D., et al.\ 2006, \pasp, 118, 1690. doi:10.1086/510099
\bibitem[Mittal et al.(2015)]{2015ApJ...798...87M} Mittal, T., Chen, C.~H., Jang-Condell, H., et al.\ 2015, \apj, 798, 87. doi:10.1088/0004-637X/798/2/87
\bibitem[Moolekamp et al.(2019)]{2019MNRAS.484.5049M} Moolekamp, F.~E., Mamajek, E.~E., James, D.~J., et al.\ 2019, \mnras, 484, 5049. doi:10.1093/mnras/stz183
\bibitem[Murakami et al.(2007)]{akari} Murakami, H., Baba, H., Barthel, P., et al.\ 2007, \pasj, 59, S369. doi:10.1093/pasj/59.sp2.S369, 
\bibitem[Nicholson et al.(2018)]{2018MNRAS.480.1754N} Nicholson, B.~A., Hussain, G.~A.~J., Donati, J.-F., et al.\ 2018, \mnras, 480, 1754. doi:10.1093/mnras/sty1965
\bibitem[Oh et al.(2018)]{2018ApJ...854..138O} Oh, S., Price-Whelan, A.~M., Brewer, J.~M., et al.\ 2018, \apj, 854, 138. doi:10.3847/1538-4357/aaab4d
\bibitem[Pecaut et al.(2012)]{2012ApJ...746..154P} Pecaut, M.~J., Mamajek, E.~E., \& Bubar, E.~J.\ 2012, \apj, 746, 154. doi:10.1088/0004-637X/746/2/154
\bibitem[Pecaut \& Mamajek(2013)]{2013ApJS..208....9P} Pecaut, M.~J. \& Mamajek, E.~E.\ 2013, \apjs, 208, 9. doi:10.1088/0067-0049/208/1/9
\bibitem[Pecaut \& Mamajek(2016)]{2016MNRAS.461..794P} Pecaut, M.~J. \& Mamajek, E.~E.\ 2016, \mnras, 461, 794. doi:10.1093/mnras/stw1300
\bibitem[Pilbratt et al.(2010)]{2010A&A...518L...1P} Pilbratt, G.~L., Riedinger, J.~R., Passvogel, T., et al.\ 2010, \aap, 518, L1. doi:10.1051/0004-6361/201014759
\bibitem[Popinchalk et al.(2021)]{2021ApJ...916...77P} Popinchalk, M., Faherty, J.~K., Kiman, R., et al.\ 2021, \apj, 916, 77. doi:10.3847/1538-4357/ac0444
\bibitem[Preibisch \& Smith(1997)]{1997A&A...322..825P} Preibisch, T. \& Smith, M.~D.\ 1997, \aap, 322, 825
\bibitem[Press et al.(1992)]{1992nrca.book.....P} Press, W.~H., Teukolsky, S.~A., Vetterling, W.~T., et al.\ 1992, Cambridge: University Press, |c1992, 2nd ed.
\bibitem[Rampalli et al.(2021)]{2021ApJ...921..167R} Rampalli, R., Ag{\"u}eros, M.~A., Curtis, J.~L., et al.\ 2021, \apj, 921, 167. doi:10.3847/1538-4357/ac0c1e
\bibitem[Rebull(2001)]{thesis} Rebull, L.~M.\ 2001, \aj, 121, 1676. doi:10.1086/319393
\bibitem[Rebull et al.(2004)]{allrot} Rebull, L.~M., Wolff, S.~C., \& Strom, S.~E.\ 2004, \aj, 127, 1029. doi:10.1086/380931
\bibitem[Rebull et al.(2006)]{orionirac} Rebull, L.~M., Stauffer, J.~R., Megeath, S.~T., et al.\ 2006, \apj, 646, 297. doi:10.1086/504865
\bibitem[Rebull et al.(2016a)]{paperI} Rebull, L.~M., Stauffer, J.~R., Bouvier, J., et al.\ 2016a, \aj, 152, 113. doi:10.3847/0004-6256/152/5/113
\bibitem[Rebull et al.(2016b)]{paperII} Rebull, L.~M., Stauffer, J.~R., Bouvier, J., et al.\ 2016b, \aj, 152, 114. doi:10.3847/0004-6256/152/5/114
\bibitem[Rebull et al.(2017)]{prae} Rebull, L.~M., Stauffer, J.~R., Hillenbrand, L.~A., et al.\ 2017, \apj, 839, 92. doi:10.3847/1538-4357/aa6aa4
\bibitem[Rebull et al.(2018)]{usco} Rebull, L.~M., Stauffer, J.~R., Cody, A.~M., et al.\ 2018, \aj, 155, 196. doi:10.3847/1538-3881/aab605
\bibitem[Rebull et al.(2020)]{taurus}Rebull, L.~M., Stauffer, J.~R., Cody, A.~M., et al.\ 2020, \aj, 159, 273. doi:10.3847/1538-3881/ab893c
\bibitem[Rebull et al.(2021)]{tauruserratum} Rebull, L.~M., Stauffer, J.~R., Cody, A.~M., et al.\ 2021, \aj, 162, 172. doi:10.3847/1538-3881/ac205f
\bibitem[Ribas et al.(2014)]{2014A&A...561A..54R} Ribas, {\'A}., Mer{\'\i}n, B., Bouy, H., et al.\ 2014, \aap, 561, A54. doi:10.1051/0004-6361/201322597
\bibitem[Ribas et al.(2015)]{2015A&A...576A..52R} Ribas, {\'A}., Bouy, H., \& Mer{\'\i}n, B.\ 2015, \aap, 576, A52. doi:10.1051/0004-6361/201424846
\bibitem[Ricker et al.(2015)]{tess} Ricker, G.~R., Winn, J.~N., Vanderspek, R., et al.\ 2015, Journal of Astronomical Telescopes, Instruments, and Systems, 1, 014003. doi:10.1117/1.JATIS.1.1.014003
\bibitem[Ripepi et al.(2015)]{2015MNRAS.454.2606R} Ripepi, V., Balona, L., Catanzaro, G., et al.\ 2015, \mnras, 454, 2606. doi:10.1093/mnras/stv2221
\bibitem[Ripepi et al.(2019)]{2019A&A...625A..14R} Ripepi, V., Molinaro, R., Musella, I., et al.\ 2019, \aap, 625, A14. doi:10.1051/0004-6361/201834506
\bibitem[Samus' et al.(2017)]{2017ARep...61...80S} Samus', N.~N., Kazarovets, E.~V., Durlevich, O.~V., et al.\ 2017, Astronomy Reports, 61, 80. doi:10.1134/S1063772917010085
\bibitem[Scholz et al.(2015)]{2015ApJ...809L..29S} Scholz, A., Kostov, V., Jayawardhana, R., et al.\ 2015, \apjl, 809, L29. doi:10.1088/2041-8205/809/2/L29
\bibitem[Scholz et al.(2018)]{2018ApJ...859..153S} Scholz, A., Moore, K., Jayawardhana, R., et al.\ 2018, \apj, 859, 153. doi:10.3847/1538-4357/aabfbe
\bibitem[Scargle(1982)]{1982ApJ...263..835S} Scargle, J.~D.\ 1982, \apj, 263, 835. doi:10.1086/160554
\bibitem[Siwak et al.(2016)]{2016MNRAS.456.3972S} Siwak, M., Ogloza, W., Rucinski, S.~M., et al.\ 2016, \mnras, 456, 3972. doi:10.1093/mnras/stv2848
\bibitem[Skiff(2014)]{2014yCat....102023S} Skiff, B.~A.\ 2014, VizieR Online Data Catalog, B/mk %https://ui.adsabs.harvard.edu/abs/2014yCat....1.2023S
\bibitem[Skrutskie et al.(2006)]{2mass} Skrutskie, M.~F., Cutri, R.~M., Stiening, R., et al.\ 2006, \aj, 131, 1163. doi:10.1086/498708 
\bibitem[Skrutskie et al.(2003)]{2massdata}Skrutskie, M.~F., Cutri, R.~M., Stiening, R., et al.\ 2003, 2MASS All-Sky Point Source Catalog, IPAC, doi:10.26131/IRSA2
\bibitem[Stauffer \& Hartmann(1987)]{1987ApJ...318..337S} Stauffer, J.~R. \& Hartmann, L.~W.\ 1987, \apj, 318, 337. doi:10.1086/165371
\bibitem[Stauffer et al.(1989)]{1989ApJ...346..160S} Stauffer, J.~R., Hartmann, L.~W., \& Jones, B.~F.\ 1989, \apj, 346, 160. doi:10.1086/167996
\bibitem[Stauffer et al.(2014)]{csi14}Stauffer, J., Cody, A.~M., Baglin, A., et al.\ 2014, \aj, 147, 83. doi:10.1088/0004-6256/147/4/83
\bibitem[Stauffer et al.(2015)]{csi15}Stauffer, J., Cody, A.~M., McGinnis, P., et al.\ 2015, \aj, 149, 130. doi:10.1088/0004-6256/149/4/130
\bibitem[Stauffer et al.(2016a)]{csi16}Stauffer, J., Cody, A.~M., Rebull, L., et al.\ 2016, \aj, 151, 60. doi:10.3847/0004-6256/151/3/60
\bibitem[Stauffer et al.(2016b)]{paperIII}Stauffer, J., Rebull, L., Bouvier, J., et al.\ 2016, \aj, 152, 115. doi:10.3847/0004-6256/152/5/115
\bibitem[Stauffer et al.(2017)]{batwing1}Stauffer, J., Collier Cameron, A., Jardine, M., et al.\ 2017, \aj, 153, 152. doi:10.3847/1538-3881/aa5eb9
\bibitem[Stauffer et al.(2018a)]{batwing2}Stauffer, J., Rebull, L., David, T.~J., et al.\ 2018, \aj, 155, 63. doi:10.3847/1538-3881/aaa19d
\bibitem[Stauffer et al.(2018b)]{binaries}Stauffer, J., Rebull, L.~M., Cody, A.~M., et al.\ 2018, \aj, 156, 275. doi:10.3847/1538-3881/aae9ec
\bibitem[Stassun et al.(2018)]{stassun18}Stassun, K.~G., Oelkers, R.~J., Pepper, J., et al.\ 2018, \aj, 156, 102. doi:10.3847/1538-3881/aad050
\bibitem[Stassun et al.(2019)]{stassun19}Stassun, K.~G., Oelkers, R.~J., Paegert, M., et al.\ 2019, \aj, 158, 138. doi:10.3847/1538-3881/ab3467
\bibitem[Strassmeier et al.(2005)]{2005A&A...440.1105S}Strassmeier, K.~G., Rice, J.~B., Ritter, A., et al.\ 2005, \aap, 440, 1105. doi:10.1051/0004-6361:20052901
\bibitem[Tajiri et al.(2020)]{2020ApJS..251...18T} Tajiri, T., Kawahara, H., Aizawa, M., et al.\ 2020, \apjs, 251, 18. doi:10.3847/1538-4365/abbc17
\bibitem[unWISE team(2019)]{unwisedata} unWISE team, 2019, unWISE catalog, IRSA, doi: 10.26131/IRSA525
\bibitem[Werner et al.(2004)]{mainspitzer} Werner, M.~W., Roellig, T.~L., Low, F.~J., et al.\ 2004, \apjs, 154, 1. doi:10.1086/422992
\bibitem[Wright et al.(2010)]{wise} Wright, E.~L., Eisenhardt, P.~R.~M., Mainzer, A.~K., et al.\ 2010, \aj, 140, 1868. doi:10.1088/0004-6256/140/6/1868
\bibitem[Wright et al.(2010)]{wisedata} Wright, E.~L., Eisenhardt, P.~R.~M., Mainzer, A.~K., et al.\ 2010, AllWISE Source Catalog, doi:10.26131/IRSA1
\bibitem[Wright \& Mamajek(2018)]{2018MNRAS.476..381W} Wright, N.~J. \& Mamajek, E.~E.\ 2018, \mnras, 476, 381. doi:10.1093/mnras/sty207
\bibitem[Zacharias et al.(2004)]{nomad} Zacharias, N., Monet, D.~G., Levine, S.~E., et al.\ 2004, \aas %https://ui.adsabs.harvard.edu/abs/2004AAS...205.4815Z/abstract
\bibitem[Zari et al.(2018)]{2018A&A...620A.172Z} Zari, E., Hashemi, H., Brown, A.~G.~A., et al.\ 2018, \aap, 620, A172. doi:10.1051/0004-6361/201834150
\bibitem[Zhan et al.(2019)]{2019ApJ...876..127Z} Zhan, Z., G{\"u}nther, M.~N., Rappaport, S., et al.\ 2019, \apj, 876, 127. doi:10.3847/1538-4357/ab158c
\bibitem[Z{\'u}{\~n}iga-Fern{\'a}ndez et al.(2021)]{2021A&A...645A..30Z} Z{\'u}{\~n}iga-Fern{\'a}ndez, S., Bayo, A., Elliott, P., et al.\ 2021, \aap, 645, A30. doi:10.1051/0004-6361/202037830
\end{thebibliography}
\end{document}